\documentclass{aa}  

\usepackage{graphicx}
\usepackage{txfonts}
\usepackage{ulem}
\usepackage{color}
\usepackage{xcolor}

\begin{document}

   \title{Astrochemical models of interstellar ices: History matters}


   \author{A. Cl\'ement\inst{1,2}, A. Taillard\inst{1}, V. Wakelam\inst{1}\thanks{valentine.wakelam@u-bordeaux.fr}, P. Gratier\inst{1}, J.-C. Loison\inst{3}, E. Dartois\inst{4}, F. Dulieu\inst{5}, J. A. Noble\inst{6}, \and M. Chabot\inst{7}
          }

   \institute{Laboratoire d'Astrophysique de Bordeaux (LAB), Univ. Bordeaux, CNRS, B18N, all\'ee Geoffroy Saint-Hilaire, 33615 Pessac, France
             \and
             Universit\'e Bordeaux, CNRS, LP2I Bordeaux, UMR 5797, F-33170 Gradignan, France
         \and
           Institut des Sciences Mol\'eculaires (ISM), CNRS, Univ. Bordeaux, 351 cours de la Lib\'eration, 33400, Talence, France
           \and
           Institut des sciences Mol\'eculaires d'Orsay, CNRS, Universit\'e Paris-Saclay, Bat 520, Rue Andr\'e Rivi\`ere, 91405 Orsay, France
         \and
         CY Cergy Paris Universit\'e, Observatoire de Paris, PSL Research University, Sorbonne Universit\'e, CNRS, LERMA, F-95000, Cergy, France
         \and 
             Physique des Interactions Ioniques et Mol\'eculaires, CNRS, Aix Marseille Univ., 13397 Marseille, France
     \and 
             Laboratoire des deux infinis Ir\`ene Joliot Curie (IJClab), CNRS-IN2P3, Universit\'e Paris-Saclay, 91405 Orsay, France
             }

   \date{Received xxxx; accepted xxxx}

 
  \abstract
  {Ice is ubiquitous in the interstellar medium.  As soon as it becomes slightly opaque in the visible, it can be seen for visual extinctions (A$_{\rm V}$) above $\sim$1.5. The James Webb Space Telescope (JWST) will observe the ice composition toward hundreds of lines of sight, covering a broad range of physical conditions in these extinct regions.}
  {We model the formation of the main constituents of interstellar ices, including H$_2$O, CO$_2$, CO, and CH$_3$OH. We strive to understand what physical or chemical parameters influence the final composition of the ice and how they benchmark to what has already been observed, with the aim of applying these models to the preparation and analysis of JWST observations. }  
{We used the Nautilus gas-grain model, which computes the gas and ice composition as a function of time for a set of physical conditions, starting from an initial gas phase composition. All important processes (gas-phase reactions, gas-grain interactions, and grain surface processes) are included and solved with the rate equation approximation.}
{We first ran an astrochemical code for fixed conditions of temperature and density mapped in the cold core L429-C to benchmark the chemistry. One key parameter was revealed to be the dust temperature. When the dust temperature is higher than 12~K, CO$_2$ will form efficiently at the expense of H$_2$O, while at temperatures below 12~K, it will not form. Whatever hypothesis we assumed for the chemistry (within realistic conditions), the static simulations failed to reproduce the observed trends of interstellar ices in our target core. In a second step, we simulated the chemical evolution of parcels of gas undergoing different physical and chemical situations throughout the molecular cloud evolution and starting a few $10^7$~yr prior to the core formation (dynamical simulations). We obtained a large sample of possible ice compositions. The ratio of the different ice components seems to be approximately constant for A$_{\rm V}$>5, and in good agreement with the observations. Interestingly, we find that grain temperature and low A$_{\rm V}$ conditions significantly affect the production of ice, especially for CO$_2$, which shows the highest variability.
}
   {Our dynamical simulations satisfactorily reproduce the main trends already observed for interstellar ices. Moreover, we predict that the apparent constant ratio of CO$_2$/H$_2$O observed to date is probably not true for regions of low A$_{\rm V}$, and that the history of the evolution of clouds plays an essential role, even prior to their formation.}

\titlerunning{Astrochemical models of interstellar ices: history matters}
   \authorrunning{Cl\'ement et al.}

   \maketitle

\section{Introduction}

Interstellar grains are key ingredients for the formation of
molecules in space, providing a catalytic surface. Molecular hydrogen, the most abundant molecule by far, is known to form
exclusively on dust grains \citep[see][for a review and references therein]{2017MolAs...9....1W}. With H$_2$ having a low binding energy to the grain surface, only a very small fraction of the formed molecules stays on the grains and the vast majority
returns into the gas-phase. However, this is not the case for other
species. As the density increases in the cold and shielded environments of star forming regions, atoms and molecules formed
in the gas-phase (such as CO) are depleted from the gas and stick
to the surface of the grains. Some reactions take place on these
surfaces, leading to the formation of more or less complex species. Grain surfaces are very often cited as being of high importance because
complex organic molecules are expected to form there \citep{2009ARA&A..47..427H}. However, even before considering the formation of complex molecules, the chemistry of the main constituents of the ices (H$_2$O, CO$_2$, CO, and CH$_3$OH) is still challenging for astrochemical models. Observations of interstellar ices in the infrared have shown that they are mostly composed of water. In addition, CO$_2$ is also detected toward all lines of sight, with the exception of peculiar circumstellar environments such as around OH/IR stars, where pure water ice is observed \citep[see][and references therein]{2015ARA&A..53..541B}. Other molecules, such as CO, CH$_3$OH, CH$_4$, NH$_3$, and H$_2$CO have also been identified in various amounts (between a few percent to a few tens of percent with respect to H$_2$O), depending on the observed environment. In cold cores, water ice is observed for visual extinction higher than a threshold A$_{\rm V}$ of approximately 1.5 (i.e., half of the observed visual extinction threshold, considering only one side of the cloud), while CO ice is seen for A$_{\rm V}$ higher than 3 and CH$_3$OH for A$_{\rm V}$ higher than 9. Finally, CO$_2$ ice is observed with a threshold A$_{\rm V}$ similar to H$_2$O over all lines of sights.

 In interstellar clouds, the formation of water ice is easily explained by chemical models as the hydrogenation of atomic oxygen is fast (no activation barrier) and requires only the diffusion of atomic hydrogen (very mobile) on the grain surfaces at very low temperatures (T$\sim$ 10~K). The formation processes of NH$_3$ and CH$_4$ are similar to those H$_2$O (easily formed on the surfaces) but chemical models form fewer of these species because the sticking of atomic N and C is in competition with their fast reactivity in the gas-phase to form CO and N$_2$ \citep{2012PNAS..10910233D}. Specifically, CO is formed in the gas-phase and sticks to the surfaces. Although the formation of methanol from CO ices is challenging because of activation barriers for some of the hydrogenation steps and the existence of some dehydrogenation channels, chemical models can form large amounts of methanol ice \citep{2007A&A...467.1103G} even without the inclusion of photochemical or radiolysis formation pathways. Since its formation is
not efficient at low temperatures, CO$_2$ remains the most problematic icy molecule \citep{2001MNRAS.324.1054R,2011ApJ...735...15G,2017ApJ...842...33V}.

While current constraints on chemical models are based on a relatively small number of observations, data from the {\it James Webb Space Telescope (JWST)} will almost certainly call into question our current understanding. The majority of ice observations to date have been performed along single, pre-identified lines of sight. As such, our current picture of ice composition and evolution is limited to a reliance on deriving trends by comparing this small number of lines of sight \citep{2015ARA&A..53..541B}. Where multiple lines of sight in a single object have been surveyed \citep{2000ApJS..128..603M,2004A&A...426..925P,2011ApJ...729...92B,2013ApJ...775...85N,2018A&A...610A...9G} -- and, more specifically, where the lines of sight have not been selected in advance (i.e., when using slitless spectroscopy), local deviations from global trends in ice composition have been observed, at scales down to a few hundred au \citep{2017MNRAS.467.4753N}. 
In addition to the much higher sensitivity of this telescope as compared to previous satellites or ground based observations, JWST will observe the ice composition toward hundreds of line of sight, covering a large range of physical conditions of the interstellar medium, as proposed, for instance, by the IceAge project \citep{2017jwst.prop.1309M}. The first results of ice observations with JWST toward three lines of sight using the MIRI \citep{2022ApJ...941L..13Y} or a combination of MIRI, NIRSpec, and NIRCam \citep{2023NatAs.tmp...25M} instruments have already demonstrated their capacity to provide high spectral resolution, high-sensitivity spectra of ices in molecular clouds when observing lines toward embedded protostars and highly extincted background stars, respectively.

In this work, we conduct a theoretical study of the chemical ice composition in a large range of cold core physical conditions in order to understand the formation of the main ice constituents and also make predictions for the sensitivity of the ice composition to the physical conditions, by considering those along the most diffuse lines of sight as well as those in the cold core. We start from an initial composition -- gas-phase and mostly atomic -- and compute time-dependent chemistry for a set of physical parameters. In doing so, we show that more sophisticated time-dependent physical conditions that follow the formation of cold cores from the earliest stages are needed to explain the general features of observed ice. 
The chemical model used for the simulations is described in Section \ref{section-chem-model}. The static chemical simulations together with their results are given in Section \ref{static_models}. Section \ref{dynamic-models} presents the results of the dynamical simulations. The results of both sets of simulations are first compared in Section \ref{comp_models}. In Sections \ref{CO2_chemistry} and \ref{influence_Tdust}, we discuss the chemical processes involved in the chemistry of CO$_2$ ices and the influence of the dust temperature. Section \ref{variability-section} presents our predictions on the ice composition as a function of visual extinction. In Section \ref{time-evolution}, we discuss the time-dependent formation of ice. We present our conclusions in the final section. 



\section{Chemical model}\label{section-chem-model}

To make predictions on the ice composition, we used the three-phase gas-grain model, Nautilus. This numerical model computes the gas and ice composition as a function of time for a set of physical conditions (e.g., gas and dust temperature, visual extinction, density, and cosmic-ray ionization rate) starting from an initial gas phase composition. The model is flexible enough to be able to use either static physical conditions, set at the beginning of the calculation, or time-dependent ones. All important processes (gas-phase reactions, gas-grain interactions, grain surface processes) are included and solved with the rate equation approximation. Gas-phase reactions are described in \citet{2015ApJS..217...20W}. Species from the gas-phase can physisorb at the surface of dust grains with an energy that depends on the binding energy of surface species (determined for water ice surfaces). They can desorb because of the dust temperature \citep{1992ApJS...82..167H}, whole grain heating induced by cosmic-rays \citep[following][]{1993MNRAS.261...83H}, impact of UV photons \citep[photodesorption,][]{2016MNRAS.459.3756R}, exothermicity of surface reactions \citep[chemical desorption,][]{2016A&A...585A..24M}, and sputtering by cosmic-rays \citep[following][]{2021A&A...652A..63W}. Detailed descriptions of the non-thermal desorption processes included in Nautilus can be found in \citet{2021A&A...652A..63W}. The Nautilus model is used in its three-phase version, which means that the molecules at the surface of the grains are divided into two separate phases. The first phase is composed of the first few monolayers of species on top of the grains (four in our case; see below for a more detailed explanation), while the rest of the molecules below these surface layers represents the bulk of the ice. The refractory parts of the grains (below the bulk) is chemically inactive. Both the surface and the bulk are chemically active. Photodissociation by direct UV photons and secondary photons induced by cosmic-rays are efficient in both phases. The diffusion (and thus reactivity) of the surface is higher than in the bulk while the species on the surface can desorb into the gas phase, contrary to the species in the bulk. Only sputtering by cosmic-rays can directly desorb species from the bulk ice. The relevant equations and chemical processes are described in \citet{2015ApJS..217...20W}, \citet{2016MNRAS.459.3756R}, and \citet{2021A&A...652A..63W}. 

With respect to the version used in \citet{2021A&A...652A..63W}, three modifications to the code have been made. First, we implemented an automatic switch for the chemical desorption, depending on the water grain surface coverage. Since we are using the formalism of  \citet{2016A&A...585A..24M} for the chemical desorption, two prescriptions are proposed by the authors: one for bare grains and one for grains covered by water, with the first shown to be more efficient than the second. We then switch from the first prescription to the second one if the grains are covered by more than four monolayers of molecules. The choice of these few layers corresponds to the following experimental observations. On the one hand, in the measurable case of the O+O reaction, it has been shown that the effect of an N$_2$ pre-adsorbed layer makes the chemical desorption disappear once two monolayers are adsorbed \citep{Minissale2014}. On the other hand, the thickness of the active chemical layer is also from one \citep{Congiu2020}  to a few monolayers in the case of water \citep{Ioppolo2010}. Four monolayers therefore seems to be a good compromise between these various situations. The second modification of the code is the implementation of a new module to compute the grain temperature using the approximation of \citet{2017A&A...604A..58H}, which is a function of visual extinction and local UV field. 
Last, for the cosmic-ray ionisation rate ($\zeta$), we used a prescription that depends on the visual extinction to take into account the attenuation with density. This prescription was determined by fitting the figure 6 of \citet{2017ApJ...845..163N}, which represents measurements of $\zeta$ through observed column densities of H$_2$ and H$_3^+$. This fit gives the following formula:
\begin{equation}
\rm \zeta (A_{\rm V}) = 10^{-0.7 \times \log_{10}(A_{\rm V})-15.6}.
\end{equation}
For A$_{\rm V}$ smaller than 0.5, we assume a constant attenuated rate of $\zeta (0.5)$ (about $4\times 10^{-16}$~s$^{-1}$). This formula is slightly different from the one used in \citet{2021A&A...652A..63W} to be more conservative at high visual extinction and closer to the model predictions of \citet{2022A&A...658A.189P}. Using the formula given in \citet{2021A&A...652A..63W}, $\zeta$ would be $1.8\times 10^{-17}$~s$^{-1}$ while it is $5\times 10^{-17}$~s$^{-1}$ with the new formula. 
For all the simulations, we start from an initial chemical composition as listed in Table~\ref{abun-ini} (in which all elements are atoms, except hydrogen, which is entirely molecular).  

\begin{table}
\caption{Initial abundances (with respect to the total proton density). }
\label{abun-ini}
\begin{tabular}{ccc}
\hline
Elements & Abundances   & Reference \\
\hline
H$_2$        &  0.5  & \\
He        &  $9.0\times 10^{-2}$ & 1\\
N         &  $6.2\times 10^{-5}$  & 2 \\
O         &  $2.4\times 10^{-4}$  & 3 \\
C$^+$        &  $1.7\times 10^{-4}$ & 2 \\
S$^+$         &  $1.5\times 10^{-5}$ & 4 \\ 
Si$^+$         & $1.8\times 10^{-6}$ & 4 \\
Fe$^+$        & $1.0\times 10^{-8}$ & 4 \\ 
Na$^+$         & $2.3\times 10^{-7}$ & 4 \\
Mg$^+$         & $2.3\times 10^{-6}$ & 4 \\
P$^+$          & $7.8\times 10^{-8}$ & 4 \\ 
Cl$^+$         & $1.0\times 10^{-9}$ & 4 \\ 
F           & $6.68\times 10^{-9}$ & 5 \\
\hline
\end{tabular}
\\
\\
References: (1) See discussion in \citet{2008ApJ...680..371W}; (2) \citet{2009ApJ...700.1299J}; (3) see discussion in \citet{2011A&A...530A..61H}; (4) low
metal abundance from \citet{1982ApJS...48..321G},
(5) depleted value from \citet{2005ApJ...628..260N}.
\end{table}

\section{Static simulations for L429-C}\label{static_models}

We run a first set of simulations using the physical conditions as observed in the cold core L429-C with no evolution of the physical conditions during the calculation of the chemistry. 

\subsection{Presentation of the L429-C cold core region}

L429-C is a quiescent cold core located in the  Aquila Rift \citep[$\sim$ 200 pc,][]{stutz_spitzer_2009}. There is no IR heating source inside. 
Herschel observations at 353~GHz from \citet{2018ApJ...852..102S} provided dust temperature and opacity maps of the region. Based on the opacity map, \citet{taillard2022} derived a H$_2$ column density and subsequently, using a methodology from \citet{2018A&A...610A..12B}, a volume H$_2$ density. We refer the reader to \citet{taillard2022} for details on the calculation and methods used. Figure~\ref{herschel_maps} in the appendix \ref{appendix_l429} shows the H$_2$ column density (cm$^{-2}$), the H$_2$ volume density (cm$^{-3}$), the visual extinction, and the dust temperature maps of the region. The maps are 200$\arcsec$ x 200$\arcsec$ size and contain 26x26 pixels\footnote{Note: the original maps from \citet{taillard2022} contained 79x79 pixels. We resampled them to compute the chemistry over fewer spatial points.}. The dust temperature as measured by Herschel ranges from 12~K to 18~K, while the H$_2$ density goes from $5\times 10^3$~cm$^{-3}$ to $3\times 10^6$~cm$^{-3}$ and the visual extinction from 5 to more than 75. In this region, \citet{2011ApJ...729...92B} estimated the column densities of H$_2$O, CO$_2$, and CH$_3$OH ices toward four background stars using {\it Spitzer} observations (see their table 6). 

\subsection{Model results}
\subsubsection{Using the observed dust temperature}

\begin{figure*}
\includegraphics[width=0.49\linewidth]{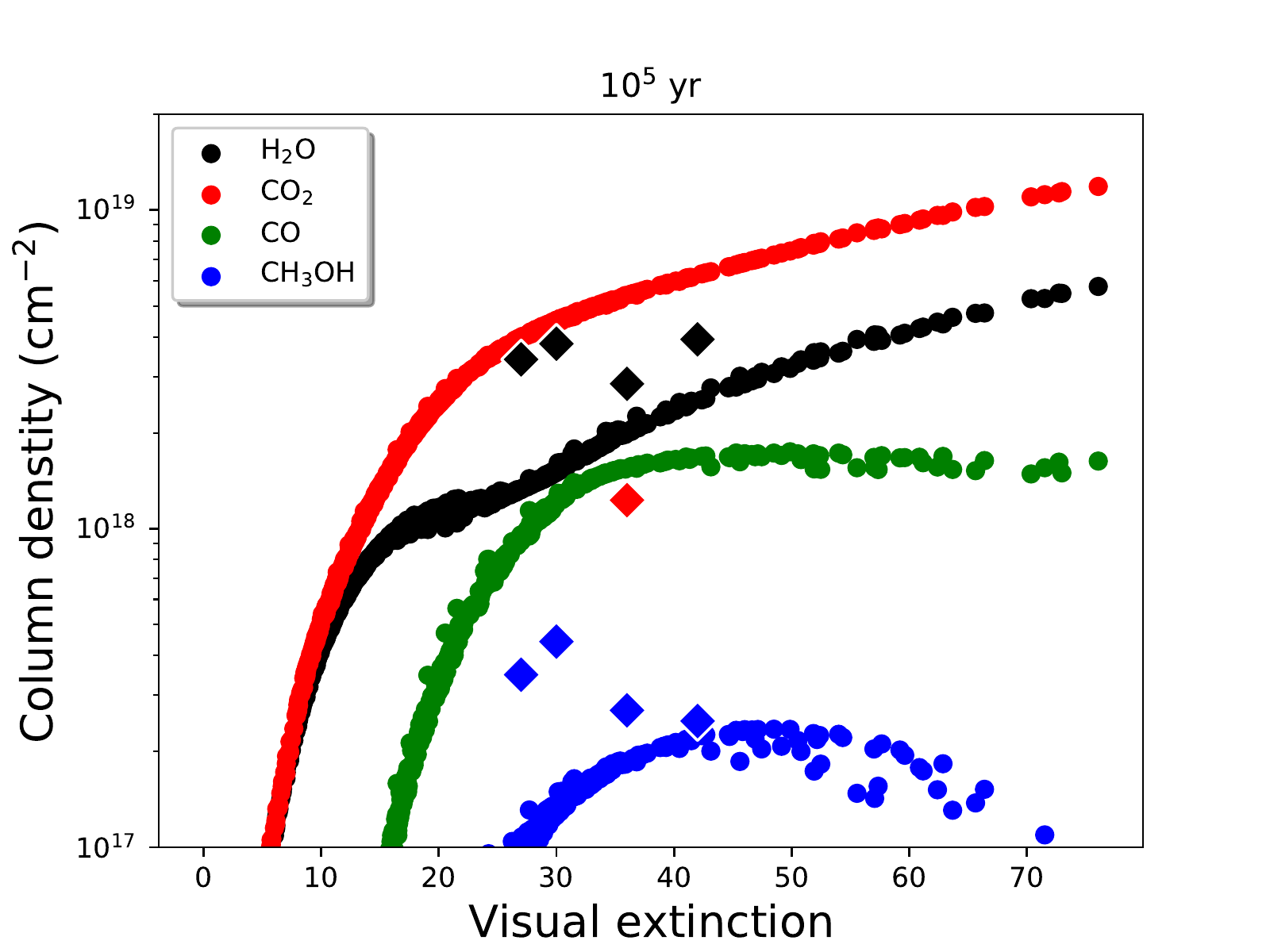}
\includegraphics[width=0.49\linewidth]{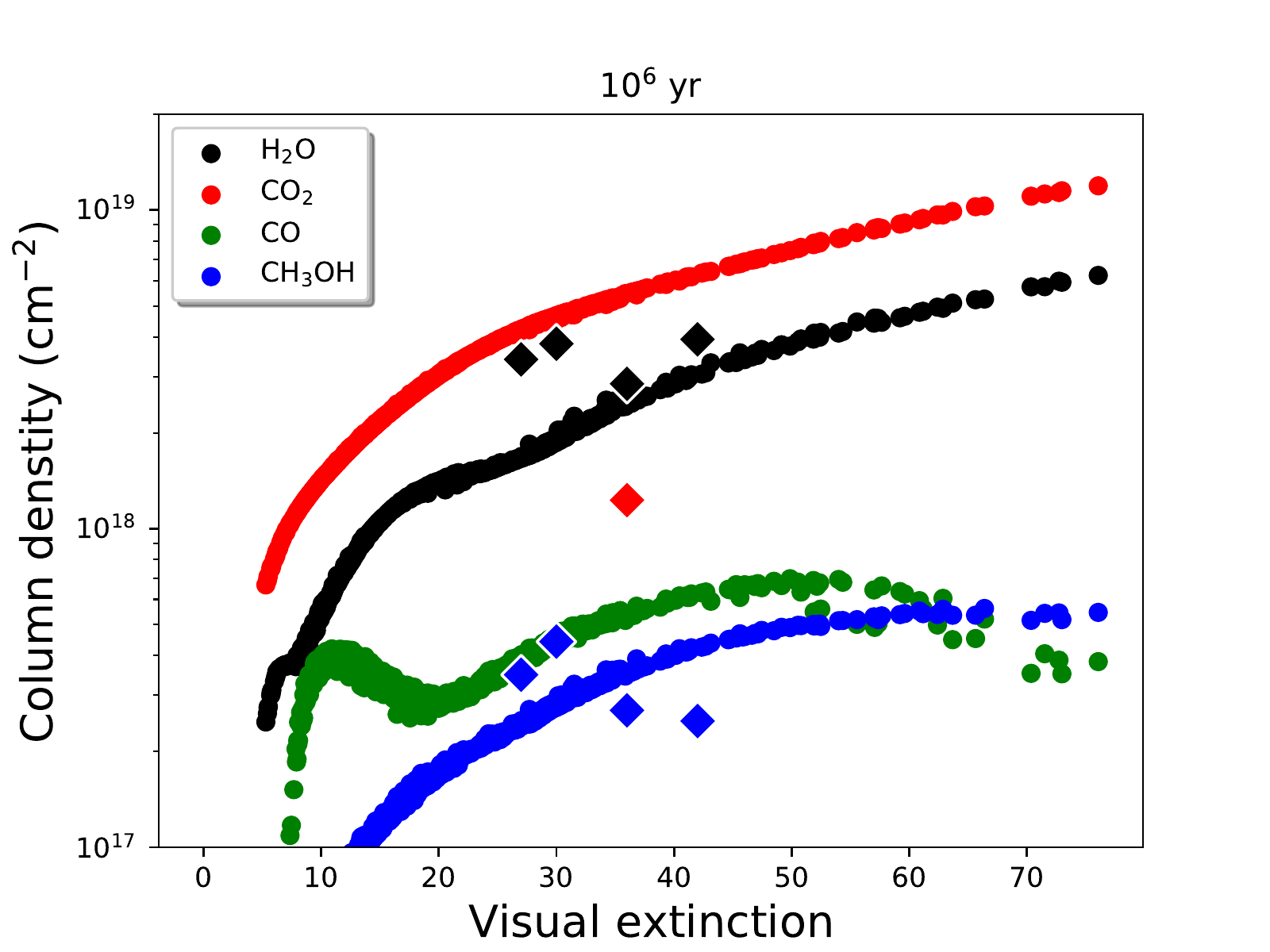}
\caption{Column density of the main ice constituents computed with Nautilus for the static physical conditions as observed in L429-C as a function of visual extinction (derived from Herschel observations). Diamonds represent the observed column densities by \citet{2011ApJ...729...92B} on specific positions of the cloud. The model result at two different times are shown: $10^5$~yr on the left and $10^6$~yr on the right. The dust temperature is equal to that observed by Herschel. \label{static_model1}}
\end{figure*}

For our first grid of chemical models, we have run Nautilus for each pixel of the maps shown in Fig.~\ref{herschel_maps}, using the observed local physical conditions: proton density, visual extinction, and temperature. In the absence of good estimates of the gas temperature, based for instance on the excitation conditions of molecules, we have set the gas temperature equal to the dust temperature measured by Herschel \citep[see also][]{taillard2022}. This represents a fair approximation at moderate to high A$_{\rm V}$. To compute the molecular column densities from the model abundances, we multiplied the modeled abundances by the observed H$_2$ column density at each position of the region, assuming that all the hydrogen was in the form of H$_2$. The surface and bulk abundances from our models are summed to obtain the total column densities of icy molecules. The model results, column density as a function of visual extinction, are shown in Fig.~\ref{static_model1} for two different times ($10^5$~yr and $10^6$~yr), that we can consider as ``early'' and ``late'' times for the main ice constituents observed in cold cores (H$_2$O, CO$_2$, CO, and CH$_3$OH). Superimposed on the figure, diamonds indicate the observed column densities of the molecules by \citet{2011ApJ...729...92B}. We note that CO ices were not targeted in the observational study and there is only one point of measurement for CO$_2$. Whatever the time, whenever the A$_{\rm V}$ is higher than about 10, using this static modeling approach, the CO$_2$ ices are much more abundant than H$_2$O -- which is not not in line with what has been observed \citep[see also][]{2015ARA&A..53..541B}. \\
In these simulations, as in most astrochemical simulations of cold cores, we start from atoms. This means that, as a function of time, there is a competition between the sticking of the gas-phase oxygen atoms onto the grains -- that will be hydrogenated to form water -- and their reactivity in the gas-phase to form CO. Once formed, CO will itself stick to the grains and can produce CO$_2$ if some atomic oxygen is still available, and either CO or O can diffuse. One key ingredient in these processes is the dust temperature. In these simulations, the dust temperature is that observed by Herschel and shown in Fig.~\ref{dust_temperature} as a function of visual extinction (see also Fig.~\ref{herschel_maps}). The values range from 12 to 18~K. For temperatures above 12~K, the amount of hydrogen on the surface is rather low, preventing the efficient hydrogenation of oxygen and favoring the formation of CO$_2$. The problem is of course not that simple because when looking at the early time in Fig~\ref{static_model1}, we can see that at A$_{\rm V}$ lower than 10, when the dust temperature is higher than 16~K, water is on the same order as CO$_2$. At these low Av, with higher UV irradiation and lower density, the formation of CO is slower than the sticking of atomic oxygen onto the grains. As a consequence, water formation can occur but, as time goes on, CO in the gas-phase is formed and CO$_2$ formation on the surface becomes more efficient, overtaking the water abundance. \\
It might be worth mentioning here that the representation of the results as column density as a function of visual extinction might be misleading. The increasing column density of H$_2$O and CO$_2$ with A$_{\rm V}$ is an effect of an increase of H$_2$ column density and not an increase of abundance of the species. In fact, the abundance of water and CO$_2$ reaches a maximum at an A$_{\rm V}$ of 10 (at $10^6$~yr) of $\sim 4\times 10^{-5}$ (with respect to H) for H$_2$O and $\sim 8\times 10^{-4}$ for CO$_2$. 

\subsubsection{Computing the dust temperature}

\begin{figure}
\includegraphics[width=1\linewidth]{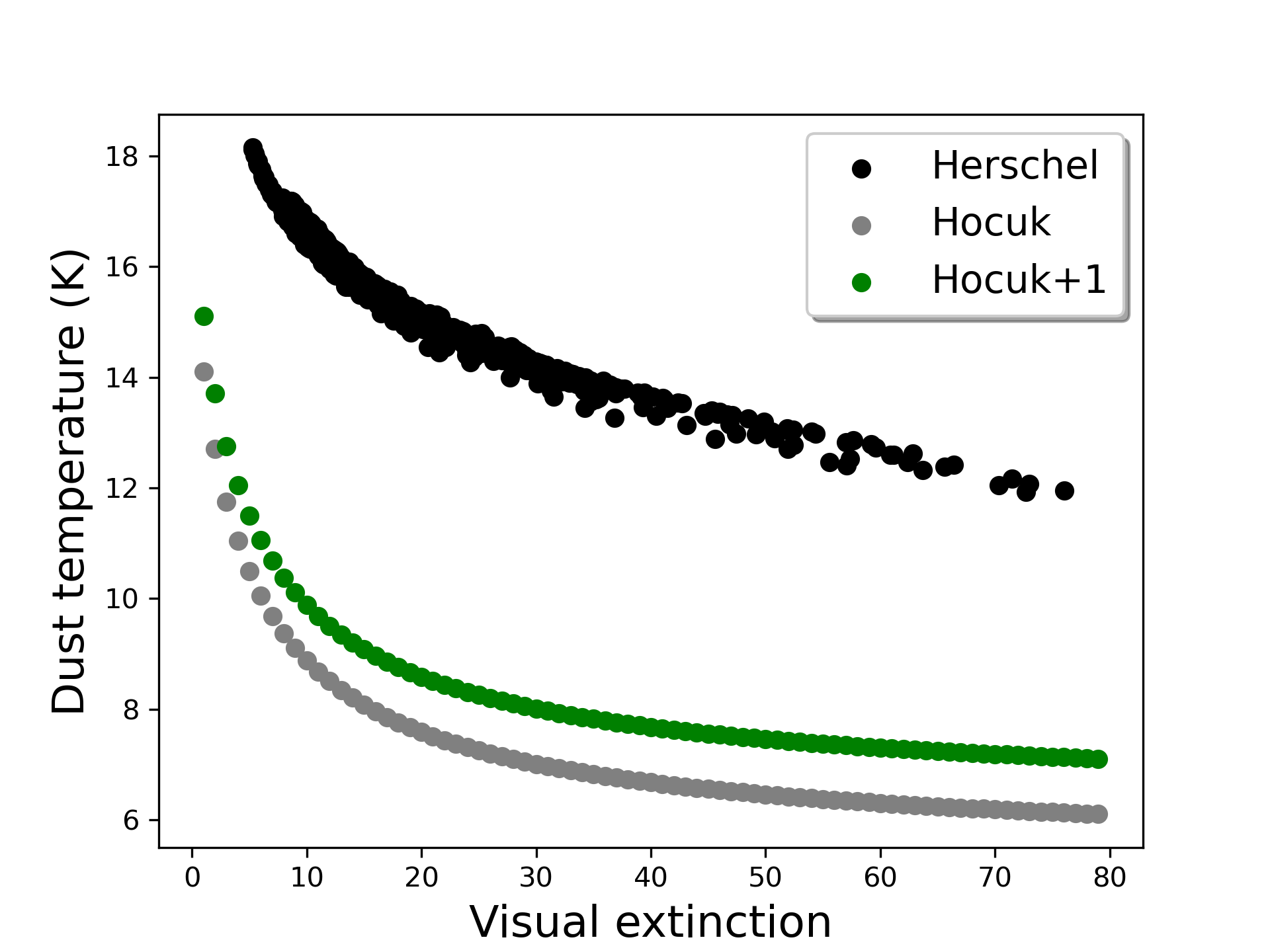}
\caption{Dust temperature as a function of visual extinction used for the models. 
In black: Dust temperature measured by Herschel. In gray: Dust temperature computed with \citet{2017A&A...604A..58H} approximation. In green: Dust temperature computed with \citet{2017A&A...604A..58H} approximation plus one Kelvin (``Hocuk+1''). \label{dust_temperature}}
\end{figure}

\begin{figure*}
\includegraphics[width=0.49\linewidth]{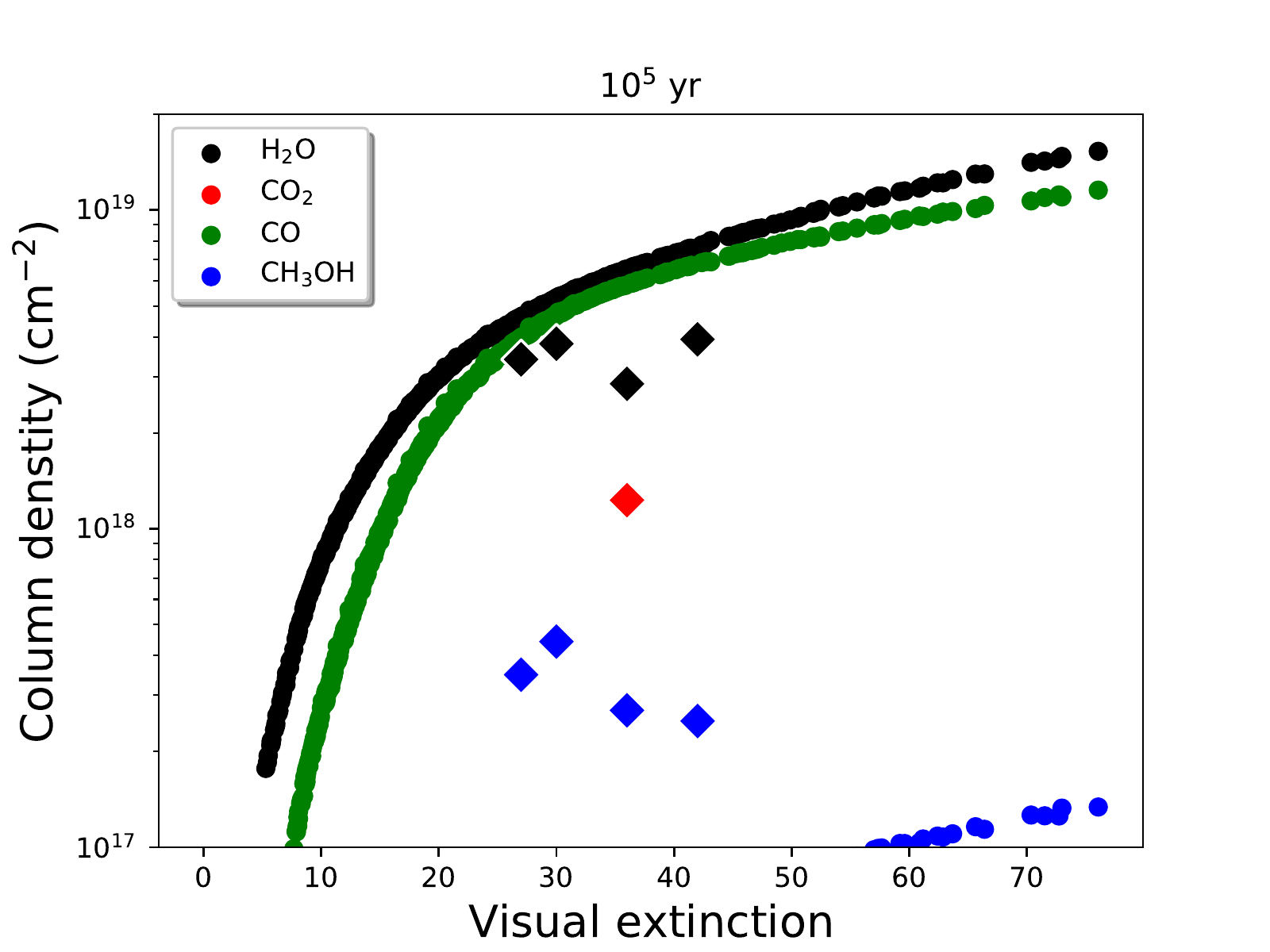}
\includegraphics[width=0.49\linewidth]{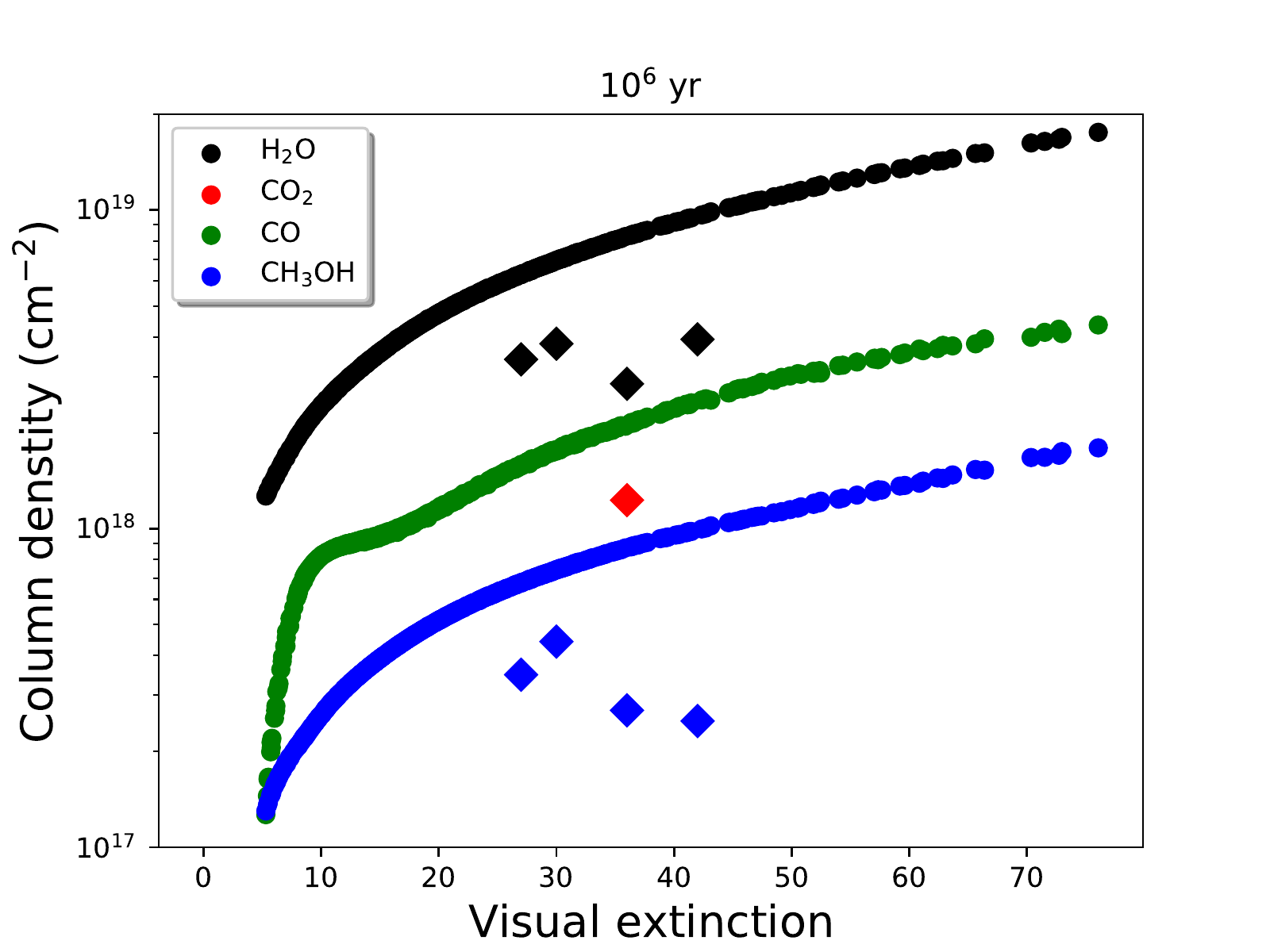}
\caption{Same as Fig.~\ref{static_model1} but for the models in which the dust temperature is computed with the \citet{2017A&A...604A..58H} approximation (rather than the temperature derived from Herschel observations). \label{static_model2}}
\end{figure*}

To test the impact of the adopted dust temperature on the ice composition, we ran the same set of models but computed the dust temperature using the prescription of \citet{2017A&A...604A..58H}, which is a function of visual extinction and local UV field. The temperature obtained with this prescription is compared to the measured one in Fig.~\ref{dust_temperature}. In the same figure, we show the dust temperature if one Kelvin is added to this prescription (called ``Hocuk+1'' in the rest of the paper, see discussion in Section~\ref{influence_Tdust}). Hocuk's prescription is proportional to the Draine UV
field strength \citep{1978ApJS...36..595D} to the power 1/5.9. In this work, we have used a value of one for this parameter in the absence of additional constraint. Our case Hocuk+1 can be obtained for 2 Draine UV field strength. For higher values, the dust temperature would also be shifted toward higher values. The predicted dust temperature also depends on the grains compositions. Hocuk's relation was obtained by assuming carbonaceous-silicate mixtures. At a high visual extinction (approximately larger than 10), grain growth or icy mantles introduce larger uncertainties in the computed dust temperature \citep{2017A&A...604A..58H}. 
For the entire range of visual extinctions considered here, the dust temperature always remains below 11~K using Hocuk's formula. For the gas temperature, in the absence of better estimates, we still used the temperature measured by Herschel. We note that within this range of values, the ice abundances will not be sensitive to this latter approximation. \\
Figure~\ref{static_model2} shows the results of this model at two different times. In this case, because of the low temperature, CO$_2$ is never efficiently produced. Water becomes the main ice constituent after $10^5$~yr while CO is transformed into CH$_3$OH with time. Again, with this modeling, we reproduce the observed general trends neither in L429-C nor in the other observed regions as summarized by \citet{2015ARA&A..53..541B}. While the abundances of water and (at late times) methanol seem overall to be well reproduced, the CO$_2$ abundance is highly underestimated and the CO abundance overestimated. The dust temperature seems to be a key ingredient in reproducing the observed ice composition at different visual extinctions.  \\
The choice of the initial chemical composition affects some of the results. First, the amount of hydrogen already converted into H$_2$ at the beginning of the simulations is an unknown, although observations show that the abundance of atomic hydrogen in dense regions should not be above a few $10^{-3}$. Starting the static simulations with this initial abundance of atomic hydrogen does not change the model results \citep[see also][for discussions]{2021A&A...652A..63W}. The CO molecule can also potentially form very early during the formation of the cold core \citep{2004ApJ...612..921B}. Starting with some CO already formed can change the ice composition, as CO would have already started sticking earlier on the grains. We redid our static simulations starting with half of the carbon in the form of CO (decreasing in the appropriate amount the initial atomic oxygen abundance). At $10^5$~yr, the CO ice column density is unchanged, but the water ice column density is slightly less at A$_{\rm V}$ larger than 10 (similar to CO). The methanol column density is the most impacted as it is already at the same level than at $10^6$~yr. At $10^6$~yr, both simulations (starting from some molecular CO or only atoms) give similar results.

\subsection{Comparison with previous model}\label{comp-models}
Using a different three-phase gas-grain model, \citet{2011ApJ...735...15G} also studied the formation of ices. With their model, they were able to produce larger quantities of CO$_2$ ices than we managed to do at 10~K. There are a number of differences between our model and theirs that can explain this difference. First, in their model, they form CO$_2$ through the surface reaction CO$_{\rm ice}$ + OH$_{\rm ice}$ $\rightarrow$ CO$_{\rm 2 ice}$ + H$_{\rm ice}$. However, this reaction has a moderate activation barrier \citep[which they assumed to be 80~K based on][]{2001MNRAS.324.1054R}, they also added a direct formation of CO$_2$ -- when an OH molecule is formed on top of a CO molecule on the surface. This mechanism can be seen as a two-step process. First, the formation of a van der Waals complex O...CO$_{\rm ice}$ is more likely via a direct landing of an oxygen atom on top of an already physisorbed CO molecule. The hydrogenation of this complex is without barrier and leads to the formation of CO$_2$ ice.  
We discuss  this process further in Section~\ref{CO2_chemistry}.
\citet{2011ApJ...735...15G} are likely to have used different diffusion rates of molecules on the surface. Even though lower diffusion results in a higher probability of reaction for reactions with activation barriers, it reduces the efficiency of encounter on a surface. These authors have lower binding energies for atomic oxygen \citep[800~K for them and 1660~K for us][]{2015ApJ...801..120H}, OH  \citep[2850~K for them and 4600~K for us,][]{2017MolAs...6...22W}, CO \citep[1150~K for them and 1300~K for us,][]{2017MolAs...6...22W}, and HCO \citep[1600~K for them and 2400~K for us,][]{2017MolAs...6...22W}. For the binding energy of atomic oxygen, the experiments carried out by \citet{2012MNRAS.425.1264W} and \citet{2016A&A...585A.146M} also found high values.
We note that the values used in our simulations are in good agreement with the recent review by \cite{Minissale2022}. For the ratio of the diffusion energy versus binding energy, we use 0.4 for the surface layer (which is four monolayers), while \citet{2011ApJ...735...15G} used a ratio that depends on the coverage of H$_2$. In our case, we do not have depletion of H$_2$ onto the grains because we include the encounter-desorption reaction from \citet{2015A&A...574A..24H}.  In our simulations, the key factor is the diffusion of atomic oxygen. As also pointed out by \citet{2011ApJ...735...15G}, the new numerical treatment of the competition between diffusion and reaction for chemical reactions with activation barriers \citep{2007A&A...469..973C,2011ApJ...735...15G} strongly diminishes the effect of the barriers. In our simulations, CO$_2$ formation is only restrained by the possibility of the reactants to move on the surface. The very low CO$_2$ ice that we obtain is only due to the very high binding energy of atomic oxygen, which in our case is based on recent calculations. Using the older value of 800~K instead of 1660~K, without changing anything in our code, we can form CO$_2$ as abundantly as CO at 10~K. With the high binding energy of 1660~K, whatever the test (changing the activation energy of the reaction CO$_{\rm ice}$ + OH$_{\rm ice}$ $\rightarrow$ CO$_{\rm 2 ice}$ + H$_{\rm ice}$; changing the other binding energies; using the O...CO$_{\rm ice}$ complex), we are not able to form CO$_2$ efficiently at 10~K. In our model, the reaction CO$_{\rm ice}$ + OH$_{\rm ice}$ $\rightarrow$ CO$_{\rm 2 ice}$ + H$_{\rm ice}$ specific channel is not very efficient at 10~K because OH is hydrogenated too efficiently.

\section{Dynamical simulations}\label{dynamic-models}

In the two sets of chemical models presented in the previous section, we used static physical conditions that represent a snapshot of an observed region. In reality, before forming this cold core, the interstellar matter has travelled through the Galaxy and the dust has experienced different physical conditions that impacted the gas-phase abundances of molecules found in molecular clouds \citep{2018A&A...611A..96R}. 

\subsection{Physical model}

To study the ice formation during the formation of cold cores from the diffuse medium, we have used the 3D Smoothed particle hydrodynamics (SPH) simulations from \citet{2013MNRAS.430.1790B}. These simulations compute the time dependent volume density and gas temperature of the interstellar matter in a galactic potential including spiral arms. This model thus provides a history of the physical conditions for cells of material in 3D that will form approximately twelve cold cores in a galactic arm. The time-dependent physical conditions for each of these cells (or trajectories) are then used in Nautilus to compute the time dependent chemistry as a post-process. With these simulations, we have already been able to study the impact of history on the gas-phase composition of cold cores \citep{2018A&A...611A..96R}, the non-detection of O$_2$ in cold cores \citep{2019MNRAS.486.4198W}, and the elemental depletion in dense regions \citep{2020MNRAS.497.2309W}. The visual extinction is not an output of the SPH model. To get an estimation of this parameter at each time step, first the total proton column density (N$_{\rm H}$) is computed by multiplying the volume density of the cell by its smoothing length \citep[see][]{2018A&A...611A..96R}. Then, A$_{\rm V}$ is computed by multiplying N$_{\rm H}$ by $5.34\times 10^{-22}$ \citep{1989MNRAS.237.1019W}.
The dust temperature is not computed by the SPH model. As such, we computed for each time step and each cell, the dust temperature using the approximation of \citet{2017A&A...604A..58H}. This dust temperature is a function of visual extinction and not consistent with the gas temperature computed by the SPH model. The initial conditions are the same as in the static models (Table~\ref{abun-ini}). The cosmic-ray ionisation rate is computed as a function of visual extinction as described in Section~\ref{section-chem-model}.   \\
In our simulations, we have twelve identified cold cores for which we have run the simulations \citep[see also ][]{2019MNRAS.486.4198W}. Some of them have only a few tens of cells while others have more than two hundred. 
In the next section, we first show the model result for one cloud as an example before going on to further discuss the diversity obtained for the other clouds.

\subsection{Model results:\ Core 0}\label{results_clump0}

\begin{figure}
\includegraphics[width=1\linewidth]{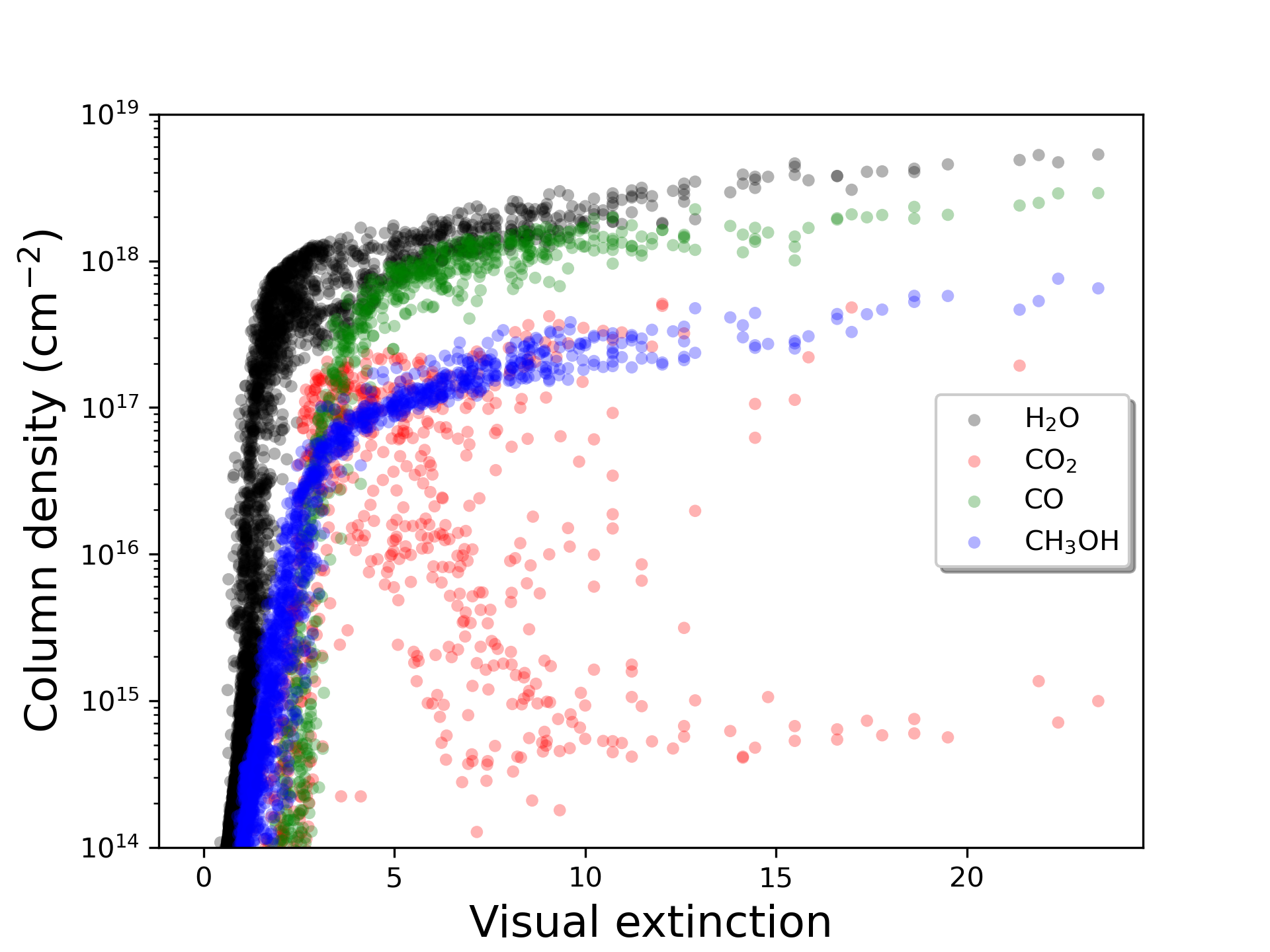}
\includegraphics[width=1\linewidth]{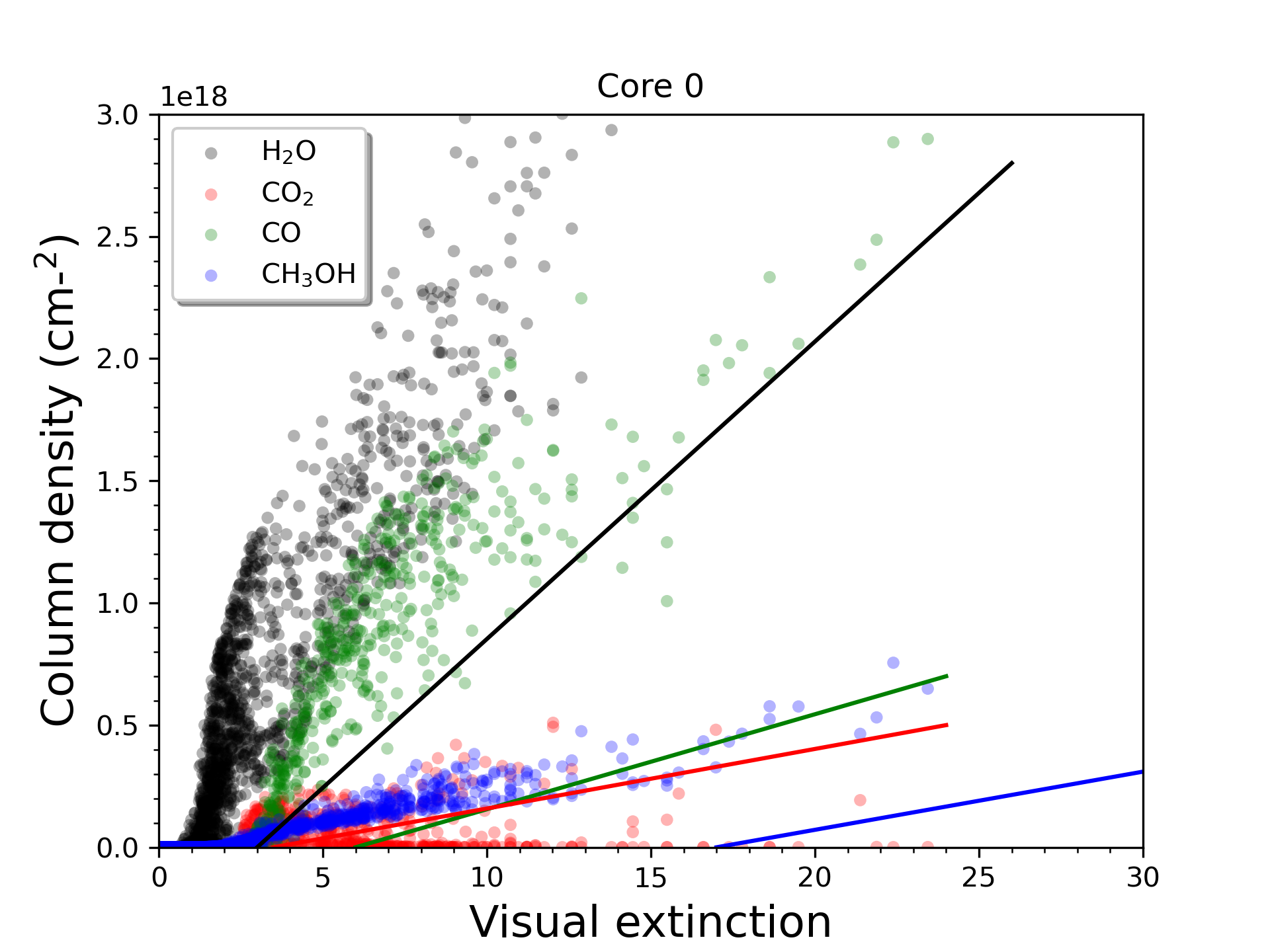}
\caption{Column densities of the main ice constituents computed with the dynamical model as a function of column density for core 0, shown in the upper panel. Lower panel shows the same figure but with an axis setting similar to Fig. 7 of \citet{2015ARA&A..53..541B}. The straight lines are the observed linear relations found by \citeauthor{2015ARA&A..53..541B}. \label{ice_cold_dens_clump0_all}}
\end{figure}

The model results for our core 0 is shown in Fig.~\ref{ice_cold_dens_clump0_all} (upper panel). In these simulations all times up to the formation of the cold core are considered. As such, the low A$_{\rm V}$ compositions represent both the edges of the cores but also the diffuse lines of sights that will form the cold cores, while the high A$_{\rm V}$ represent the center of the cores. The lower panel is the same figure but with axis setting similar to Fig. 7 of \citet{2015ARA&A..53..541B} for a better comparison with the observational trend. For a more direct comparison, we overplotted on the figure the linear relations of column density versus the A$_{\rm V}$ values found by \citeauthor{2015ARA&A..53..541B}.
We restrain the comparison with the observational trends to qualitative aspects rather quantitative ones for two reasons. First, the visual extinctions computed in the model may not reflect the exact observed physical conditions. Second, the observed linear relations highlighted by  \citeauthor{2015ARA&A..53..541B} may be biased by a lack of lines of sight and may depend on the observed cloud. 

Our first result is that with these dynamical simulations, we can reproduce the general characteristics of the ice observations.
Water dominates the compositions at all visual extinctions, while CO$_2$ is the next molecule to be formed on the grains, followed by CO and methanol. The CO molecule becomes more abundant than CO$_2$ when A$_{\rm V}$ is larger than approximately four while methanol remains low. However, the amount of CO$_2$ ice, with respect to water (in this example) is  still low compared to the observations. In such a comparison, the steepness of the column densities increase are much larger in our simulations meaning that the ices grow faster than in the observations. The visual extinctions at which the species column densities become larger than about $10^{17}$~cm$^{-2}$ (i.e., the  apparent observation threshold 
 of ice formation is not the same as in the observations). In our simulations, we seem to form large quantities of CO and CH$_3$OH ices at much smaller A$_{\rm V}$. The CH$_3$OH ice column density, for instance, becomes greater than $10^{17}$~cm$^{-2}$ for A$_{\rm V}$ values higher than about 4 in our model -- compared to 9 in the observations. The model results of the dynamical simulations do not depend on the initial chemical composition because the simulations start with physical conditions for a  diffuse medium and any molecule would be destroyed very quickly \citep{2020MNRAS.497.2309W}. \\

\begin{figure*}
\begin{center}
\includegraphics[width=0.47\linewidth]{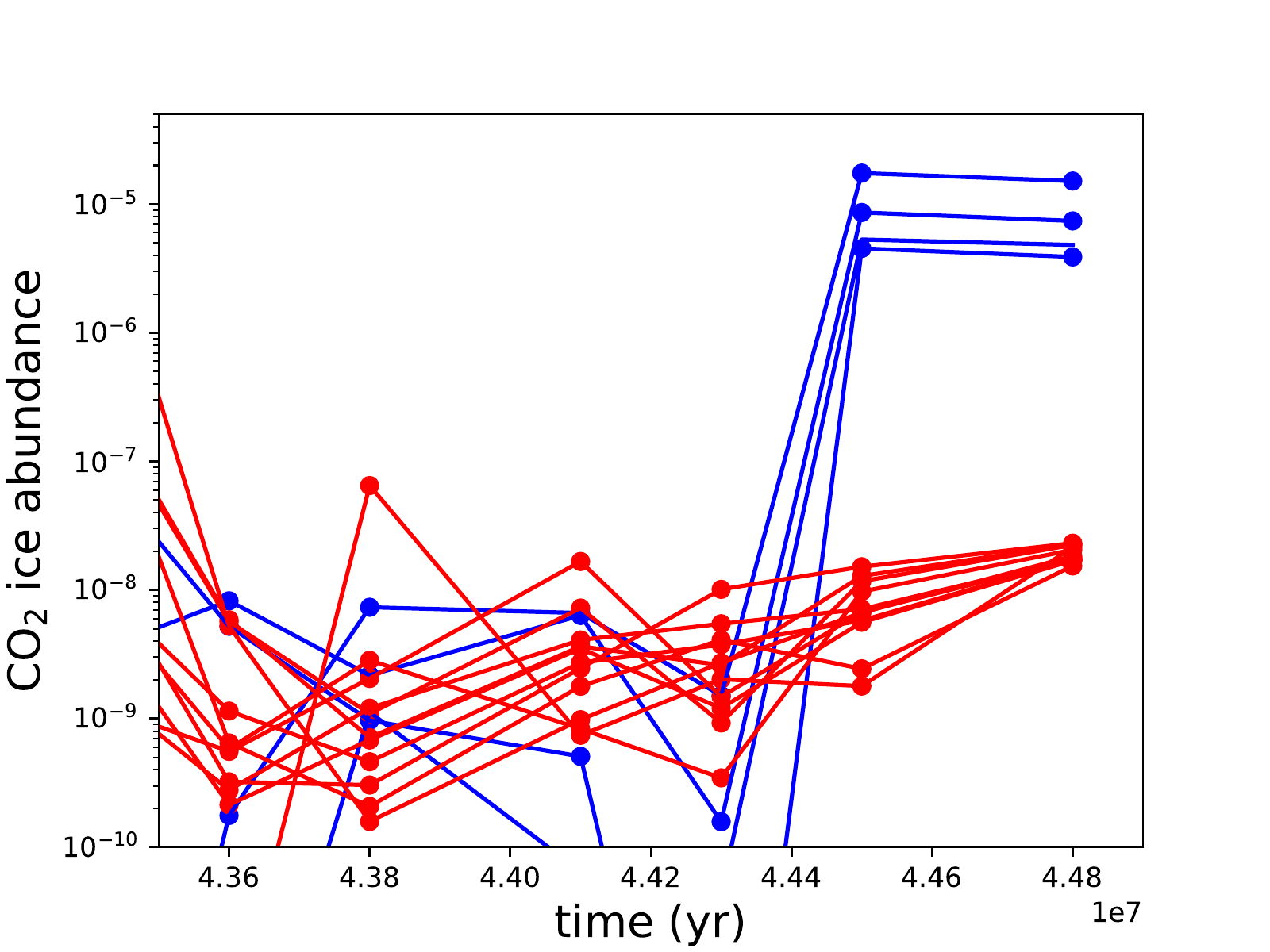}
\includegraphics[width=0.47\linewidth]{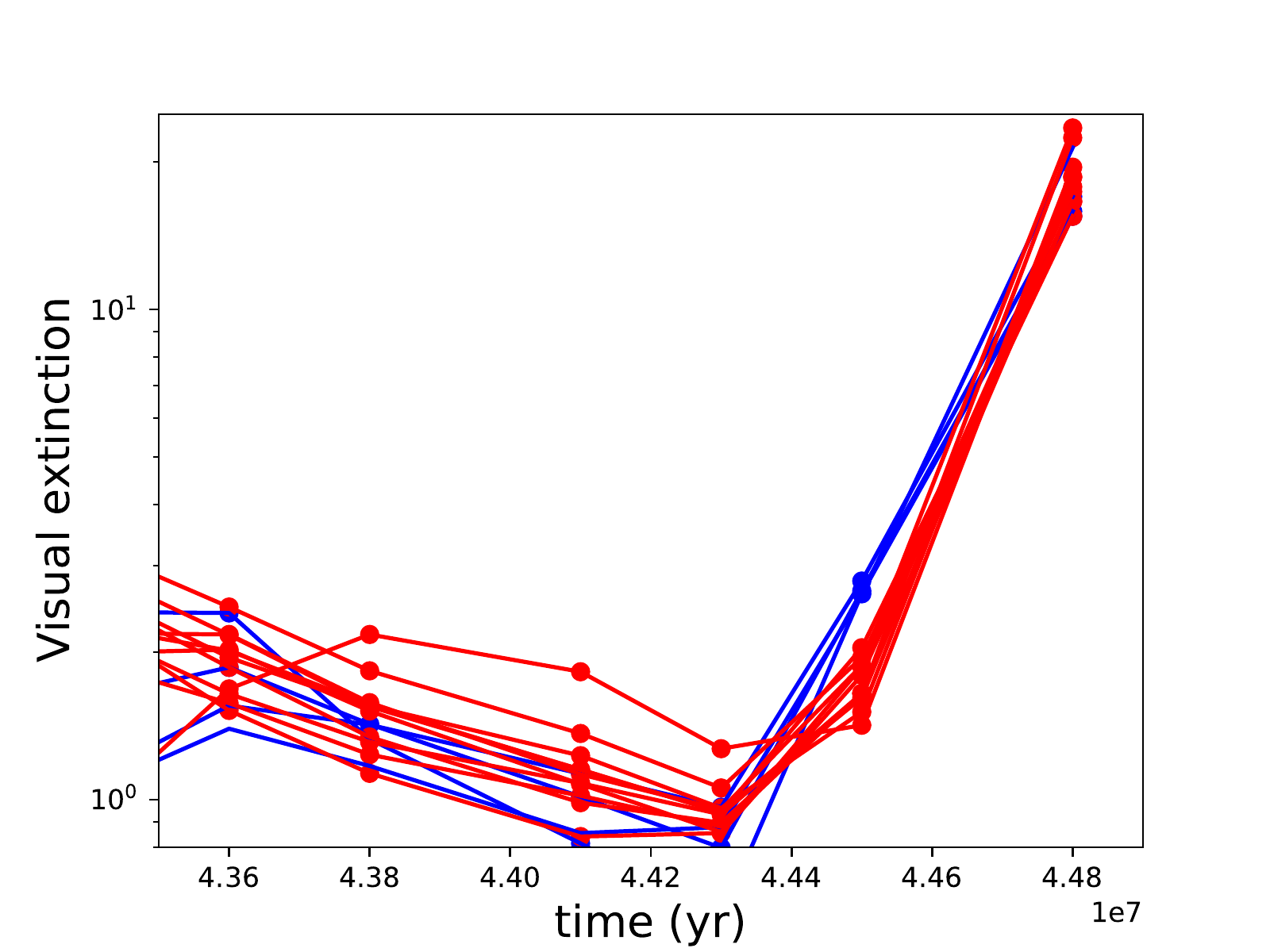}
\includegraphics[width=0.47\linewidth]{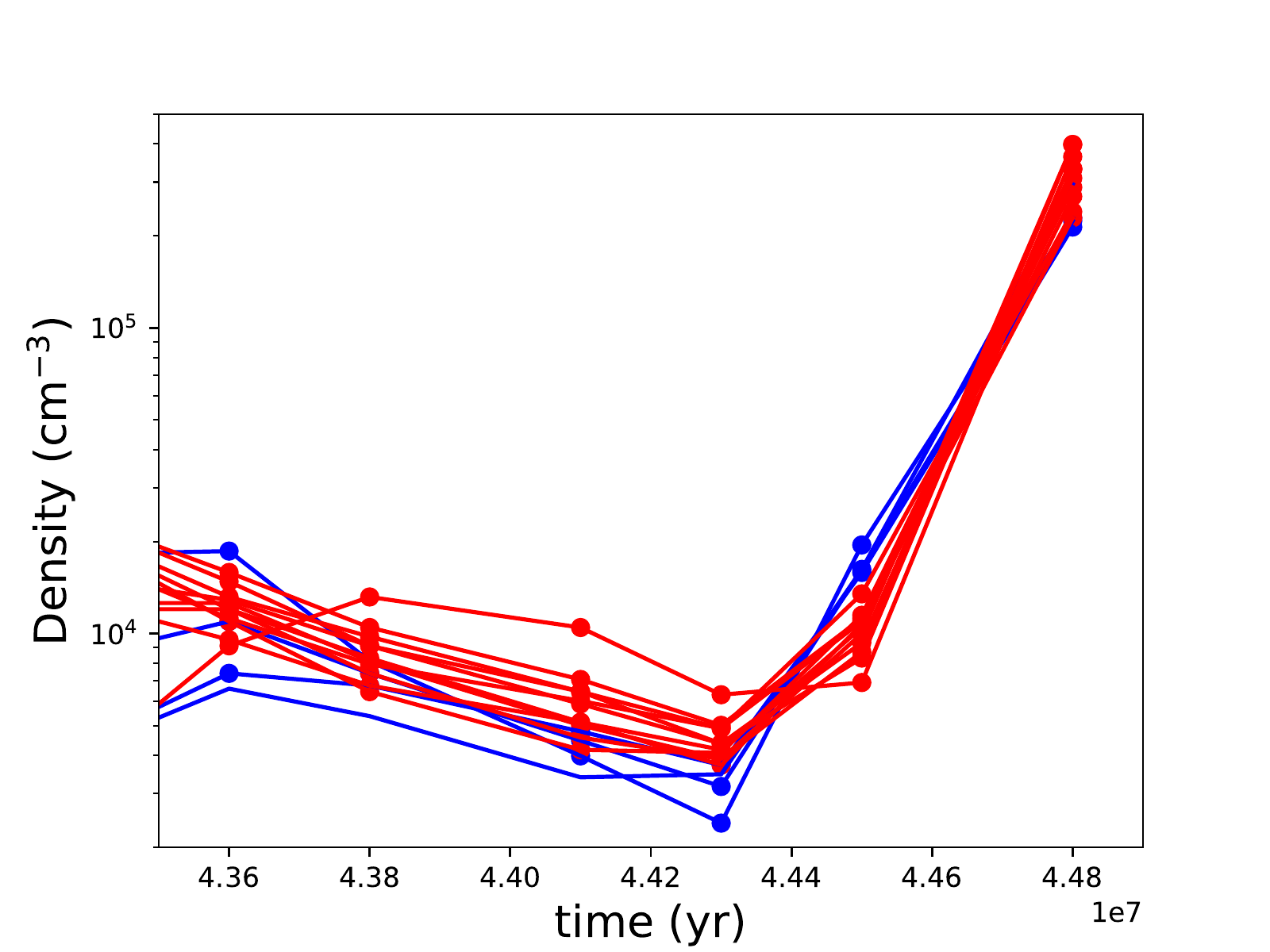}
\includegraphics[width=0.47\linewidth]{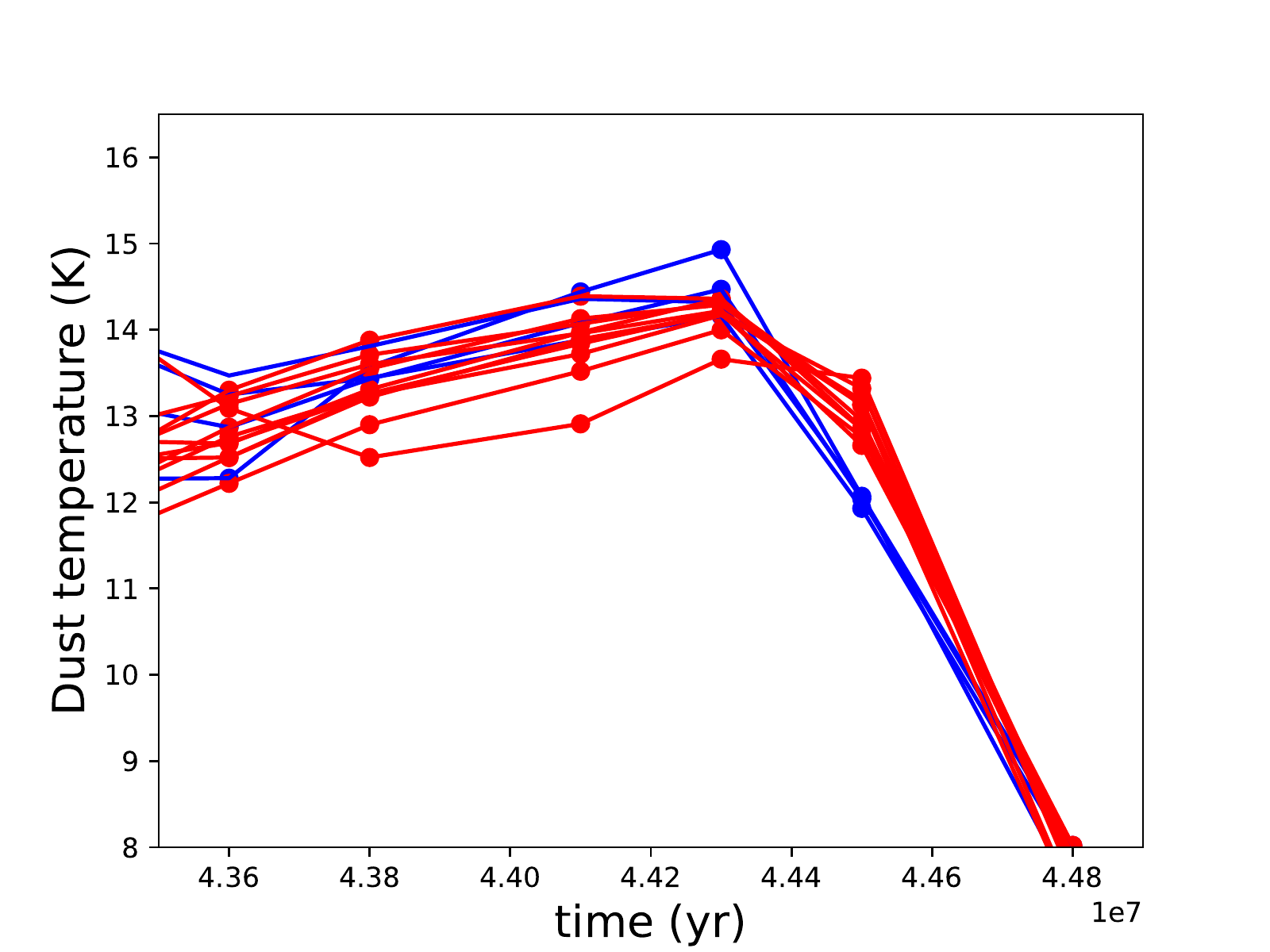}
\caption{CO$_2$ ice abundance (upper left panel), visual extinction (upper right panel), H$_2$ density (lower left panel), and dust temperature (lower right panel) as a function of time for a few cells. The blue curves present the results for the "high CO$_2$ column density cases" while the red ones are the "low CO$_2$ column density cases." \label{CO2_ice_time}}
\end{center}
\end{figure*}

One prediction of this model is the spread in ice compositions for one single core. In particular, many cells of material show a low abundance of CO$_2$ ice (with column densities below $10^{17}$~cm$^{-2}$ at A$_{\rm V}$ larger than 10). To understand the difference between the "high CO$_2$ column density cases" and the "low CO$_2$ column density cases," we looked at the chemical evolution, as well as the evolution of the physical conditions, of the individual computed trajectories. To this end, we first identified in our simulations the cells responsible for both cases and visualized them as a function of time. In Fig.~\ref{CO2_ice_time}, we show the CO$_2$ ice abundance as a function of time for the low (in red) and high (blue) CO$_2$ productions, as well as some of the physical parameters (dust temperature, density, and visual extinction). In these simulations, the formation of the core occurs at $4.48\times 10^7$ yr (where we have the maximum density and the lowest temperature). In the case of the cells where CO$_2$ ices are abundant, the formation of the molecule starts earlier (at $4.45\times 10^7$~yr). At that time, the blue models are slightly  denser, have a slightly higher visual extinction and so a slightly lower grain temperature. To understand which is the crucial parameter from a chemical point of view, we ran several tests taking the physical conditions of the blue curves at $4.45\times 10^7$~yr and replacing them one by one with the red ones. These tests showed that the key parameter is the visual extinction. All the red curves have a visual extinction lower than two while all the blue curves have a visual extinction around 2.6-2.8. Changing this parameter in the red curve leads to CO$_2$ abundances as high as the blue one, which survives at the next time step. The lower visual extinction produces a higher destruction of CO$_2$ ices. Here the dust temperature is larger than 12~K in both cases. This example illustrates the fact that CO$_2$ ices are built on the grains when the dust temperature is greater than 12~K and the visual extinction higher than approximately 2 (in our simulations). The ubiquity of CO$_2$ ices observed on interstellar grains seems to indicate that most interstellar material forming cold cores has experienced such conditions. However, this result is  model dependent. If a faster diffusion of atomic oxygen and/or a smaller activation energy is assumed for the reaction O + CO, then the formation of CO$_2$ could be more efficient at lower temperatures. 

\section{Static versus dynamical models}\label{comp_models}
The two sets of models (static and dynamical) present very different approaches. The static simulations are based on the observed conditions in a specific cloud (L429-C) and compared to the ice composition observed toward a few lines of sight in this region.
The dynamical simulations are based on time dependent physical conditions computed  using a SPH model and compared to observations gathered in the literature for various clouds. As such the physical conditions (density and gas temperature) have been computed along evolutionary paths leading to a variety of local density and
gas temperatures for a given visual extinction (see Fig.~\ref{physics-2simus}).
The observed spread at a single visual extinction is thus not linked to model uncertainty but related to the various possible chemical trajectories as discussed below. The densities in the dynamical simulations is higher than in the static models, while the gas temperature is smaller. We note that the gas temperature (within the ranges considered here) has little impact on the ice composition. For the static and dynamical simulations using Hocuk's approximation, the dust temperature is the same at a given A$_{\rm V}$. The range of visual extinctions scanned for both sets of simulations is not the same. The static models only probe visual extinctions larger than 5 (up to 76) while the dynamical simulations start at A$_{\rm V}$ of $3\times 10^{-4}$ (only up to 25). The main difference between the two types of simulations are however the time evolution of the physical conditions (non existent for the static models). For this reason, the times of both simulations do have the same meaning. To compare the column densities obtained from both sets of simulations, we plotted them in the same figure (Fig.~\ref{coldens-2simus}) for the common range of A$_{\rm V}$ (5 to 25). The two sets of models compared here have the same dust temperature as a function of A$_{\rm V}$ and the same initial composition. The results of the static model are shown for two times: $10^5$ and $10^6$~yr. The dynamical model is shown for the final time and core 0.  \\
The first obvious difference is the spread of molecular column densities obtained from the dynamical simulations for a single value of A$_{\rm V}$, while the static simulations all give the same column density. This spread is particularly important for CO$_2$. The late time ($10^6$~yr) static models seem to fall within the dynamical simulations for H$_2$O and CH$_3$OH. The static simulations always produce too small amounts of CO$_2$ ices (similar or even below some of the dynamical simulations) whatever the time considered. For CO ice, the static simulations produce smaller amounts except at high A$_{\rm V}$ ($> 15$) for which the static simulations are similar to the dynamic ones at late times. Similar conclusions can be drawn when comparing to the dynamical simulations for other cores. 

\section{Chemistry of CO$_2$}\label{CO2_chemistry}

In these simulations, we have not included the formation of CO$_2$ ices though energetic processes such as the irradiation of CO ices by photons, charged particles, or electrons \citep{1996A&A...312..289G,1998A&A...334..247P,2006ApJS..163..184J,2009A&A...493.1017I,2014ApJ...791L..21Y}. These processes are particularly efficient in relatively diffuse regions because of the high external vacuum ultraviolet (VUV) flux and so, if anything, they would strengthen our conclusion that CO$_2$ ices are already built before the formation of the cold cores.


In our chemical model, we have included the following production reactions of CO$_2$ ices:
\begin{align}
\rm HCO_{ice} + O_{ice} & \rm \rightarrow CO_{2 ice} + H_{ice}, & (50\%), \label{reaction1}\\
& \rm \rightarrow CO_{ice} + OH_{ice}, & (50\%), \label{reaction2}
\end{align} 
\begin{align} 
\rm CO_{ice} + O_{ice} \rightarrow CO_{2 ice},  \label{reaction3}
\end{align} 
\begin{align} 
\rm CO_{ice} + OH_{ice} & \rm \rightarrow CO_{2 ice} + H_{ice}, & (50\%), \label{reaction4} \\
 & \rm \rightarrow HOCO_{ice}, & (50\%), \label{reaction5}
\end{align} 
\begin{align} 
\rm H_2CO_{ice} + O_{ice} \rightarrow CO_{2 ice} + H_{2 ice}, \label{reactionH2CO}
\end{align} 
\begin{align} 
\rm O_{ice} + HOCO_{ice} \rightarrow CO_{2 ice}  + OH_{ice}, \label{reaction7}
\end{align} 
\begin{align} 
\rm H_{ice} + HOCO_{ice} & \rm \rightarrow CO_{2 ice}  + H_{2 ice}, & (70\%), \label{reaction8}\\
 & \rm \rightarrow CO_{ice} + H_2O_{ice}, & (20\%), \label{reaction9}\\
 & \rm \rightarrow HCOOH_{ice}, & (10\%). \label{reaction10}
\end{align} 

In astrochemical models with large networks, many chemical processes are in competition. At 10~K, reactions \ref{reaction1} and \ref{reaction2}, for instance, will end up up competing with the hydrogenation of HCO to form CH$_3$OH, which will be much faster than the reaction with atomic oxygen. Reaction \ref{reactionH2CO} \citep[studied by][]{2015A&A...577A...2M} is also in competition with the hydrogenation of H$_2$CO (much faster) leading to methanol. The primary formation of HOCO, necessary to reactions \ref{reaction7} to \ref{reaction10}, is slow. So the fastest pathways to form CO$_2$ ices will be reactions \ref{reaction3}-\ref{reaction5}. 
Reactions \ref{reaction4} and \ref{reaction5} should be more efficient than reaction \ref{reaction3} because they have a smaller activation barrier: 150~K for reactions \ref{reaction4} and \ref{reaction5} \citep{1996JChPh.105..983F} and 627~K for reaction \ref{reaction3} \citep[mean value of the barrier found by ][]{2013A&A...559A..49M}.
It should be noted that the comparison is difficult to make, as the barriers for reactions \ref{reaction4} and \ref{reaction5} have been measured only in the gas phase -- contrary to reaction  \ref{reaction3}. 
OH would however be hydrogenated to form H$_2$O much faster than it can react with CO. For this reason, our simulations do not form CO$_2$ ice when the dust temperature remains at 10~K but mostly H$_2$CO, CO, and CH$_3$OH ices, as shown in Fig.~\ref{static_model2}. 

Based on \citet{2010MNRAS.406.2213G}, \citet{2011ApJ...735...15G} tested the hypothesis that atomic oxygen, once absorbed on the dust surface, would form a special bond with CO$_{\rm ice}$ and produce a $\rm O...CO_{ice}$ complex. This complex would then be hydrogenated and quickly react to form CO$_{\rm 2 ice}$. This would mimic reaction~\ref{reaction3} with a much lower barrier. In this way, \citet{2011ApJ...735...15G} could form significant amounts of CO$_2$ ice. Our model includes such processes, as previously detailed in \citet{2015MNRAS.447.4004R}:
\begin{align} 
\rm O...CO_{ice} + H_{ice} & \rm \rightarrow CO_{2 ice}  + H_{ice}, & (60\%), \label{reaction11} \\
 & \rm \rightarrow HOCO_{ice}, & (20\%), \label{reaction12} \\
 & \rm \rightarrow CO_{ice} + OH_{ice}, & (20\%). \label{reaction13}
\end{align} 

\begin{figure*}
\includegraphics[width=0.33\linewidth]{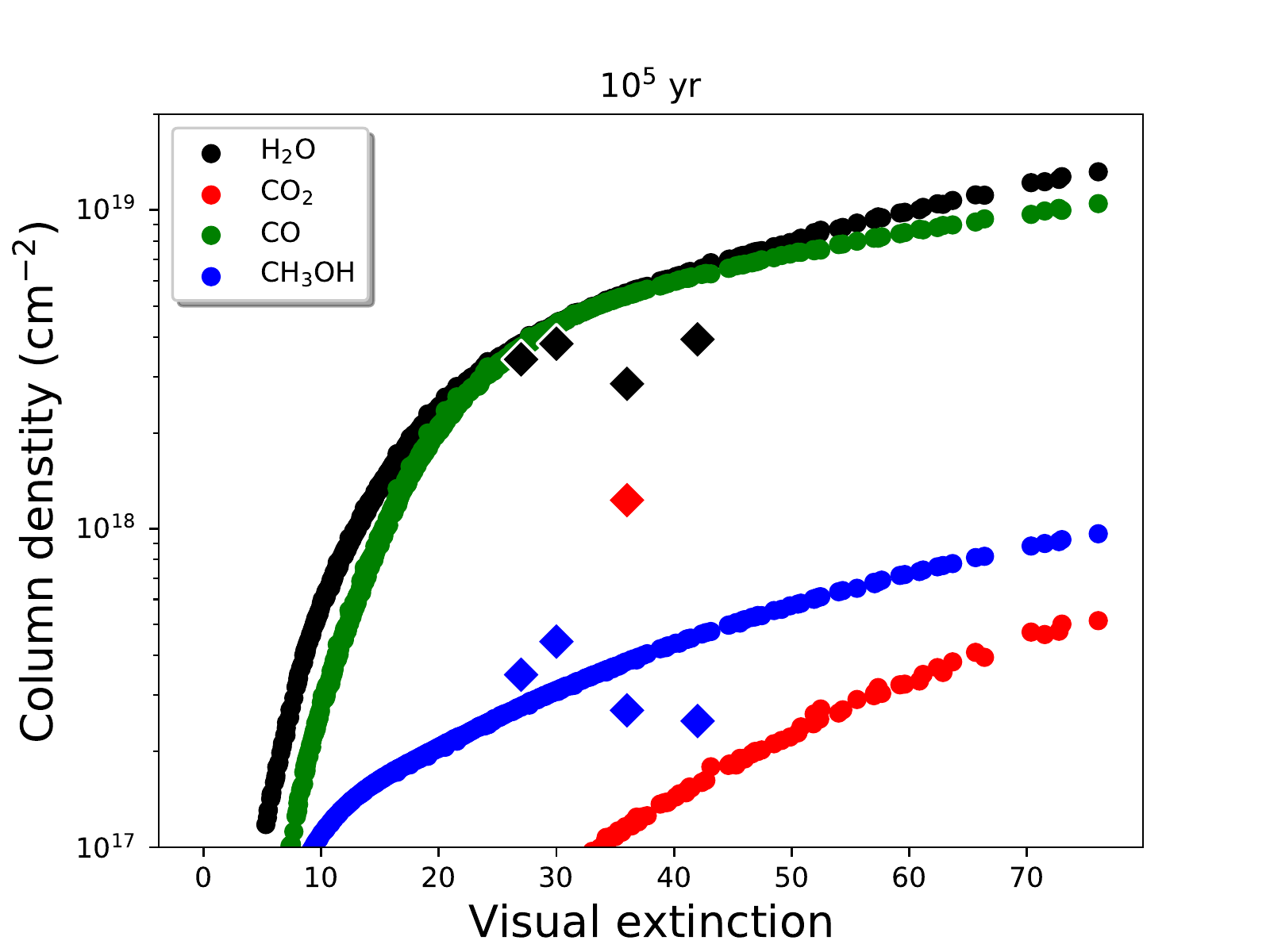}
\includegraphics[width=0.33\linewidth]{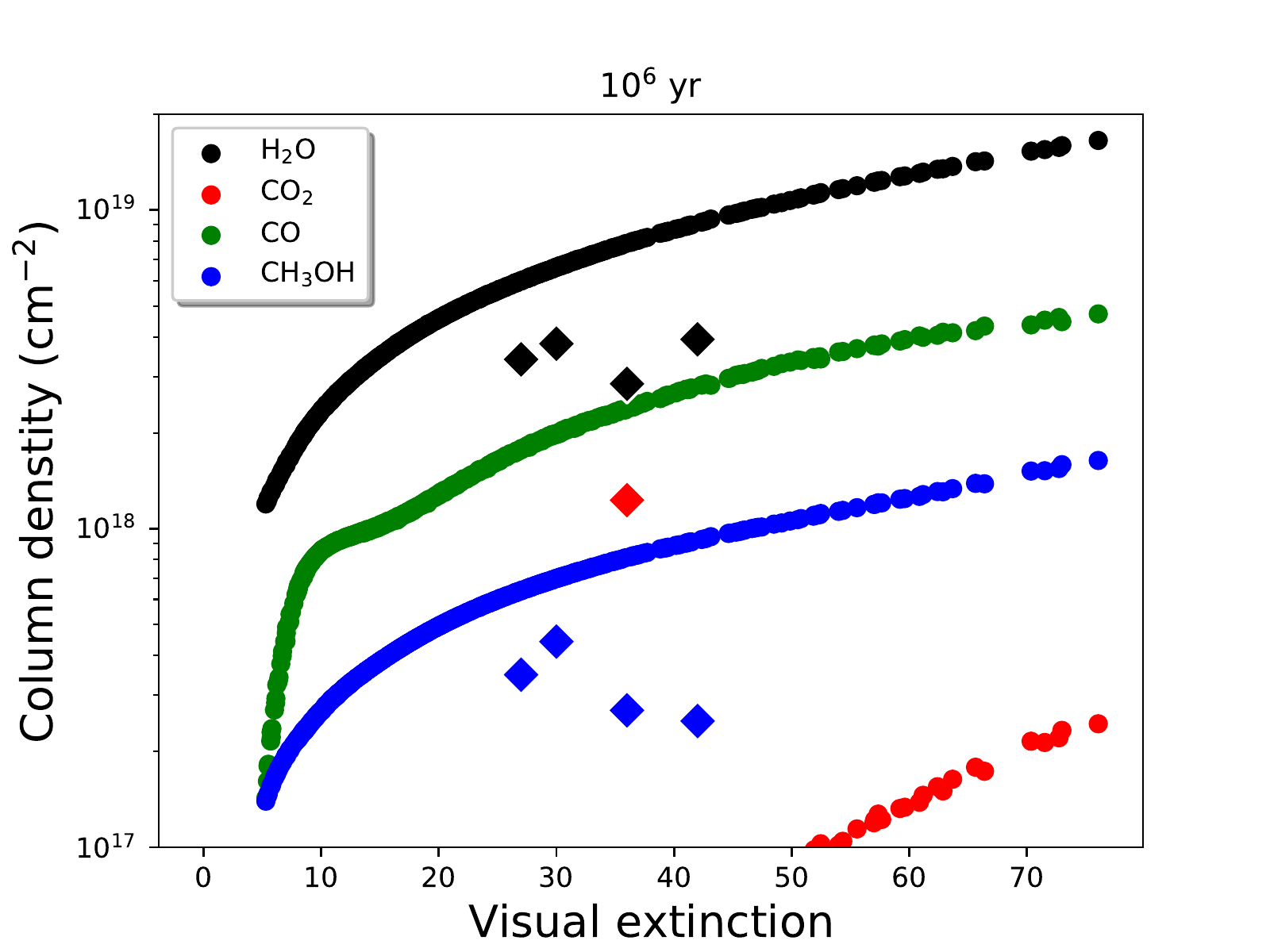}
\includegraphics[width=0.33\linewidth]{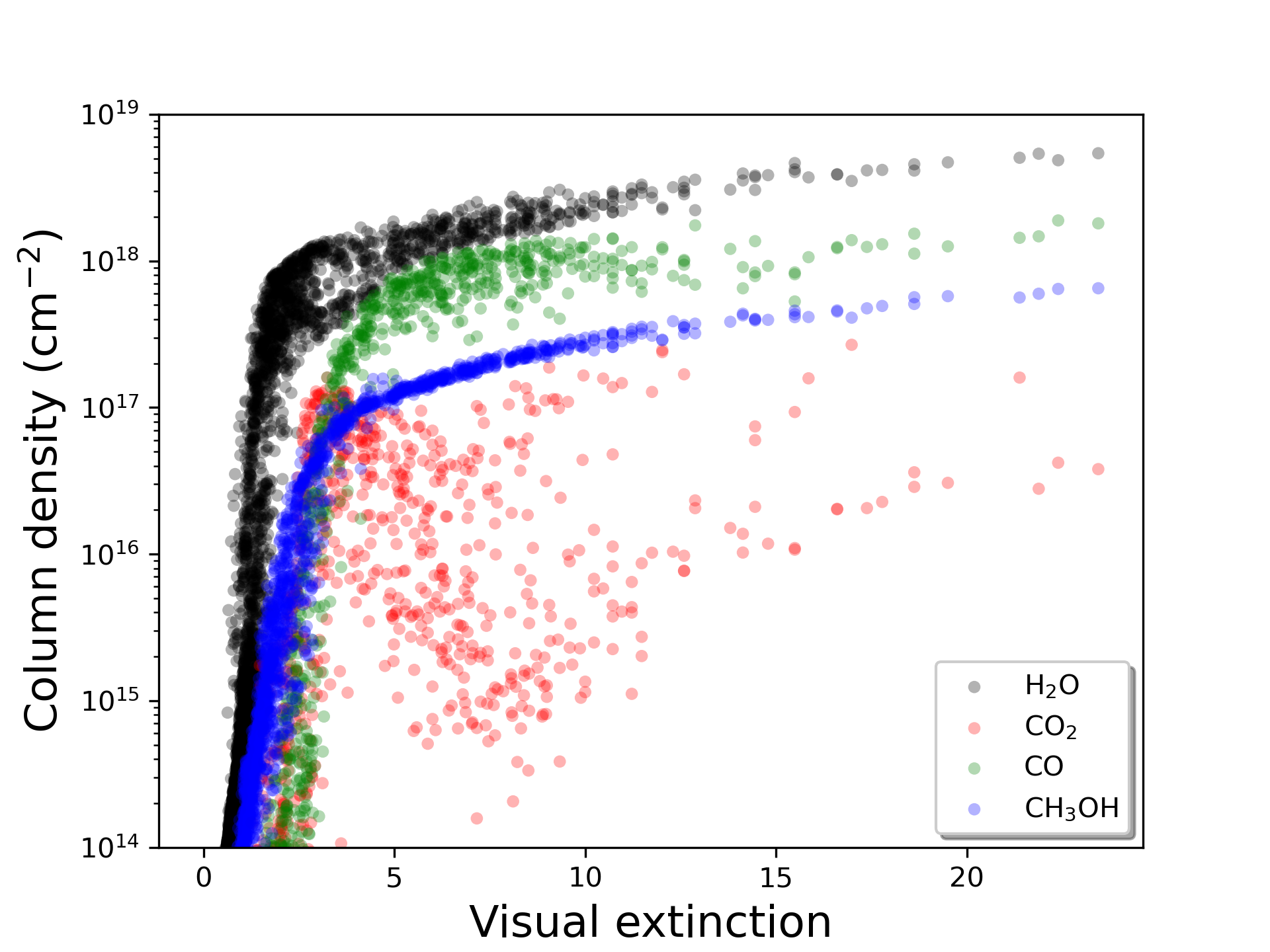}
\caption{ Column density of the main ice constituents computed with Nautilus for the static physical conditions as observed in L429-C as a function of visual extinction (left and middle). Model results are shown at $10^5$~yr in the left-hand panel and $10^6$~yr in the central panel. 
Diamonds represent the column densities observed by \citet{2011ApJ...729...92B} at specific positions in the core. Column densities of the main ice constituents computed with the dynamical model as a function of column density (right). For all figures, the dust temperature is computed following \citet{2017A&A...604A..58H} and the $\rm O...CO_{ice}$ complex from \citet{2015MNRAS.447.4004R} is included.
\label{CO2_ice_with_Eley}}
\end{figure*}

This process was switched off in the simulations shown here because of all the uncertainties in the temperature dependence of the existence of the $\rm O...CO_{ice}$ complex.
In \citet{2015MNRAS.447.4004R}, the authors found that this process could form significant amounts of CO$_2$ ice if the binding energy of the complex $\rm O...CO_{ice}$ was greater than 400~K. Their predicted abundance was, however, a few $10^{-6}$ and they did not compare quantitatively with observations. In addition, their model was only a two-phase model. In the current three-phase version of nautilus, the formation of $\rm O...CO_{ice}$ occurs only at the surface of the grains, not the bulk. We tested the effect of this process by switching it on and running the following models: (1) Static models with the physical conditions observed in L429-C except for the dust temperature computed with the \citet{2017A&A...604A..58H} approximation. The results are presented in the left and central panels  of Fig.~\ref{CO2_ice_with_Eley} and should be compared with Fig.~\ref{static_model2}. (2) Dynamical model for core 0 with the dust temperature computed with the \citet{2017A&A...604A..58H} approximation. The results are presented in the right-hand panel of Fig.~\ref{CO2_ice_with_Eley} and should be compared with Fig.~\ref{ice_cold_dens_clump0_all}.

The effect of these reactions is strong at high visual extinction and, thus, low temperature. The computed column densities of CO$_2$ are much higher in both the static and dynamical simulations. In the case of the static models, the predicted column densities of CO$_2$ are still lower than the observations by several orders of magnitude, especially at low visual extinction. The H$_2$O and CO predicted column densities are unchanged but the methanol columns densities are also increased through the following path: $\rm C_{gas} + \rm H_2O_{ice} \rightarrow  C...H_2O_{ice} \overset{H_{ice}}{\longrightarrow} CH_2OH_{ice}/CH_3O_{ice} \overset{H_{ice}}{\longrightarrow} CH_3OH_{ice}$ \citep[pathways added with the $\rm O...CO_{ice}$ complex from][]{2015MNRAS.447.4004R}. For the dynamical models, the CO$_2$ ice column densities are also increased at high visual extinction, but only in the cases giving low values. This means that the overall CO$_2$ column densities are not increased and that the main paths leading to values close to the observations are still the ones with the appropriate physical conditions at earlier times.

\section{Influence of the dust temperature}\label{influence_Tdust}

Whatever model is applied, the dust temperature is always a key parameter. To understand the sensitivity of the model results to this parameter, we reran all our simulations (static and dynamic), adding 1~K to the dust temperatures computed with the \citet{2017A&A...604A..58H} approximation (``Hocuk+1'').
 The dust temperature as a function of visual extinction in that case is shown in green in Fig.~\ref{dust_temperature}. The dust temperature of 12~K occurs for a larger visual extinction (4 instead of 2.7) compared to Hocuk value. This change extends the window within which the CO$_2$ can form (T$_{\rm dust} > 12~K$ and Av$> 2$). Although there is no proper error bar given with Hocuk's relation, a departure of one Kelvin from this parametrization is certainly reasonable while comparing the observed dust temperatures at various A$_{\rm V}$ with this relation (see their figure 1).

\begin{figure}
\includegraphics[width=1\linewidth]{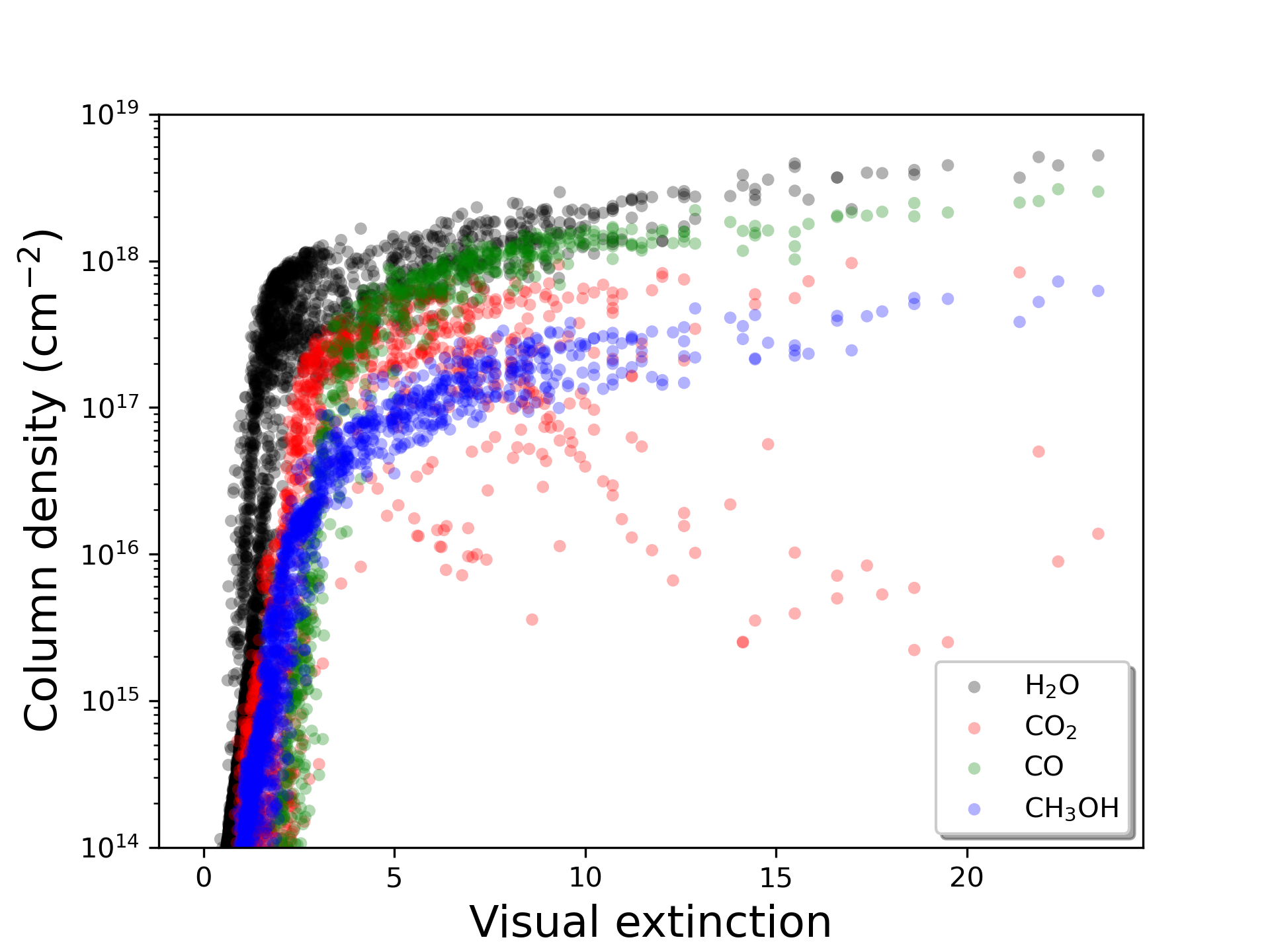}
\includegraphics[width=1\linewidth]{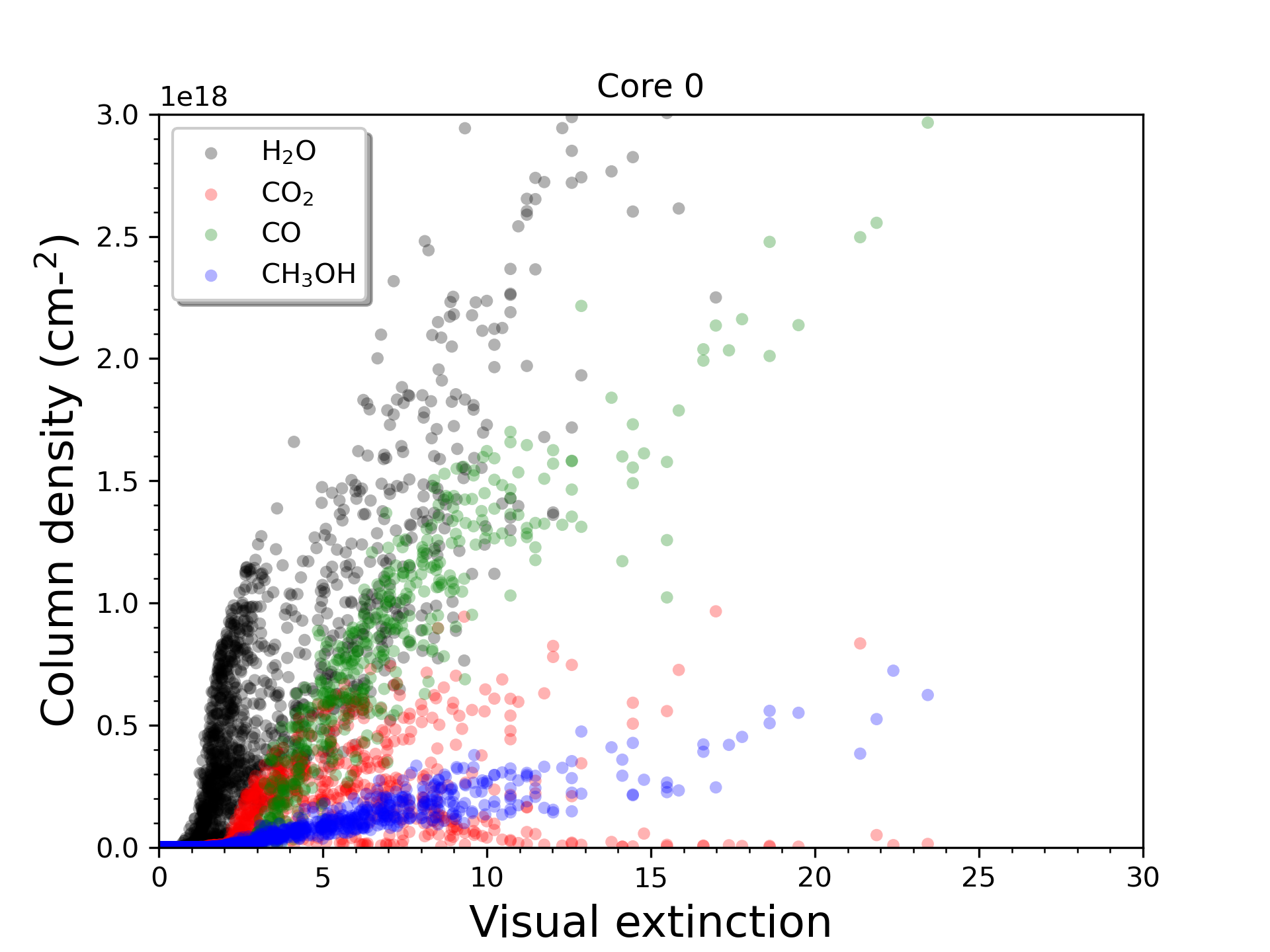}
\caption{Column densities of the main ice constituents computed with the dynamical model as a function of column density for the "Hocuk+1" model (upper panel).  Same figure but with an axis setting similar to Fig. 7 of \citet{2015ARA&A..53..541B} for the "Hocuk+1" model (lower panel). \label{ice_cold_dens_clump0_all_hocuk+1}}
\end{figure}

For the static models, results are unchanged. The  "Hocuk+1" case, including the O...CO complex, gives similar results to the case Hocuk with the O...CO complex, except that the CO$_2$ column densities are higher at very low Av, but again not at the levels constrained by the observations. The impact on the dynamical simulations are more significant. Figure~\ref{ice_cold_dens_clump0_all_hocuk+1} shows the model results in that case, which should be compared to Fig.~\ref{ice_cold_dens_clump0_all}. A larger number of trajectories produce detectable amounts of CO$_2$. Adding the O...CO complex to this model does not significantly change  the results. So for the rest of the paper, we adopt this prescription of the dust temperature. 

The effect of the dust temperature on the chemistry rises through the diffusion rates of the species on the grains. In the model, the diffusion rate of each species is proportional to $\exp{\rm (-E_{bind}/kT)}$ with T as the dust temperature, $\rm E_{bind}$ as the binding energy of the species, and k as the Boltzmann constant. \citep{2016MNRAS.459.3756R}. The higher the dust temperature or the smaller the binding energy, the higher the diffusion. As discussed in Section~\ref{comp-models}, the binding energies of physisorbed species are quite uncertain and very likely vary from one model to another. In addition, the grain surfaces are not homogeneous and the binding energies are better represented by a distribution rather than a single value -- as assumed in the models \citep[see for instance][and references therein]{2012MNRAS.421..768N,2015JChPh.143h4703D,2022ESC.....6..597M}.

\section{Predictions of the ice composition diversity}\label{variability-section}

\begin{figure*}
\includegraphics[width=0.33\linewidth]{Figs_ice_clumps_commeBoogert_Hocuk+1/ice_cold_dens_clump0_commeBoogert_Hocuk+1_up_peak.png}
\includegraphics[width=0.33\linewidth]{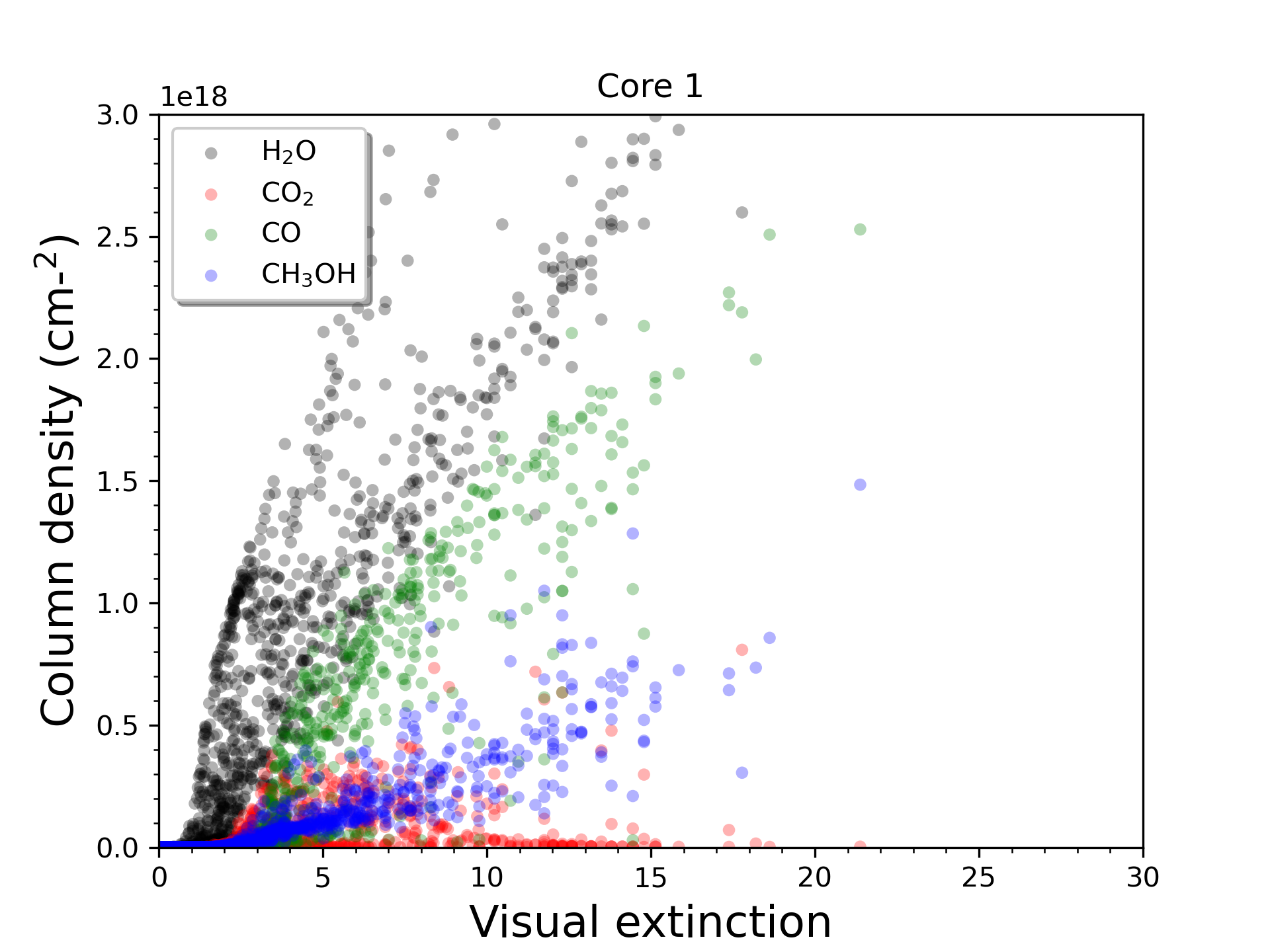}
\includegraphics[width=0.33\linewidth]{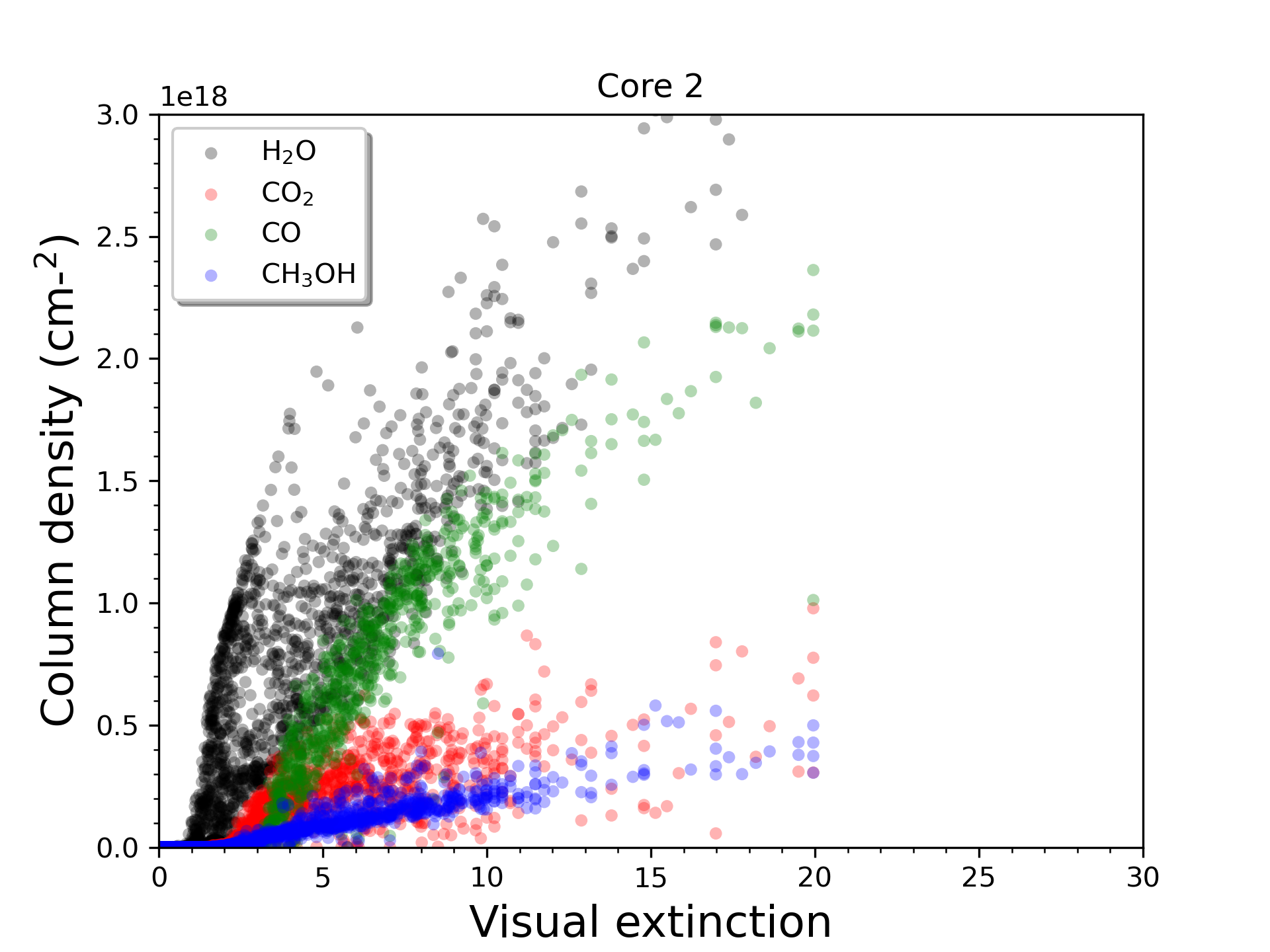}
\includegraphics[width=0.33\linewidth]{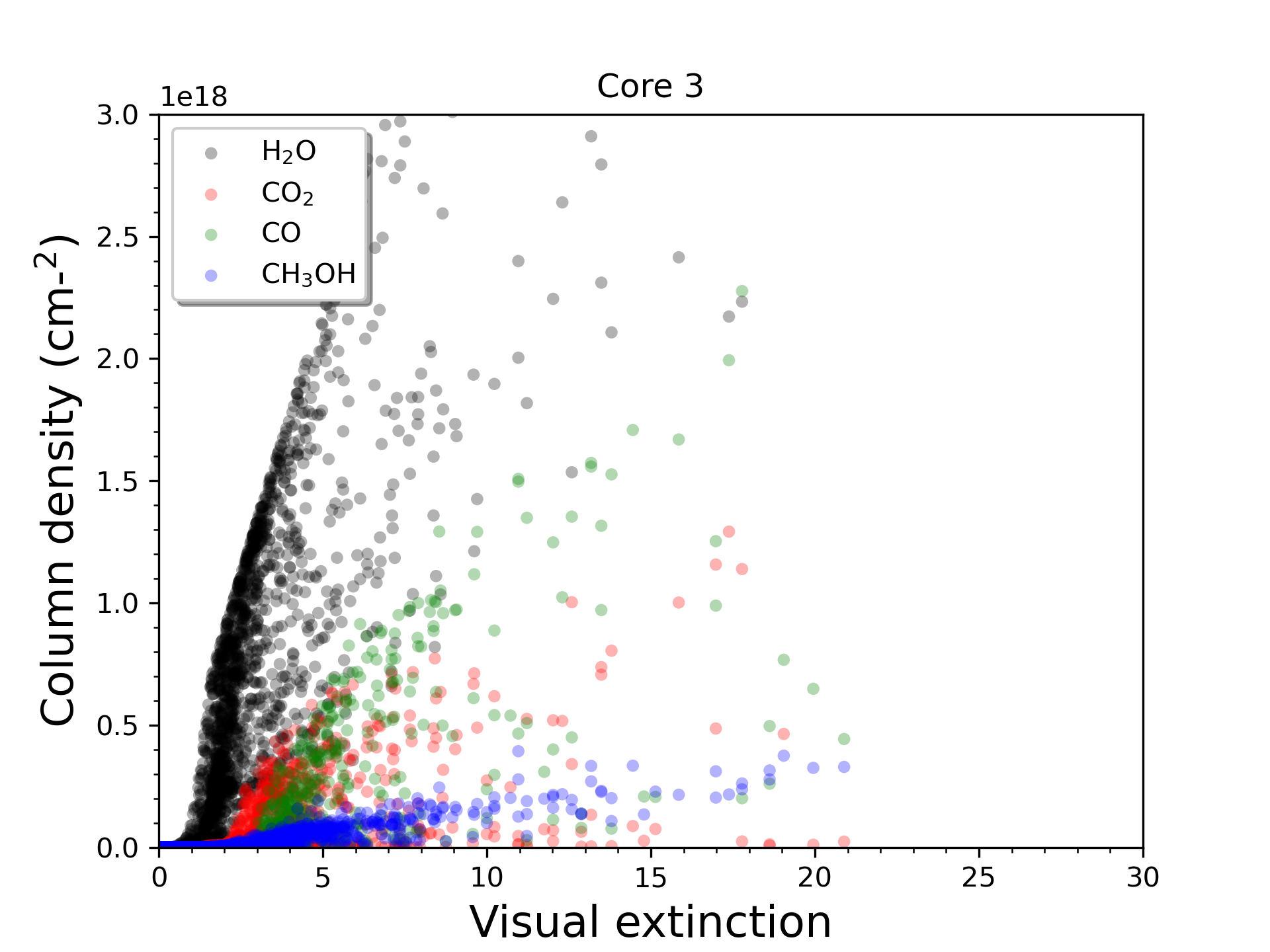}
\includegraphics[width=0.33\linewidth]{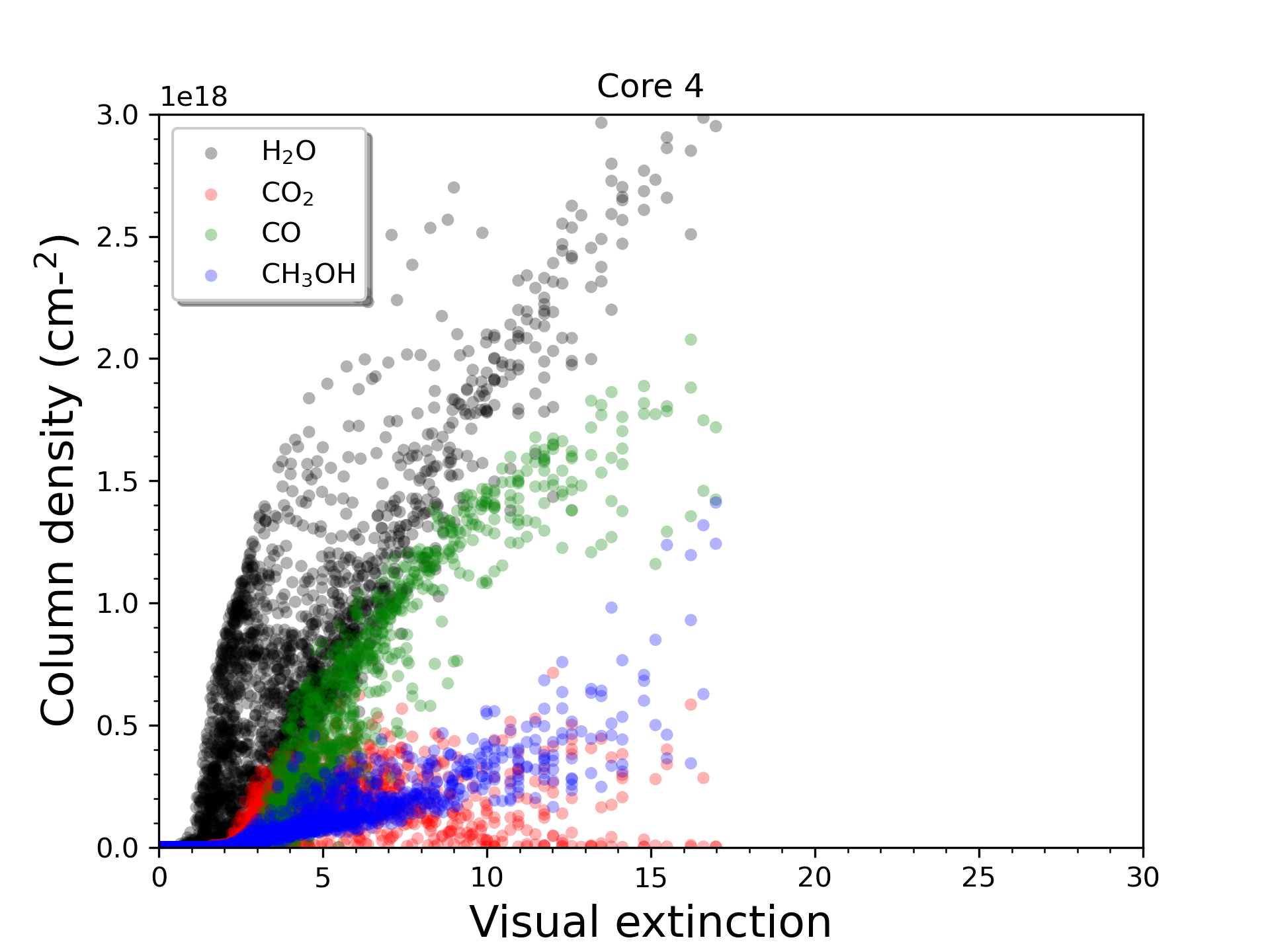}
\includegraphics[width=0.33\linewidth]{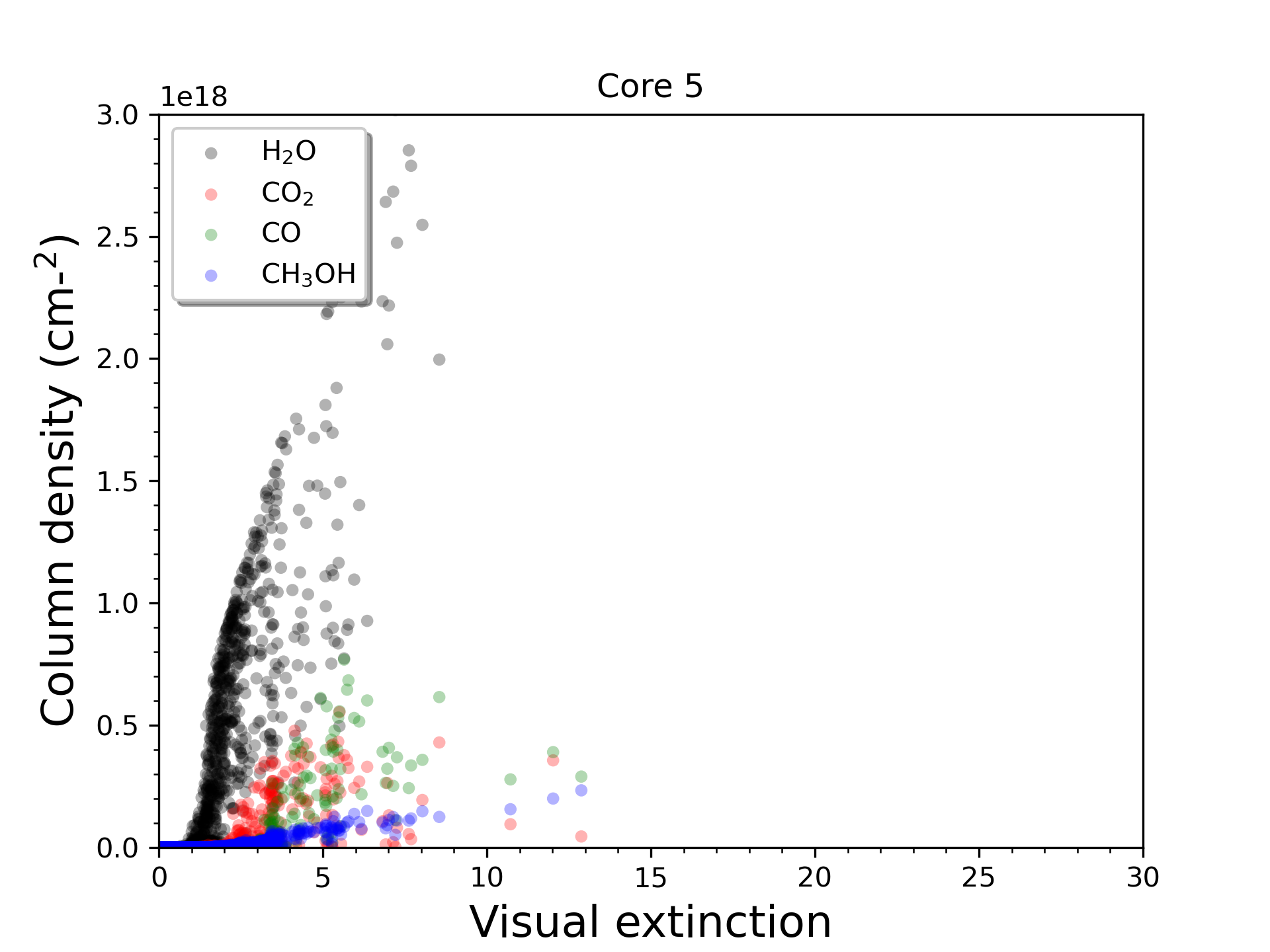}
\includegraphics[width=0.33\linewidth]{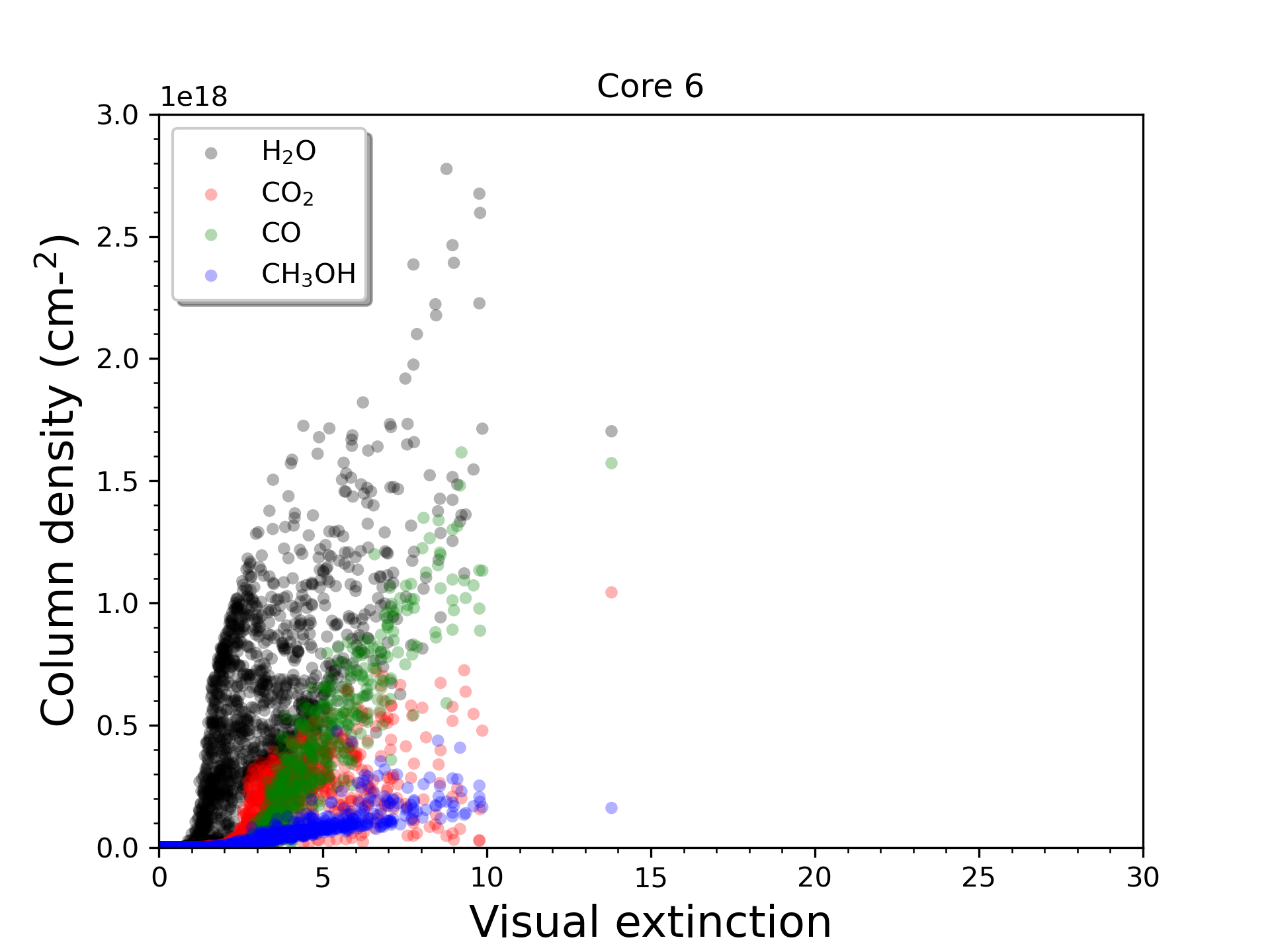}
\includegraphics[width=0.33\linewidth]{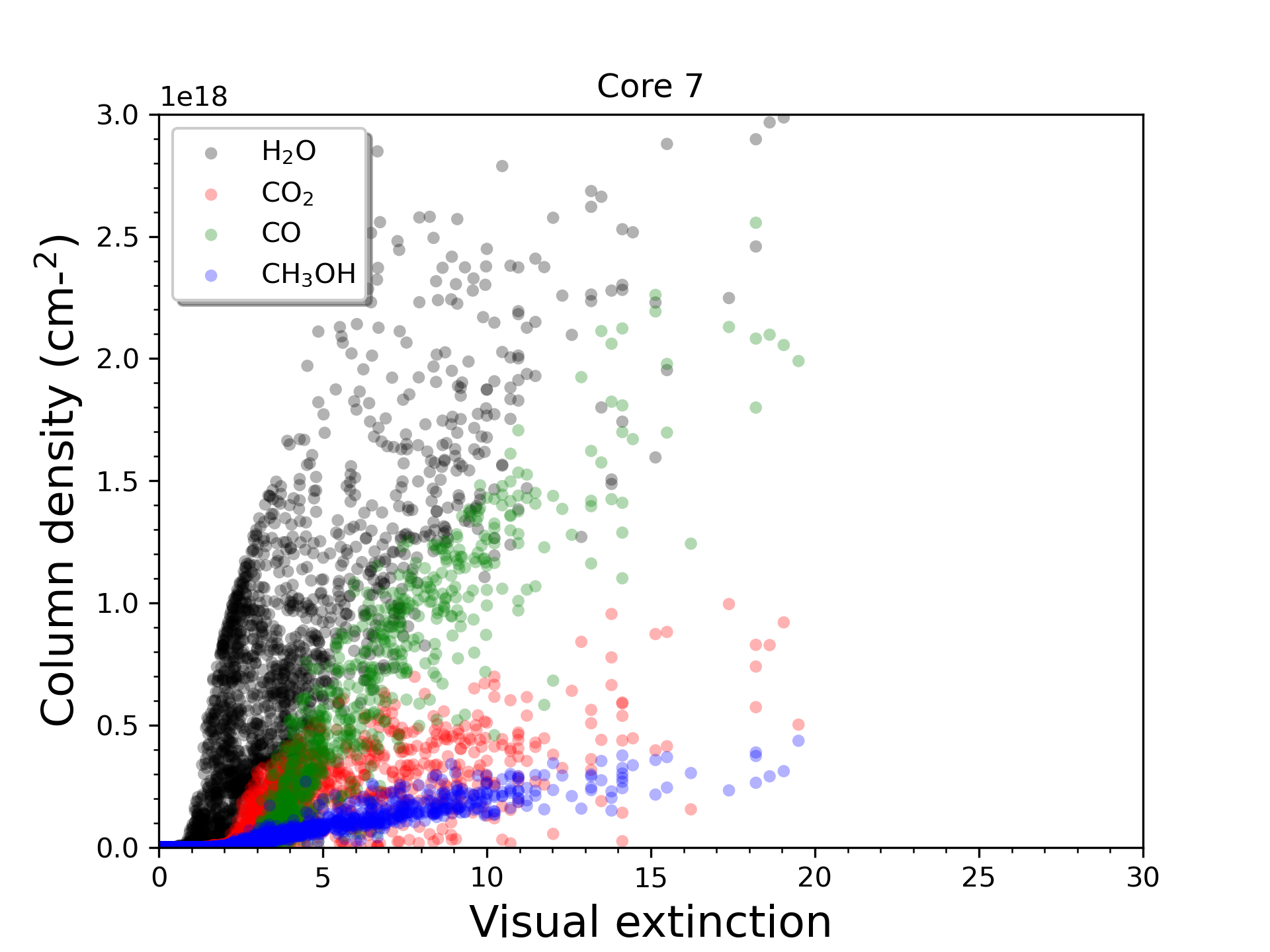}
\includegraphics[width=0.33\linewidth]{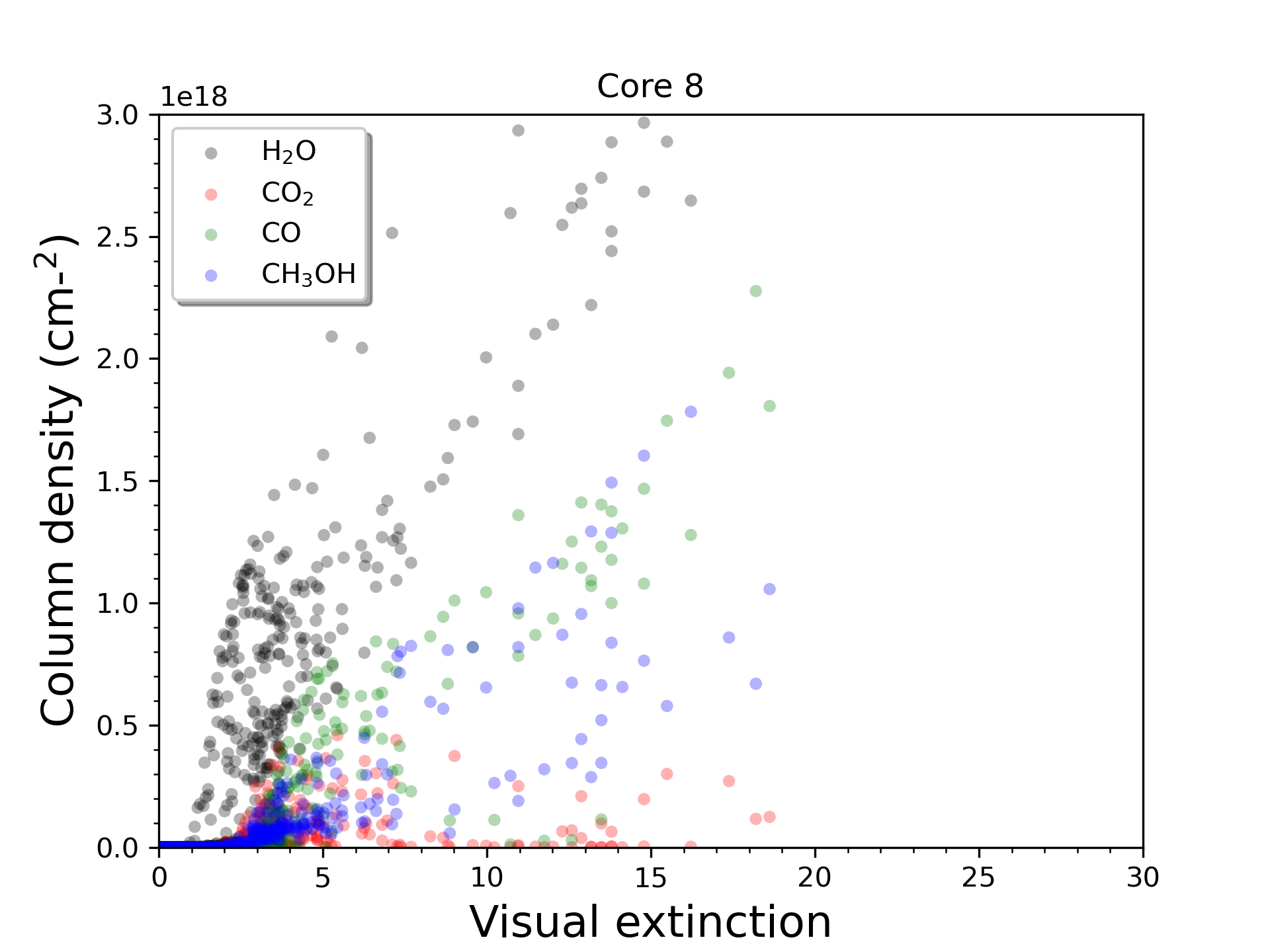}
\includegraphics[width=0.33\linewidth]{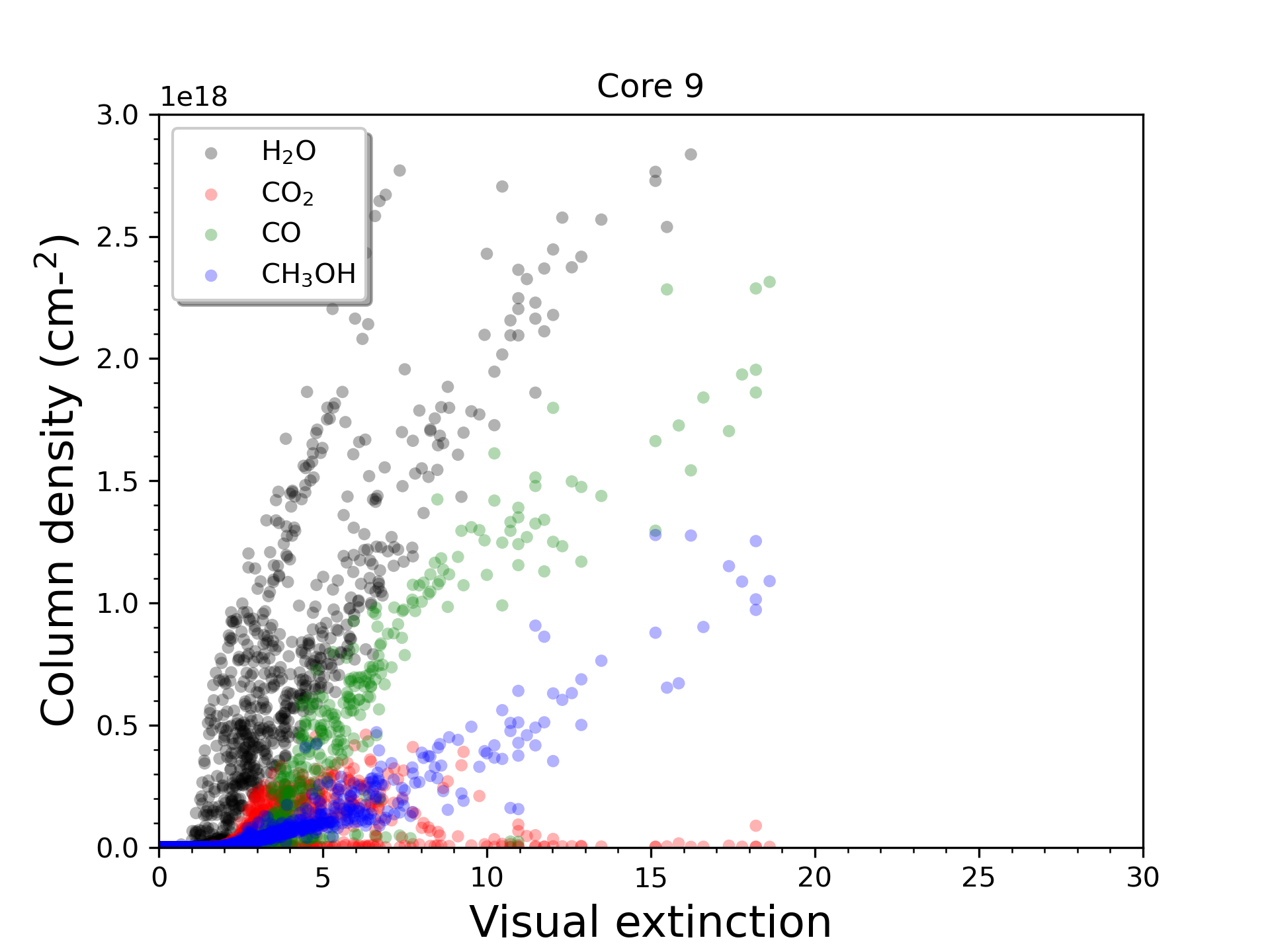}
\includegraphics[width=0.33\linewidth]{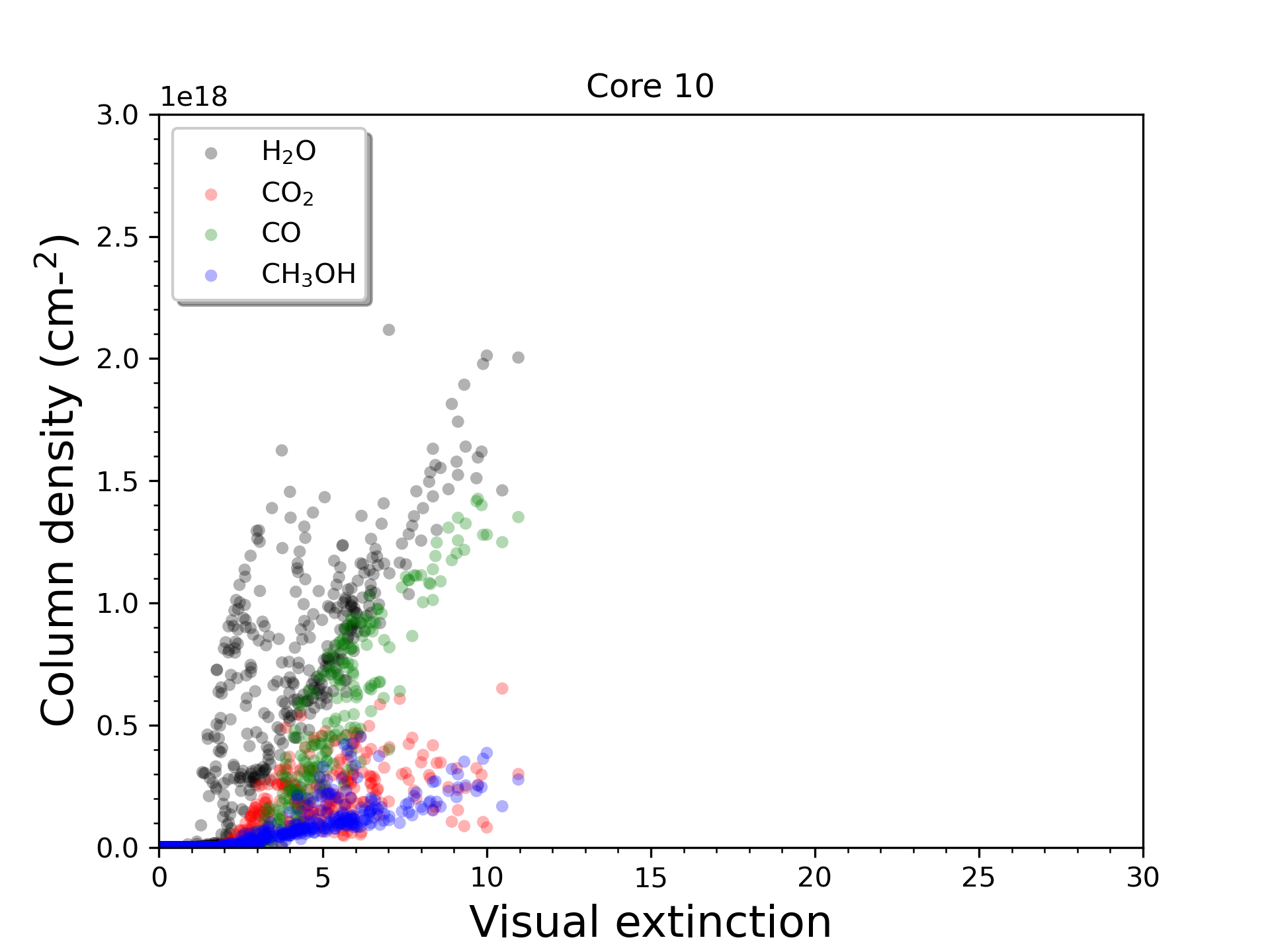}
\includegraphics[width=0.33\linewidth]{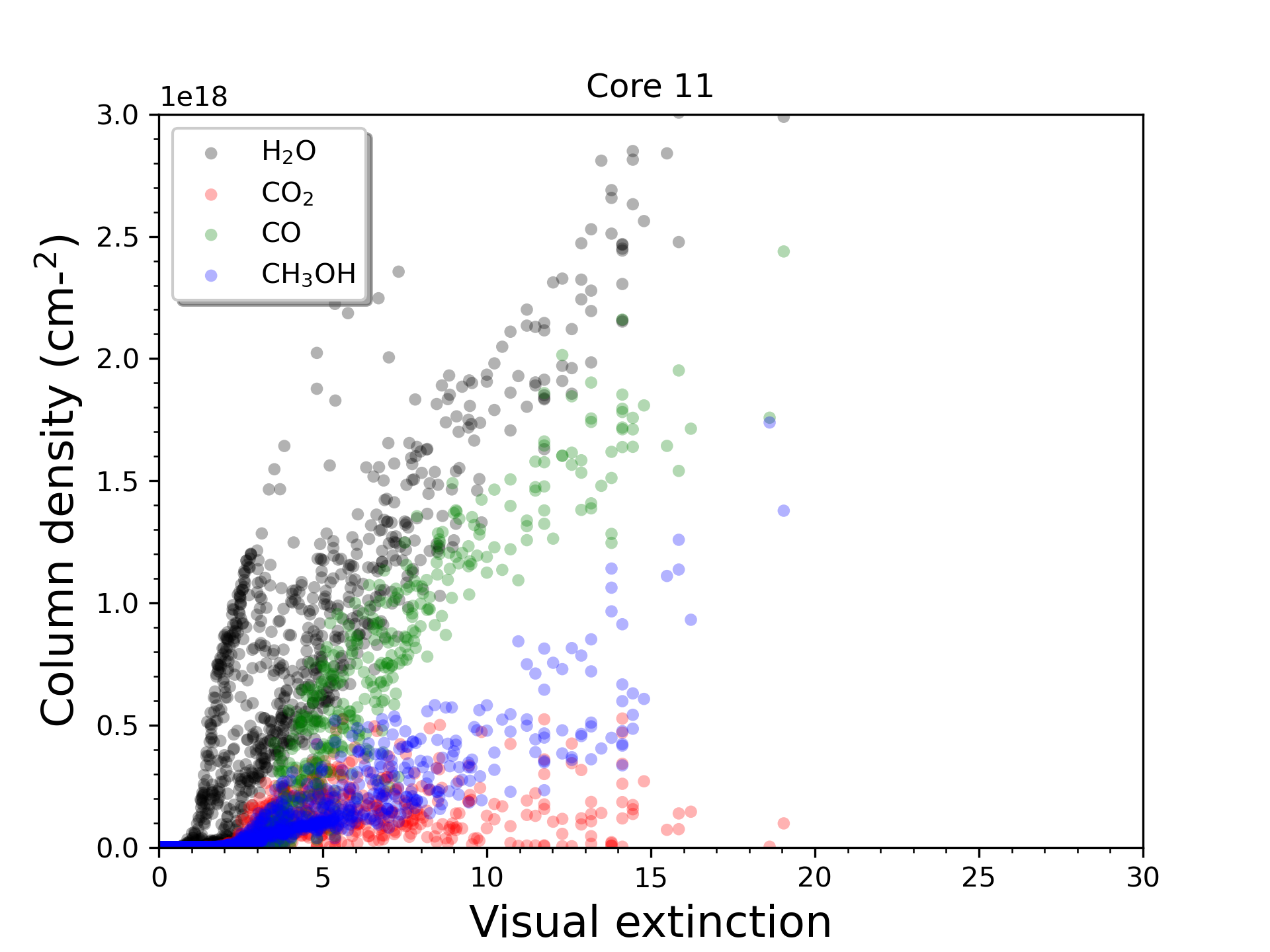}
\caption{Column densities of the main ice constituents computed with the dynamical model for 12 cores as a function of column density, using the dust temperature approximation "Hocuk+1".
\label{ice_cold_dens_allclumps_all_hocuk+1}}
\end{figure*}

Among the 12 theoretical cores we studied, we chose to present core 0 in Section~\ref{results_clump0} because it presents both trajectories leading to high and low abundances of CO$_2$ ice. In Fig.~\ref{ice_cold_dens_allclumps_all_hocuk+1}, we show the model results for all the cores (model "Hocuk+1" without the O...CO complex). Each core presents a different result. 
For instance, core 5 does not have CH$_3$OH ice column densities greater than $10^{17}$~cm$^{-2}$ whatever the visual extinction and it exhibits a low CO ice column density with respect to the other cores. Cores 8 and 9 present large amounts of CH$_3$OH ice and scarcer CO$_2$ ice when the visual extinction increases. Contrary to core 0, core 2 does not have any trajectories producing CO$_2$ column densities lower than $10^{17}$~cm$^{-2}$ for visual extinctions larger than 15, because the increase in density (and, thus, the visual extinction) and the decrease in dust temperature are smoother with time. To quantify the variability of the ice predictions, we computed  the mean values of each species and the standard deviation for
bins of visual extinctions. These values for each core are shown in Appendix \ref{std-section} and Fig.~\ref{std}. When the standard deviation (std) is high, the mean value has no real meaning. In all cores, the std is high for all species at low A$_{\rm V}$ ($< 4$). At high Av, it is mostly CO$_2$ that presents the higher std. All the other species show a small std when A$_{\rm V}$ is larger than 4 (with the exception of the final A$_{\rm V}$ for cores 5, 6, and 10 for which there are small numbers of points). Interestingly, some of the cores (core 2, 6, 7, and 10) also show a small std of CO$_2$ at high A$_{\rm V}$. 

\begin{figure*}
\includegraphics[width=0.33\linewidth]{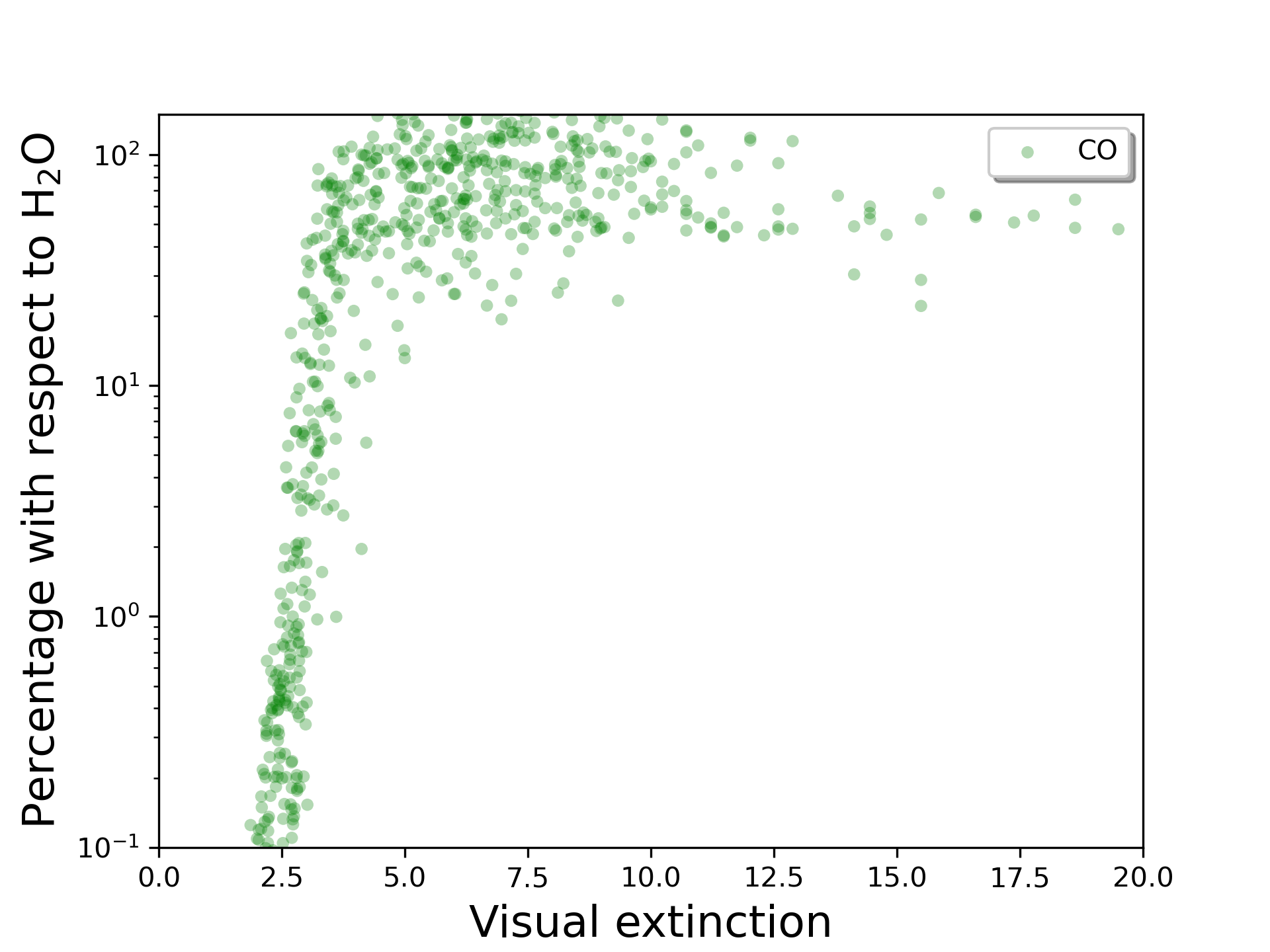}
\includegraphics[width=0.33\linewidth]{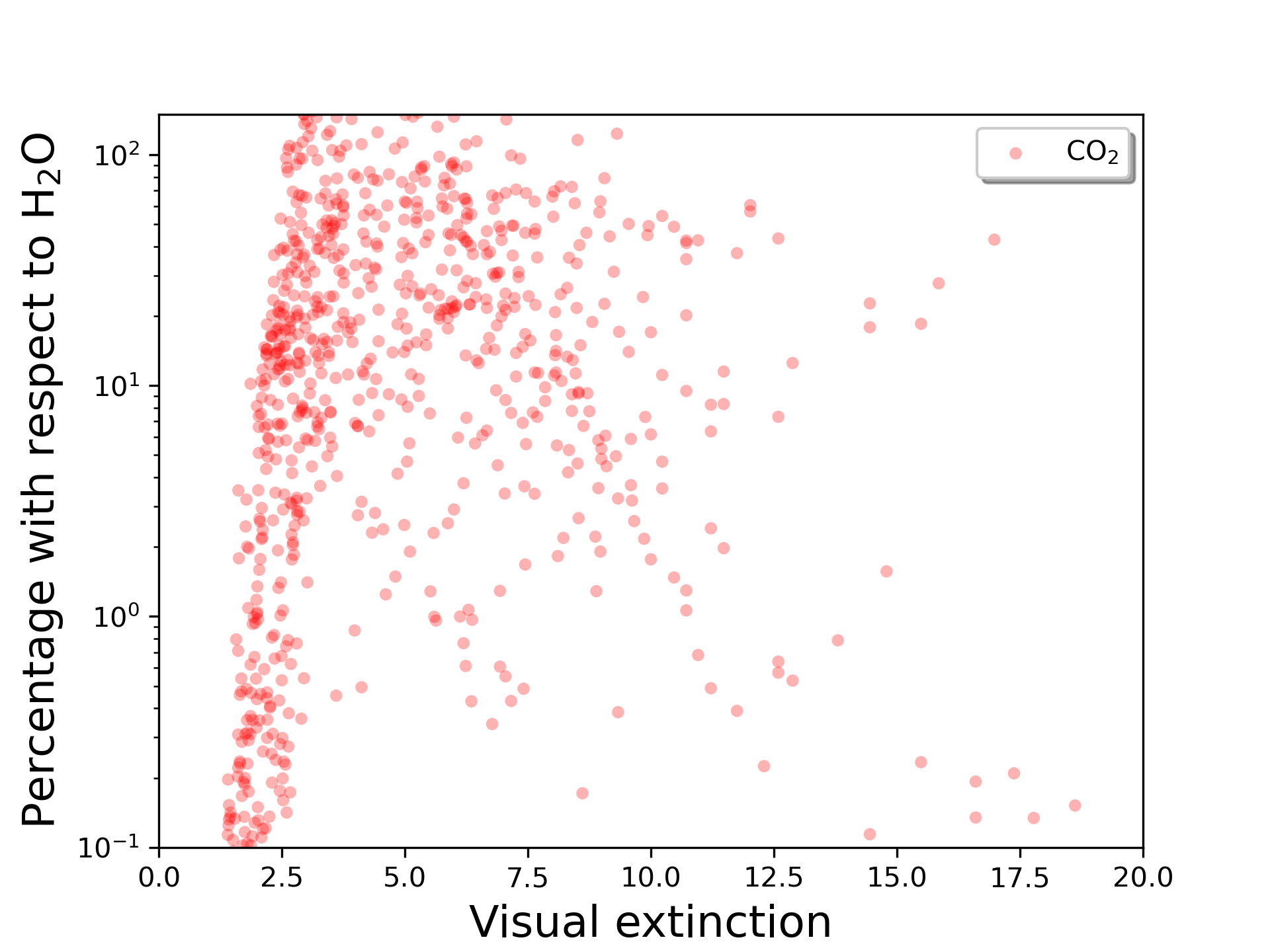}
\includegraphics[width=0.33\linewidth]{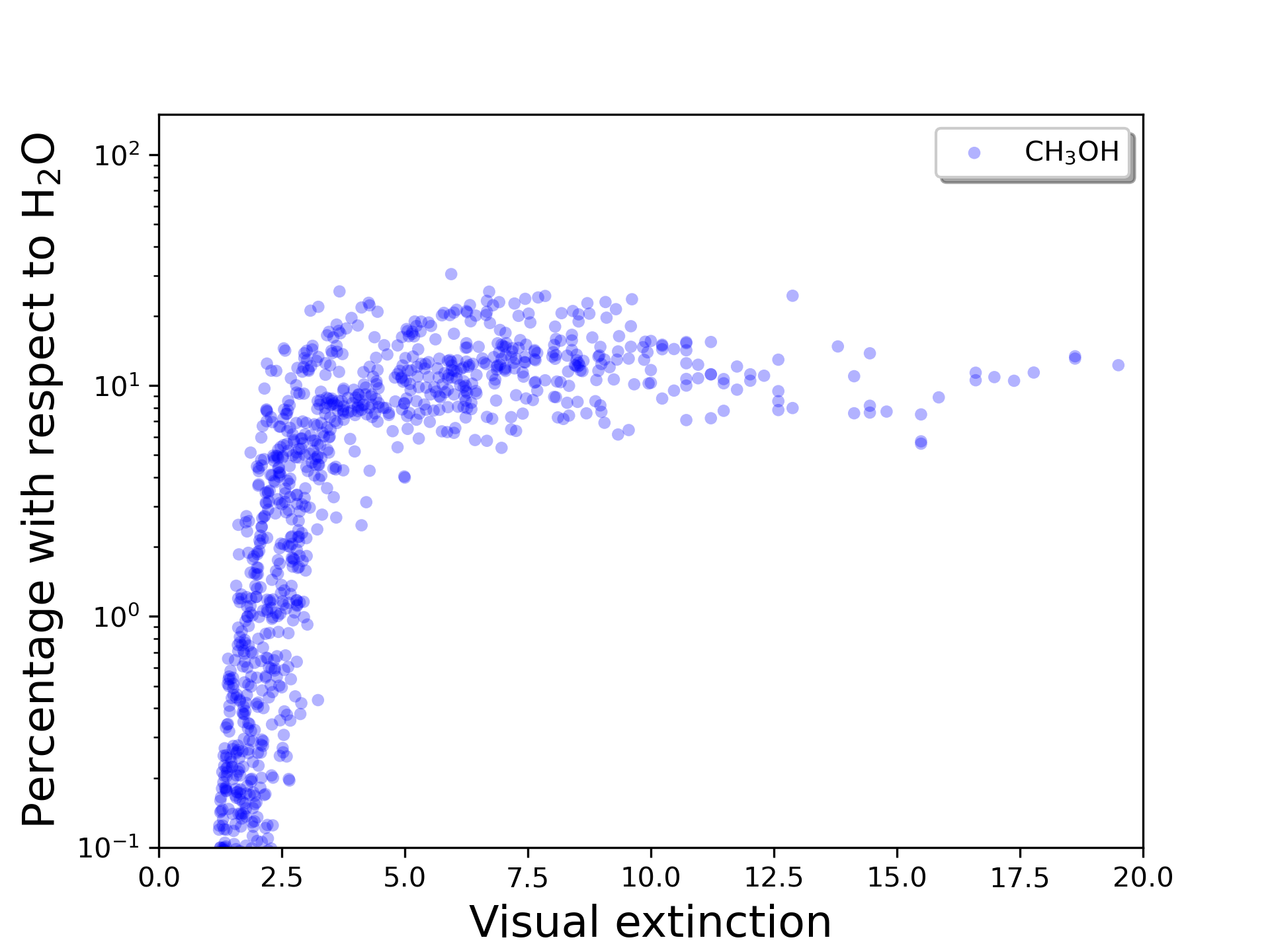}
\includegraphics[width=0.33\linewidth]{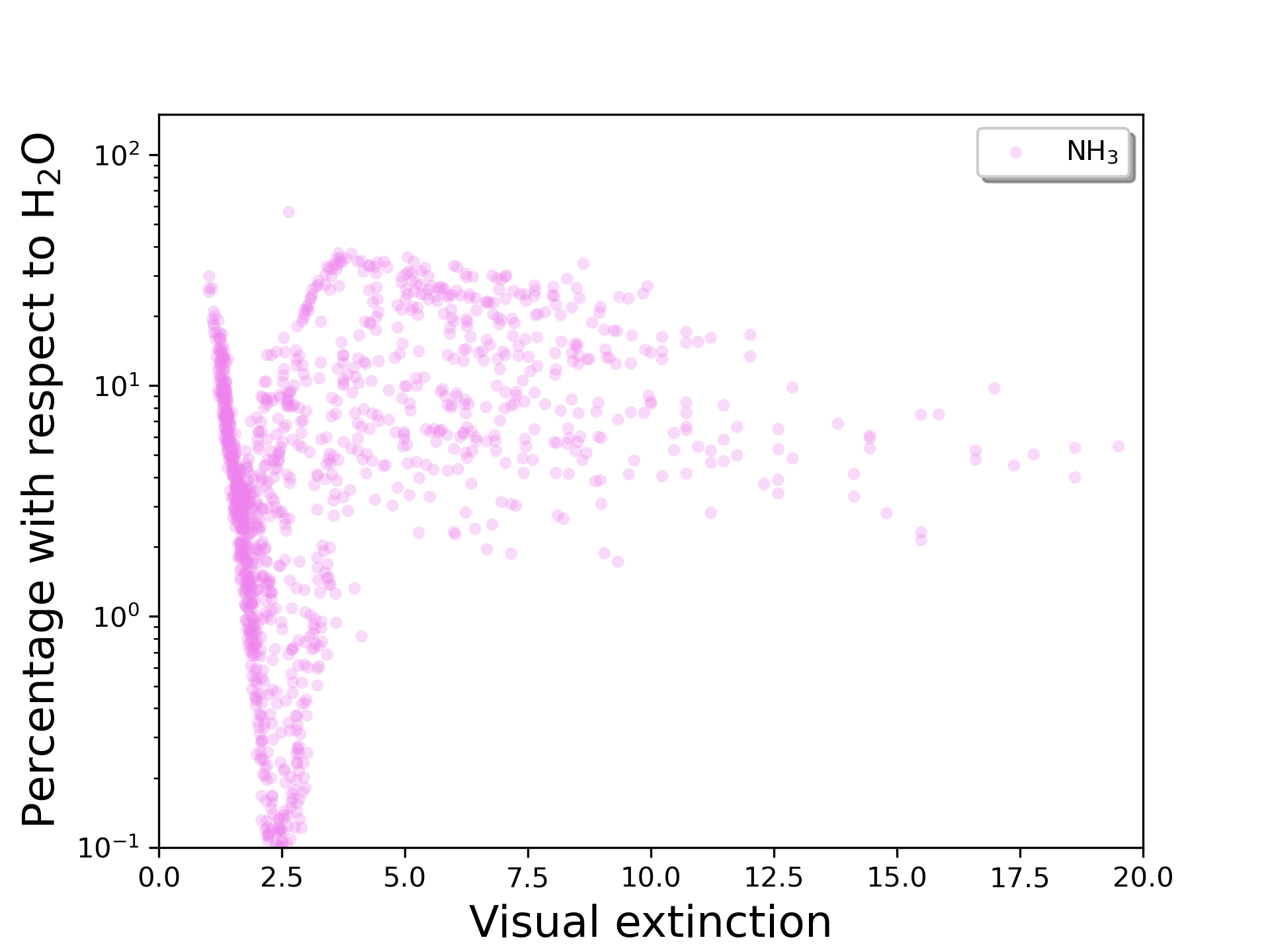}
\includegraphics[width=0.33\linewidth]{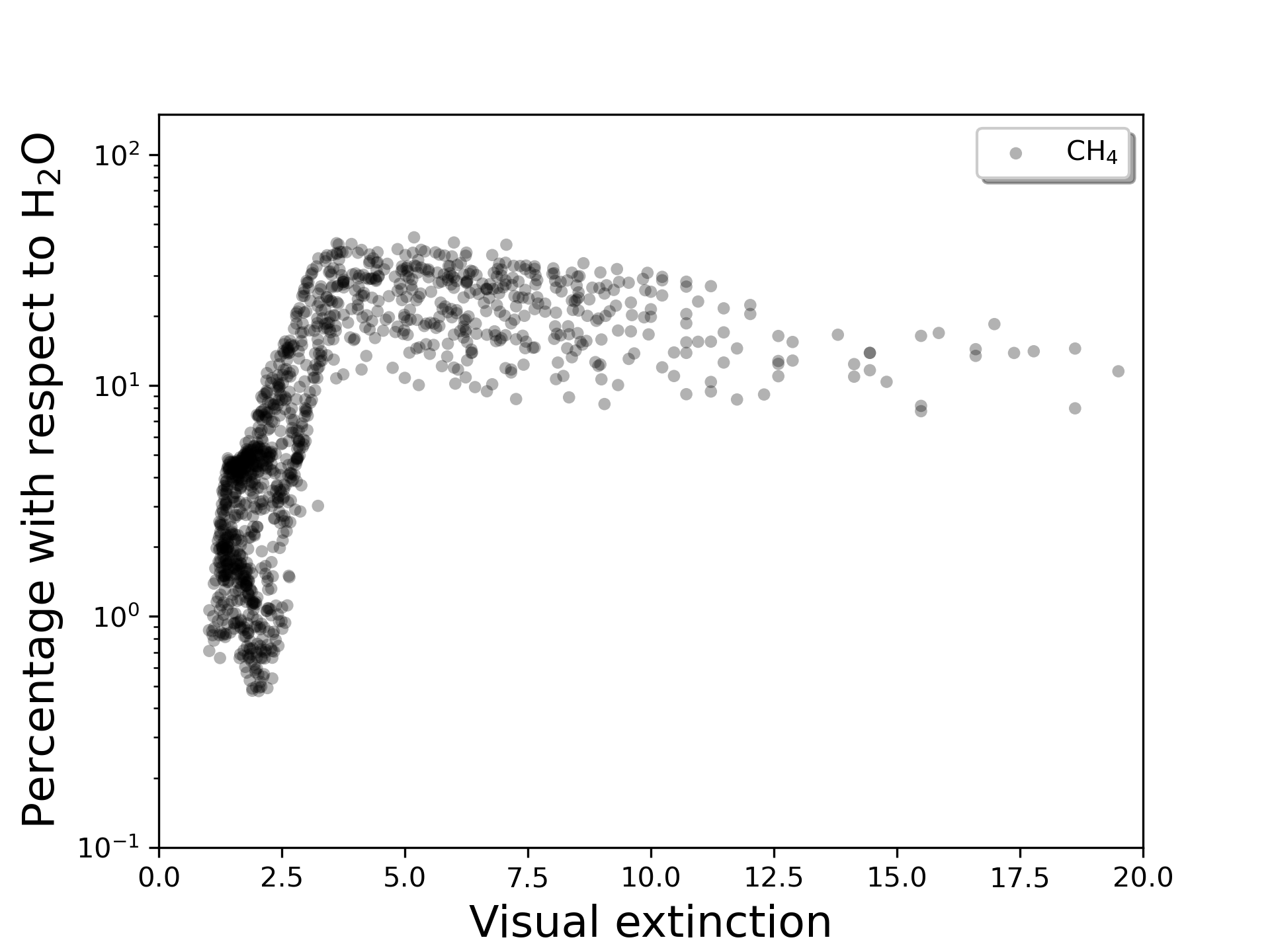}
\caption{Predicted ice composition (percentage with respect to H$_2$O ice) as a function of visual extinction for core 2 for all trajectories with a water column density greater than $10^{17}$~cm$^{-2}$.
\label{ice_percent_clump2}}
\end{figure*}

To take a different perspective on our model results, we show in Fig.~\ref{ice_percent_clump2} the percentage of the main ice constituents with respect to the water abundance as a function of visual extinction for core 2. For all species, the percentage increases rapidly with A$_{\rm V}$ until it reaches a plateau at approximately A$_{\rm V}$ = 5. Such a result is also found for the abundance of these species with respect to the total proton density. One difficulty in addressing these results is the variability of the ice composition with the local physical conditions and their history. All species present a spread in their percentage at a specific visual extinction. While this spread seems to decrease with A$_{\rm V}$ (except for CO$_2$), this could be an effect of having less data. We know that CO$_2$ is the species that presents the largest dispersion in the percentage with respect to H$_2$O. For A$_{\rm V}$ between 10 and 20, the CO$_2$ percentage is approximately between 1 and 50\%, the CO percentage between 18 and 115\% (although only a very small number of points is responsible for the smallest values), the CH$_3$OH percentage is between 6 and 20\%, the NH$_3$ percentage is between 3 and 29\%, and the CH$_4$ percentage is between 13 and 34\%. We note that the range of percentages are in good agreement with the observed trends, except for CO, which is predicted to be more abundant than water in some cases. These percentages are rather constant with the visual extinction, although variations can exist due to the spread in values. One limitation of our work is that none of our simulations go beyond an A$_{\rm V}$ of 25 because the physical simulations do not include self-gravity. So extrapolations to high visual extinctions are not possible. Figure 7 of \citet{2015ARA&A..53..541B}, shows an increase in all ice components that appears to be constant with A$_{\rm V}$. Our model beyond A$_{\rm V}$ 5 reproduces this feature well. Indeed, having a similar proportion of each component results in a linear evolution of the column density as a function of the visual extinction.

\section{Time-dependent formation of ices}\label{time-evolution}

\begin{figure}
\includegraphics[width=1\linewidth]{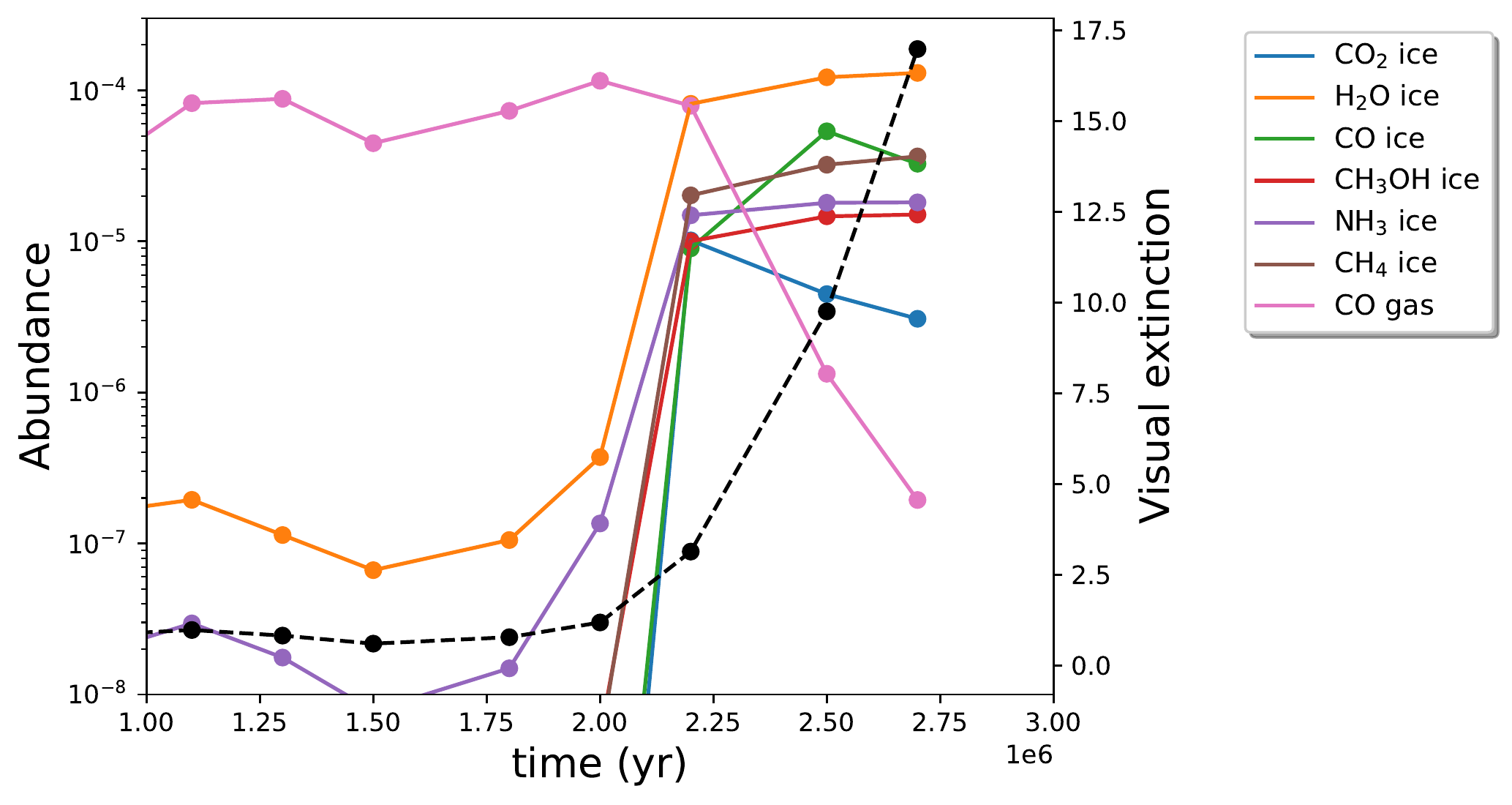}
\caption{Abundances (with respect to H) of the main ice constituents and gas-phase CO as a function of time. The model is core 2, cell 8 (a trajectory that produced large amounts of CO$_2$ ices). The visual extinction for this simulation is shown in black (dotted line). \label{ab_time_clump2}}
\end{figure}

In all the figures of column density as function of visual extinction shown up to now for the dynamical models, several timelines are mixed. In this section, we discuss the ice composition as a function of time for one trajectory of core 2 forming large amounts of CO$_2$ ice at high A$_{\rm V}$. In Fig.~\ref{ab_time_clump2}, we  show the abundance of the main ice constituents as a function of time for this cell, zoomed on the time axis to cover the maximum density peak. On the same figure, we show the increase of visual extinction with time. To compare the formation of the ices with the depletion of gas-phase CO, we also plot the CO gas-phase abundance as a function of time. In this example, ice formation becomes efficient within one time step (of $2\times 10^5$~yr, when the visual extinction becomes larger than two). All of the ice constituents but CO form mainly prior to the CO catastrophic freeze-out onto the grains. Then, in the next time step, CO freezes out onto the grains producing a large amount of CO ice. The other ice constituents also increase slightly during this time, except for CO$_2$, whose abundance decreases. This result is in agreement with the proposed evolution sequence of ices by \citet{2011ApJ...740..109O}.

\section{Conclusions}\label{conclusion}

We conducted a theoretical study of the formation of interstellar ices at various visual extinctions. Our goal was to reproduce features currently revealed by the observations as well as to make predictions for future JWST results. We carried out two types of simulations. The first was a classic application of an astrochemical model (static model). We used the observed physical conditions from a specific region (cold core L429, probing visual extinctions from 7 to 75) and ran the chemical model for these fixed conditions for a period of time representative of cold cores. One key parameter was revealed to be the dust temperature. When the dust temperature is higher than 12~K, CO$_2$ forms efficiently to the detriment of H$_2$O, while at temperatures below 12~K, CO$_2$ does not form. Whatever our hypothesis on the chemistry, the static simulations failed to reproduce the observed trends of interstellar ices.\\

When considering the time-dependent physical conditions experienced by interstellar matter forming cold cores (dynamical simulations), we were able to qualitatively reproduce the observations. For these sets of models, we computed the chemistry using time dependent physical conditions from a 3D SPH model of core formation. We studied the formation trajectories of 12 cores, each sampled by tens to hundreds of cells of independent material. We found that a large fraction of the ice was built very early during the formation of the core (especially the CO$_2$ ice) but also that the ice fraction keeps evolving until the density of the core is reached. Large amounts of H$_2$O, CH$_3$OH, NH$_3$, and CH$_4$ are formed on the grains before the catastrophic freeze-out of CO and their abundance keeps increasing afterward. On the contrary, CO$_2$ seems to be mostly formed prior to freeze-out and can even decrease when large amounts of CO stick to the grains. One important result of this study is that efficient formation of CO$_2$ ices requires very specific conditions: a dust temperature greater than 12~K and a visual extinction higher than 2. In our simulations, these conditions are met for a limited number of trajectories and some of our cores never experience them. Considering the ubiquity of CO$_2$ ices, this would appear to be a strong constraint on physical models of the evolution of interstellar matter.\\
From a chemical point of view, we investigated the various formation pathways of CO$_2$ ice (including the formation of O...CO complex on interstellar grains) and concluded that the CO$_{\rm ice}$ + O$_{\rm ice}$ reaction at the surface of the grain remains the dominant pathway, despite the slow diffusion of atomic oxygen and the activation barrier to this reaction. \\
When comparing the static and dynamical simulations, we found that they could produce similar H$_2$O and CH$_3$OH column densities for an integration time of $10^6$~yr of the static models and visual extinction between 5 and 25 (the range of conditions probed by both sets of simulations). For early times, the static models give much smaller column densities. The dynamical simulations produce larger column densities of CO for A$_{\rm V}$ smaller than approximately 15 than the static ones while they give similar results than the static model for $10^6$~yr for larger A$_{\rm V}$. Last, the static model produces low column densities of CO$_2$ ices, similarly to the low CO$_2$ cases of the dynamic models. This comparison underlines again the need to follow the chemistry during the formation of the clouds, which should happen smoothly. One limitation of our dynamical simulations is that they do not probe the high (A$_{\rm V}$ > 25) visual extinction zone. This is an intrinsic limitation of the physical model that we are using. We however found that percentage of CO, CH$_3$OH, NH$_3$, and CH$_4$ with respect to water was constant for A$_{\rm V}$ larger than 5. Considering\ that the observational results are based on a very small statistical sample and that our simulations are representative -- but still do not reproduce the exact formation trajectories of each observed regions -- we conclude that our simulations  reproduce the observations well. The large statistical sample, which will be provided by JWST will test the model results and provide new constraints to iteratively improve the simulations. \\

\begin{acknowledgements}
The authors acknowledge the CNRS program "Physique et Chimie du Milieu Interstellaire" (PCMI) co-funded by the Centre National d'Etudes Spatiales (CNES). The authors are grateful to Ian Bonnell for providing the SHP numerical simulations. 

\end{acknowledgements}

\bibliographystyle{aa}
\bibliography{bib}

\appendix
\section{Observed physical conditions in L429-C}\label{appendix_l429}

\begin{figure*}[h]
\includegraphics[width=0.49\linewidth]{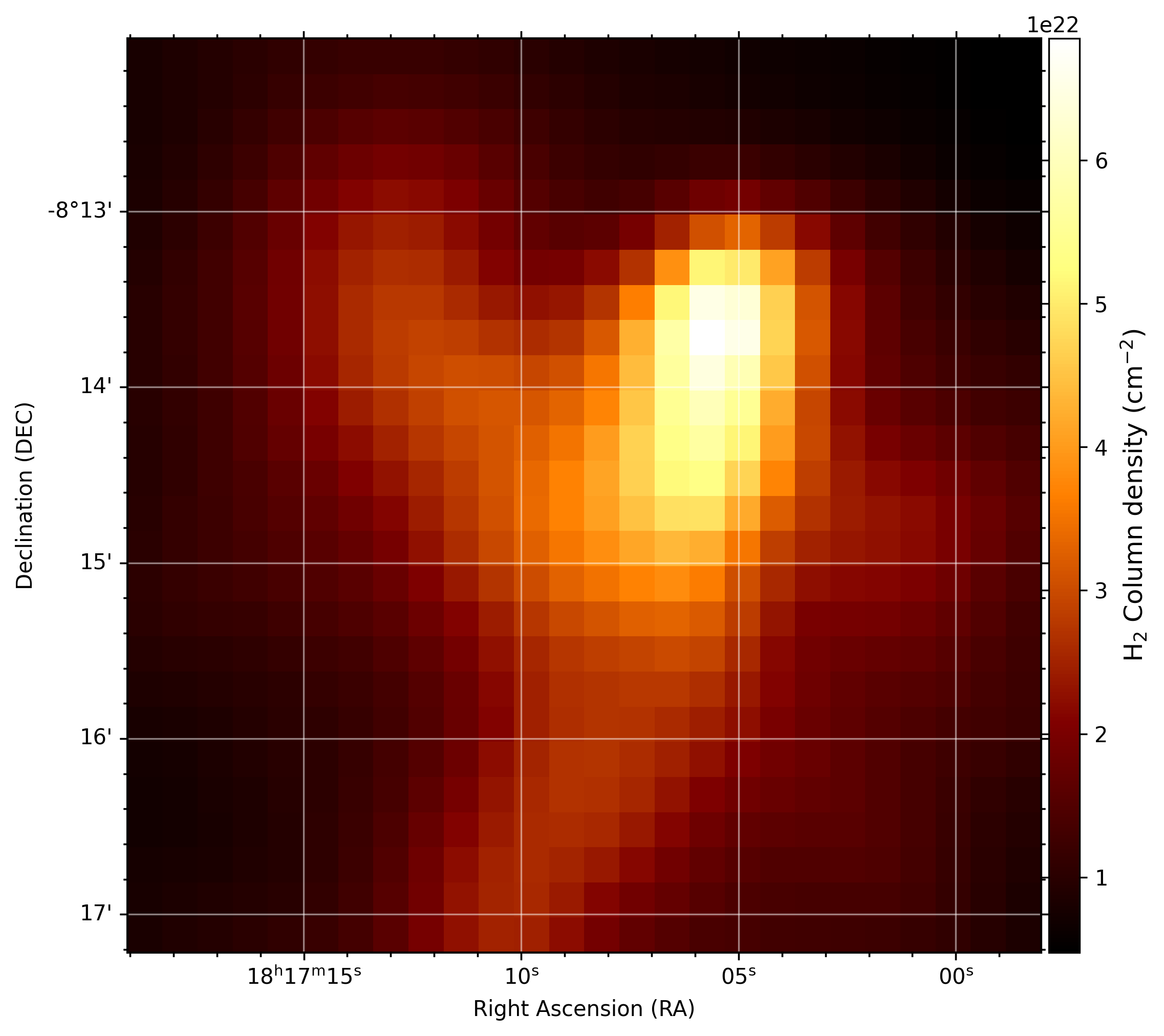}
\includegraphics[width=0.49\linewidth]{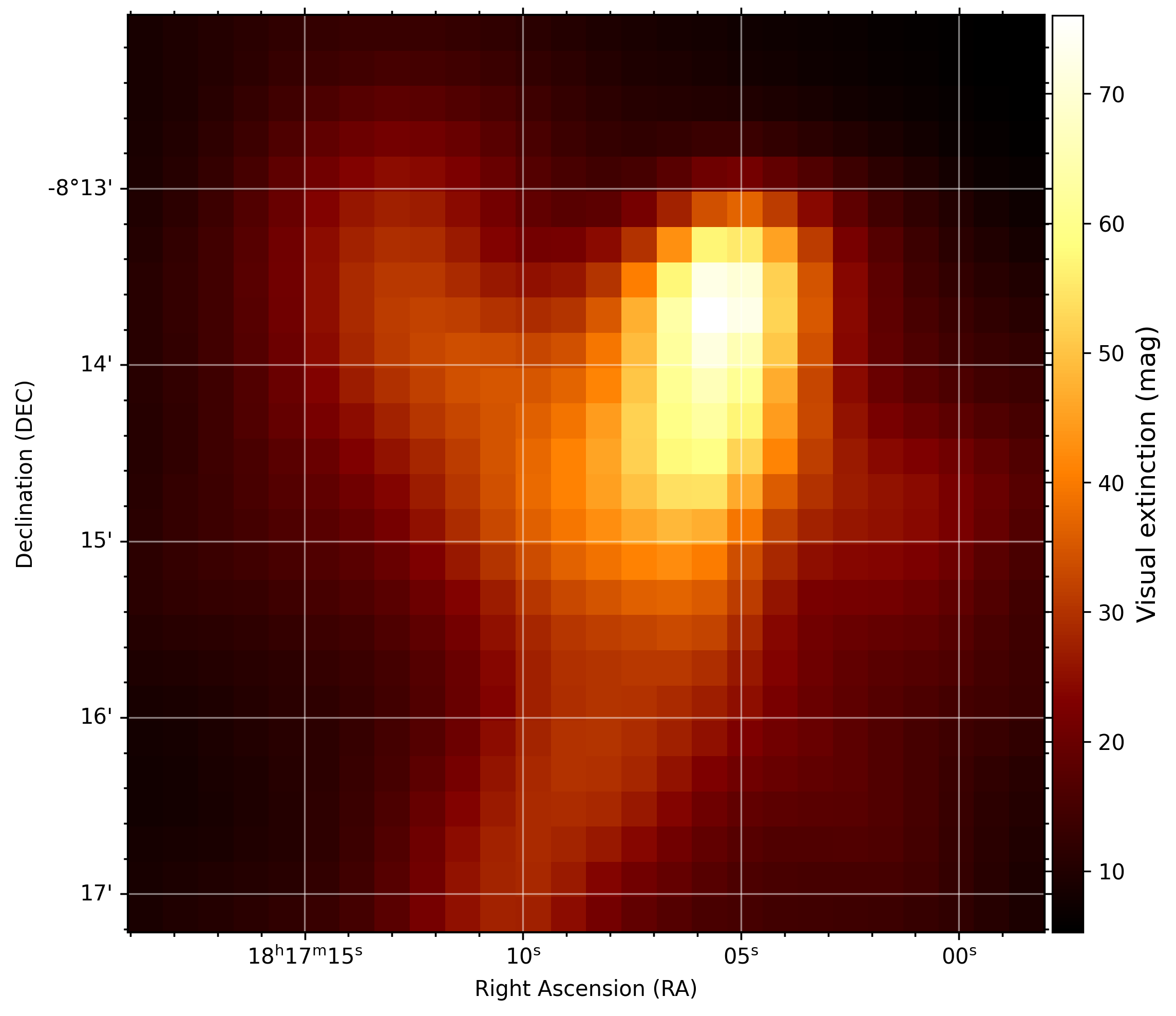}
\includegraphics[width=0.49\linewidth]{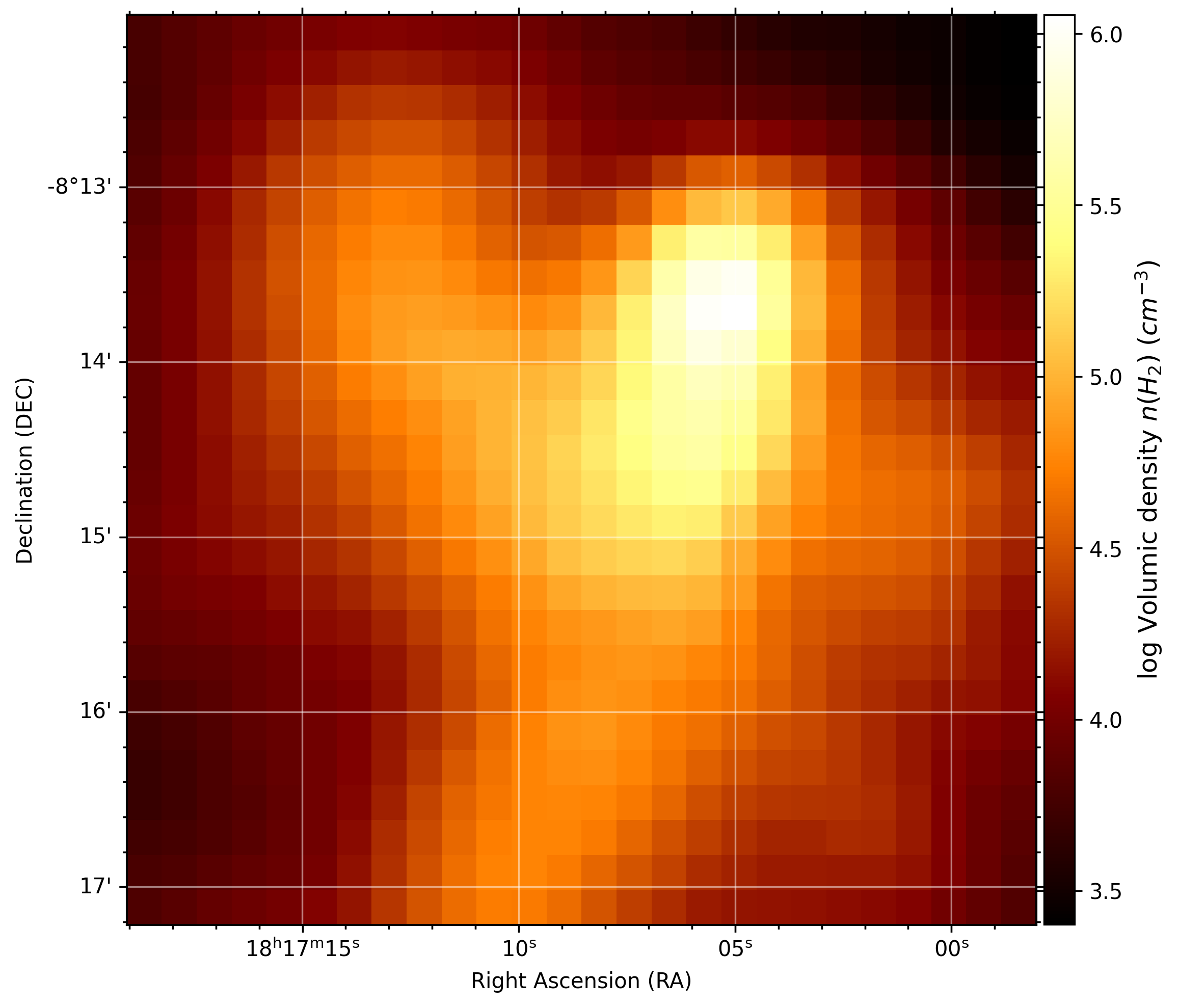}
\includegraphics[width=0.49\linewidth]{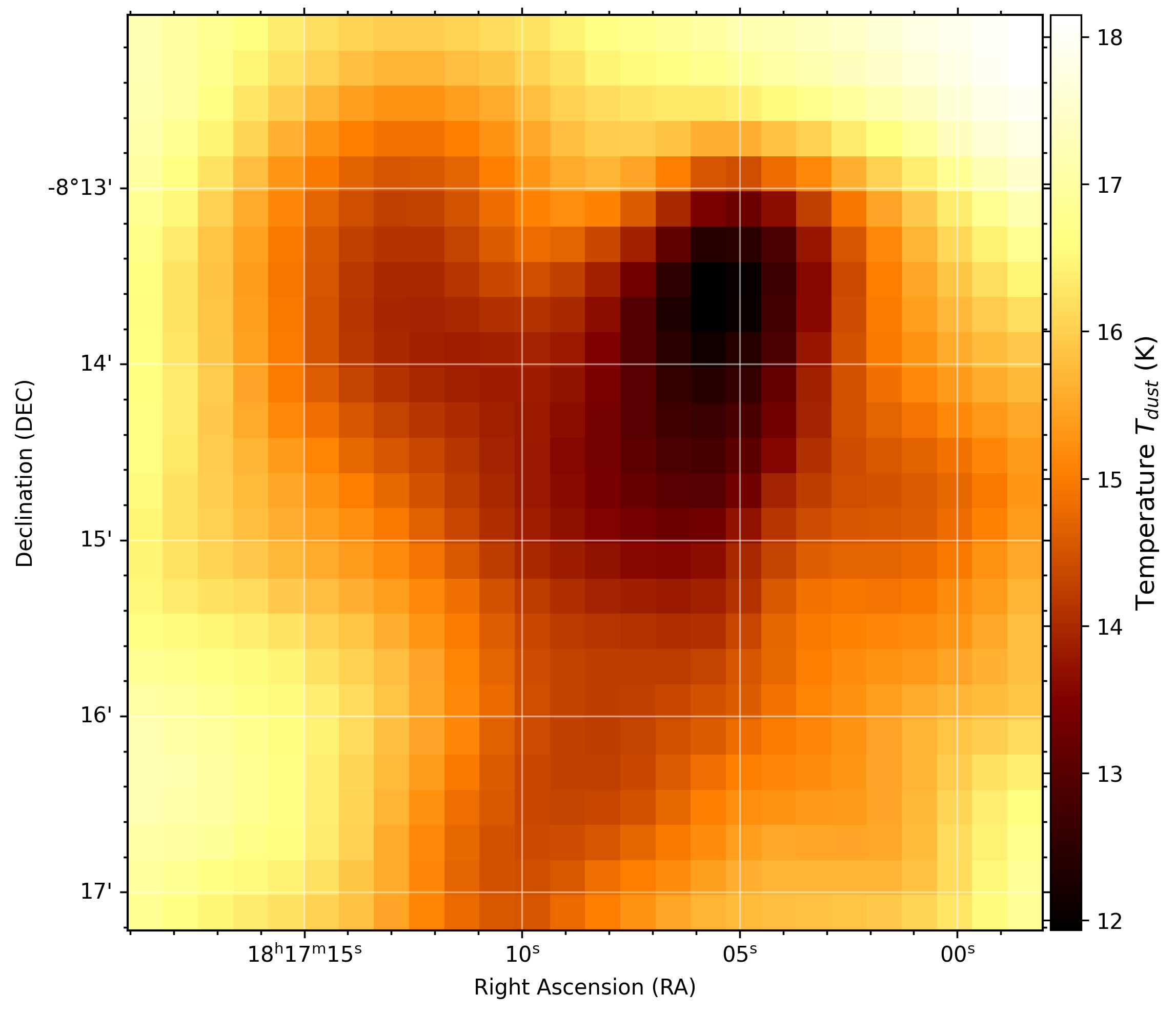}
\caption{H$_2$ column density (cm$^{-2}$, H$_2$ volume density (cm$^{-3}$),  visual extinction, and dust temperature maps of the L429-C region. \label{herschel_maps}}
\end{figure*}

In Fig~\ref{herschel_maps}, we present the physical conditions for the cold core L429 that we used in the static simulations shown in Section~\ref{static_models}.

\clearpage
 
\section{Physical conditions and column densities comparison between static and dynamic simulations}\label{stat-vs-dyn-section}

\begin{figure*}[h]
\includegraphics[width=0.48\linewidth]{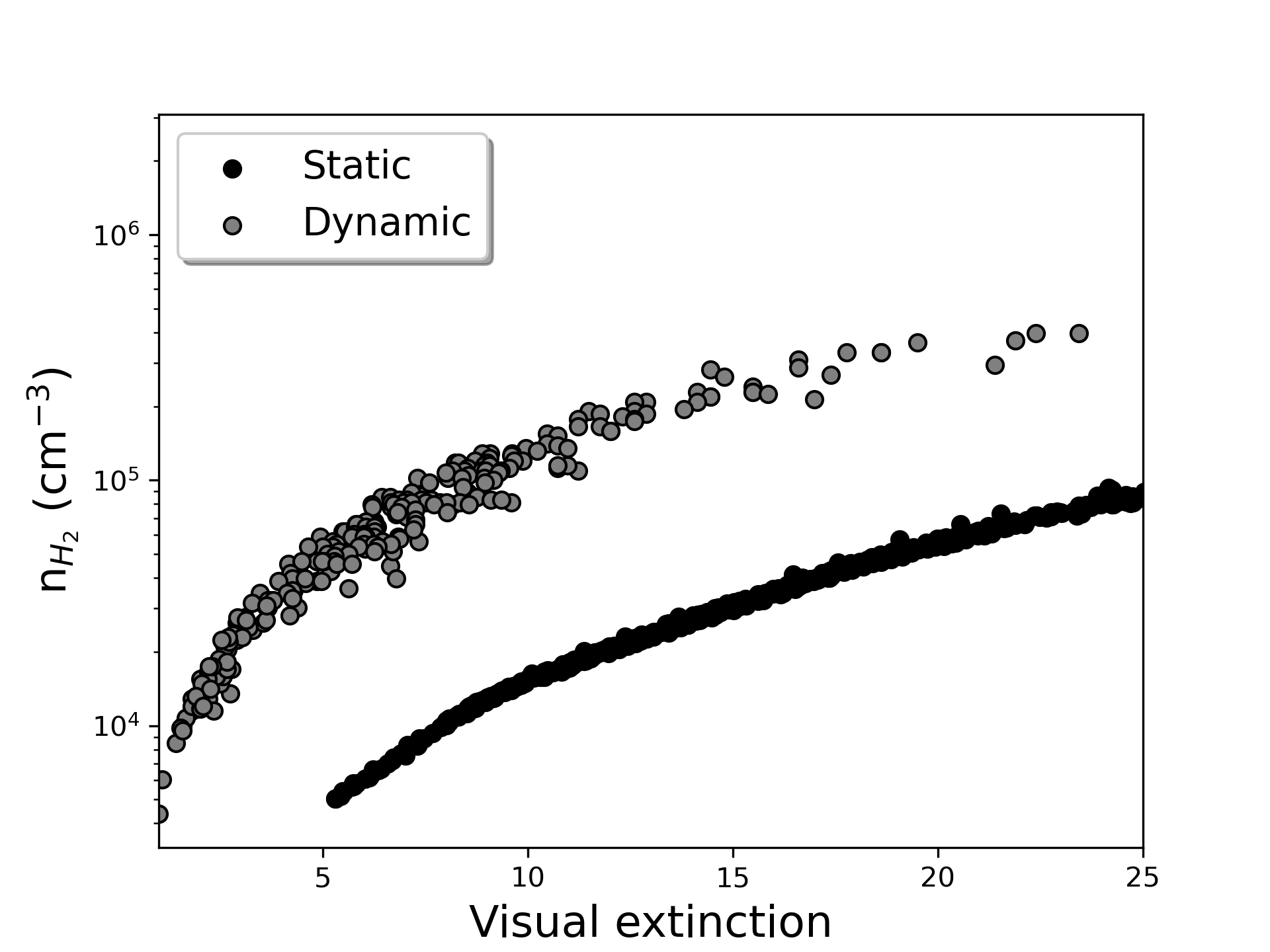}
\includegraphics[width=0.48\linewidth]{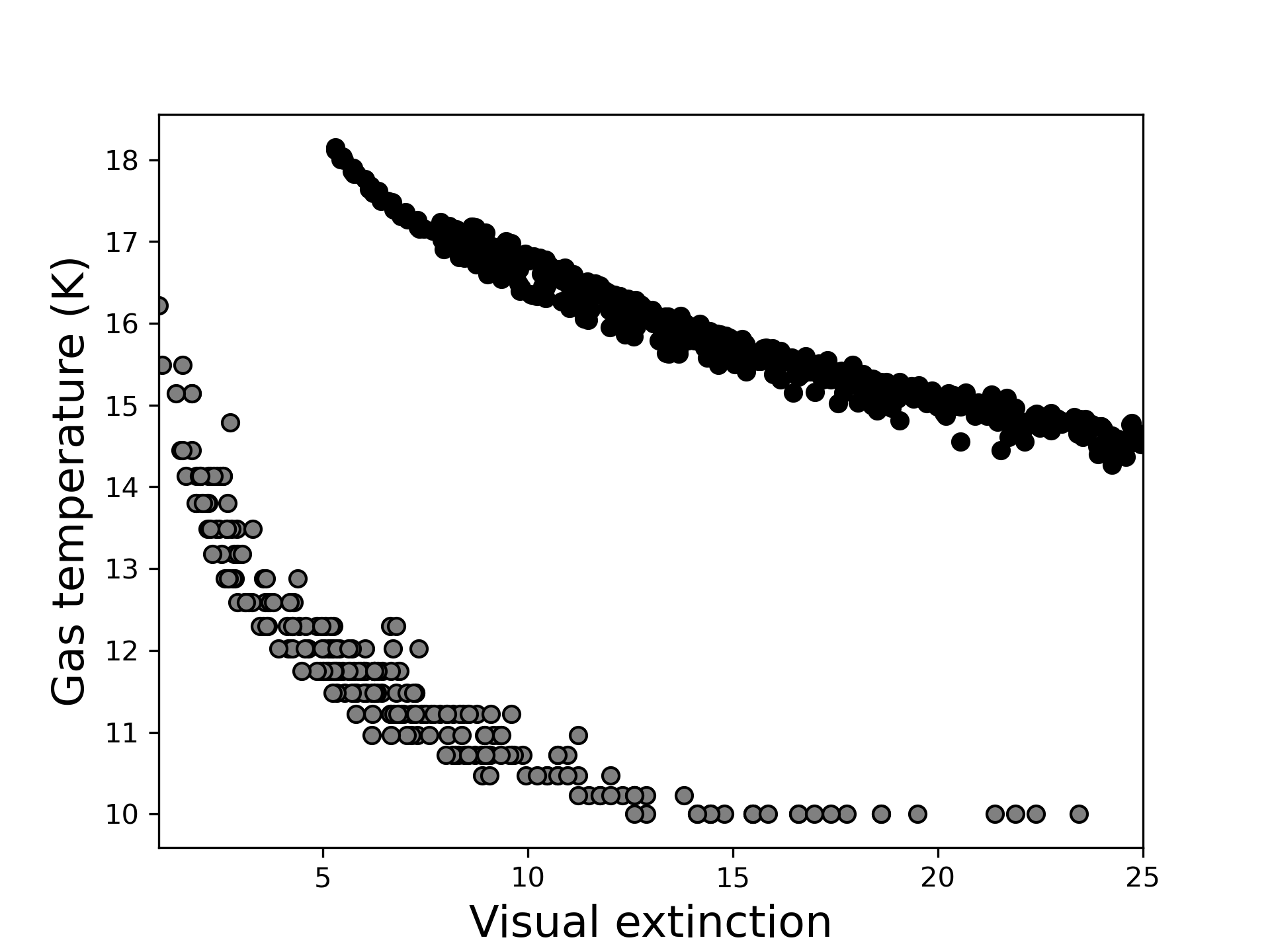}
\caption{H density and gas temperature as a function of visual extinction for the static model (same as Fig.~\ref{static_model2}, black dots) and for the dynamical model of core 0 (same as Fig.~\ref{ice_cold_dens_clump0_all}, gray dots). For both simulations, the grain temperature is computed with Hocuk's formula as a function of visual extinction.
\label{physics-2simus}}
\end{figure*}

\begin{figure*}[h]
\begin{center}
\includegraphics[width=0.48\linewidth]{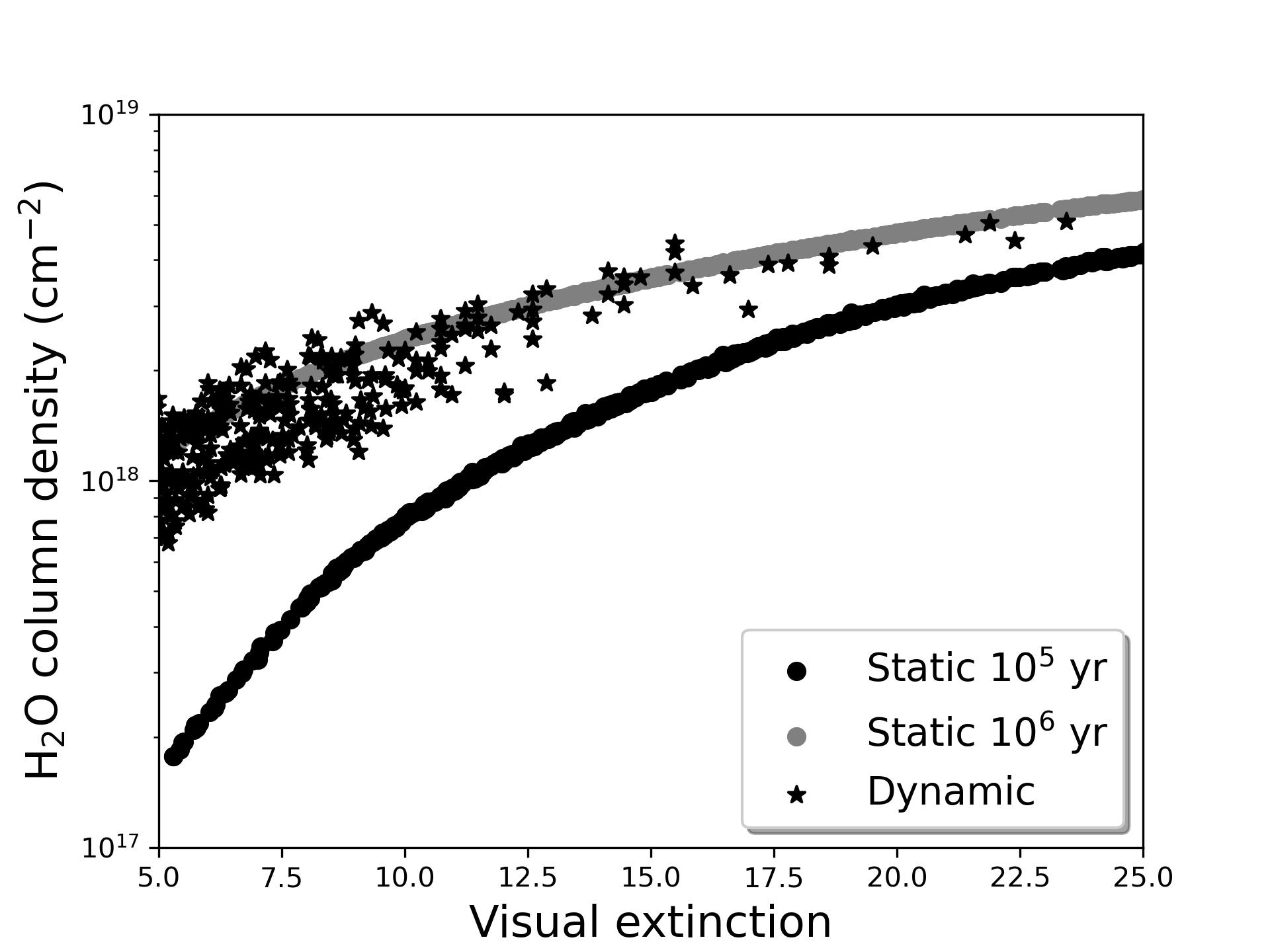}
\includegraphics[width=0.48\linewidth]{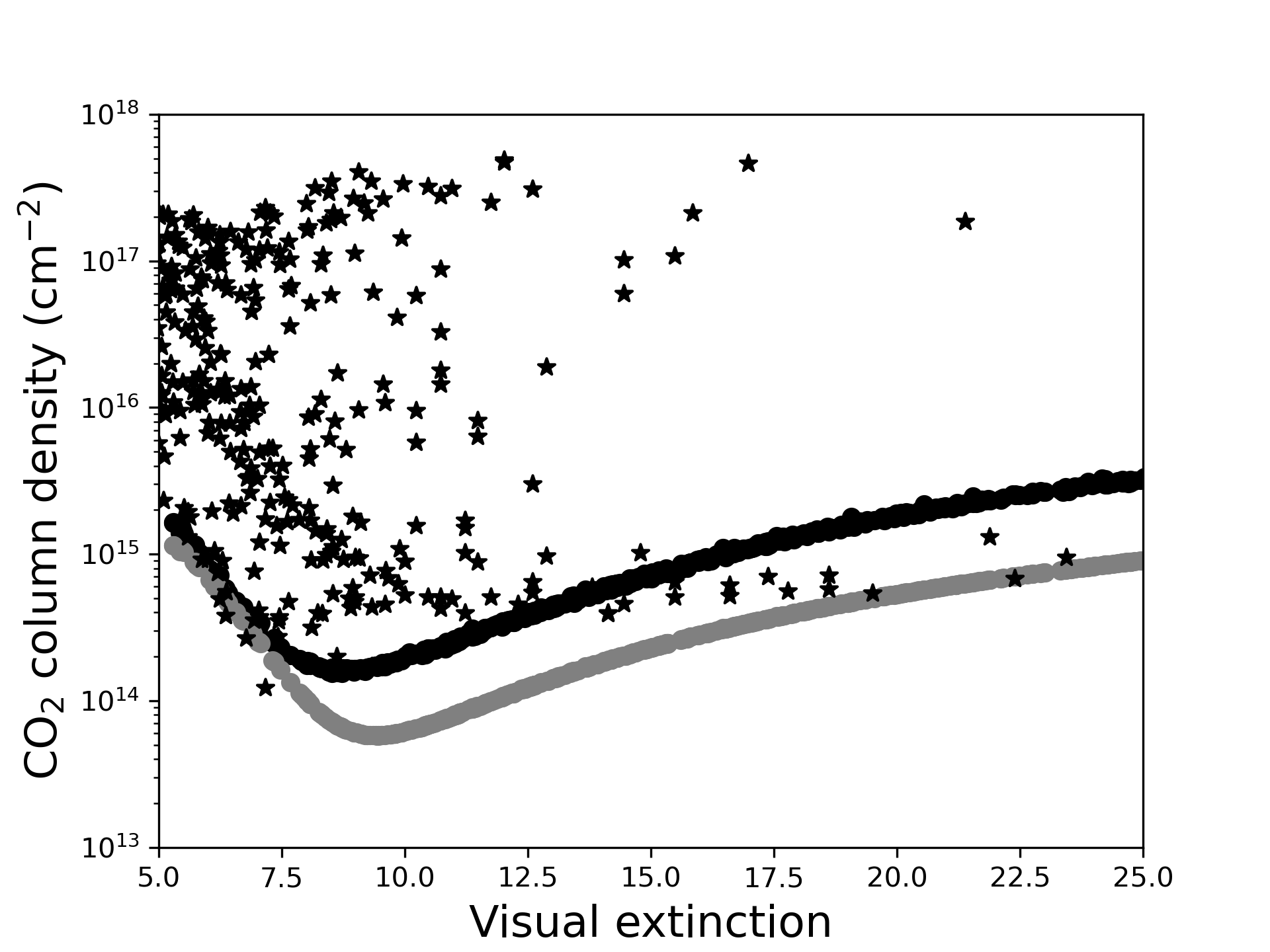}
\includegraphics[width=0.48\linewidth]{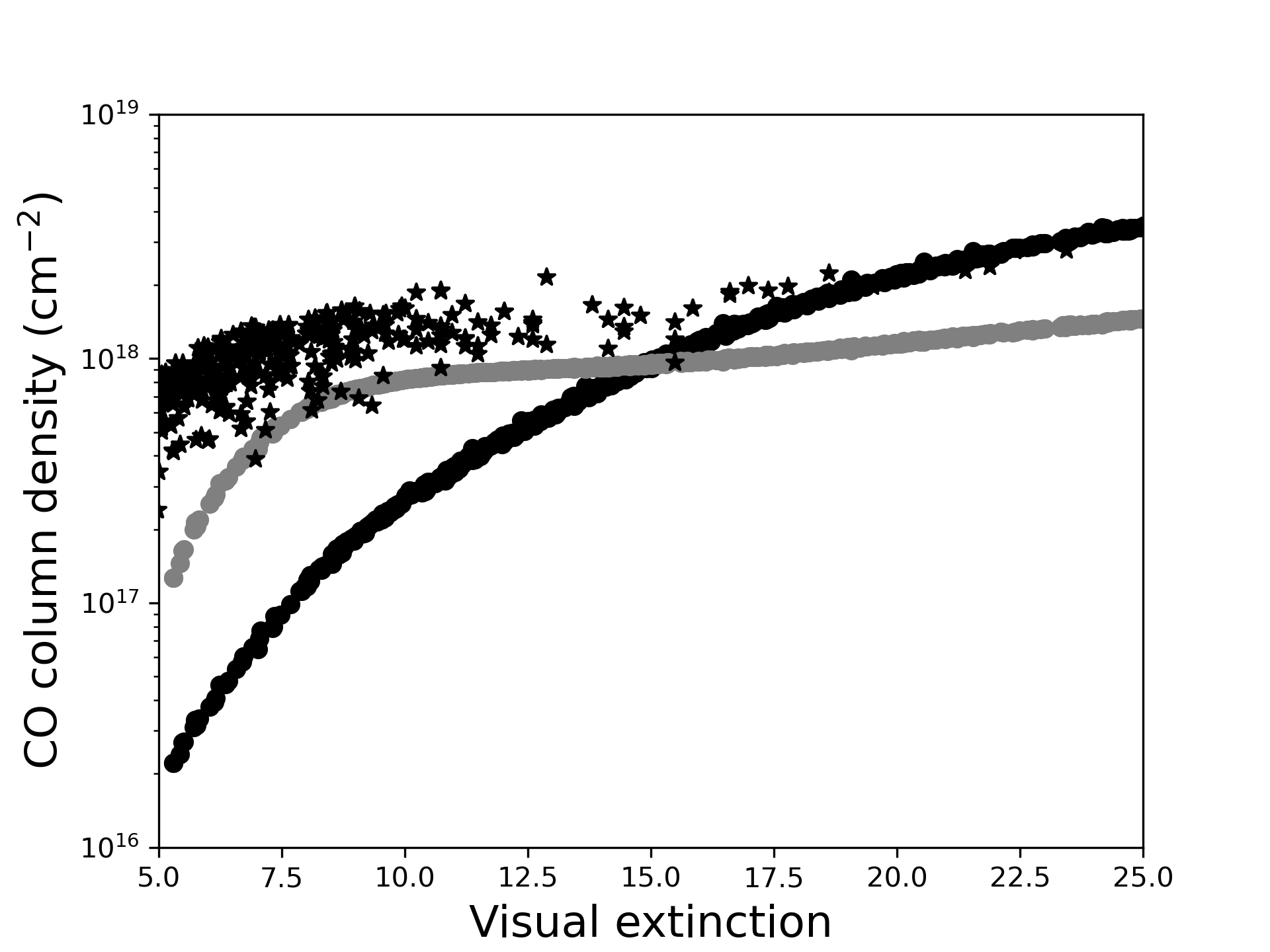}
\includegraphics[width=0.48\linewidth]{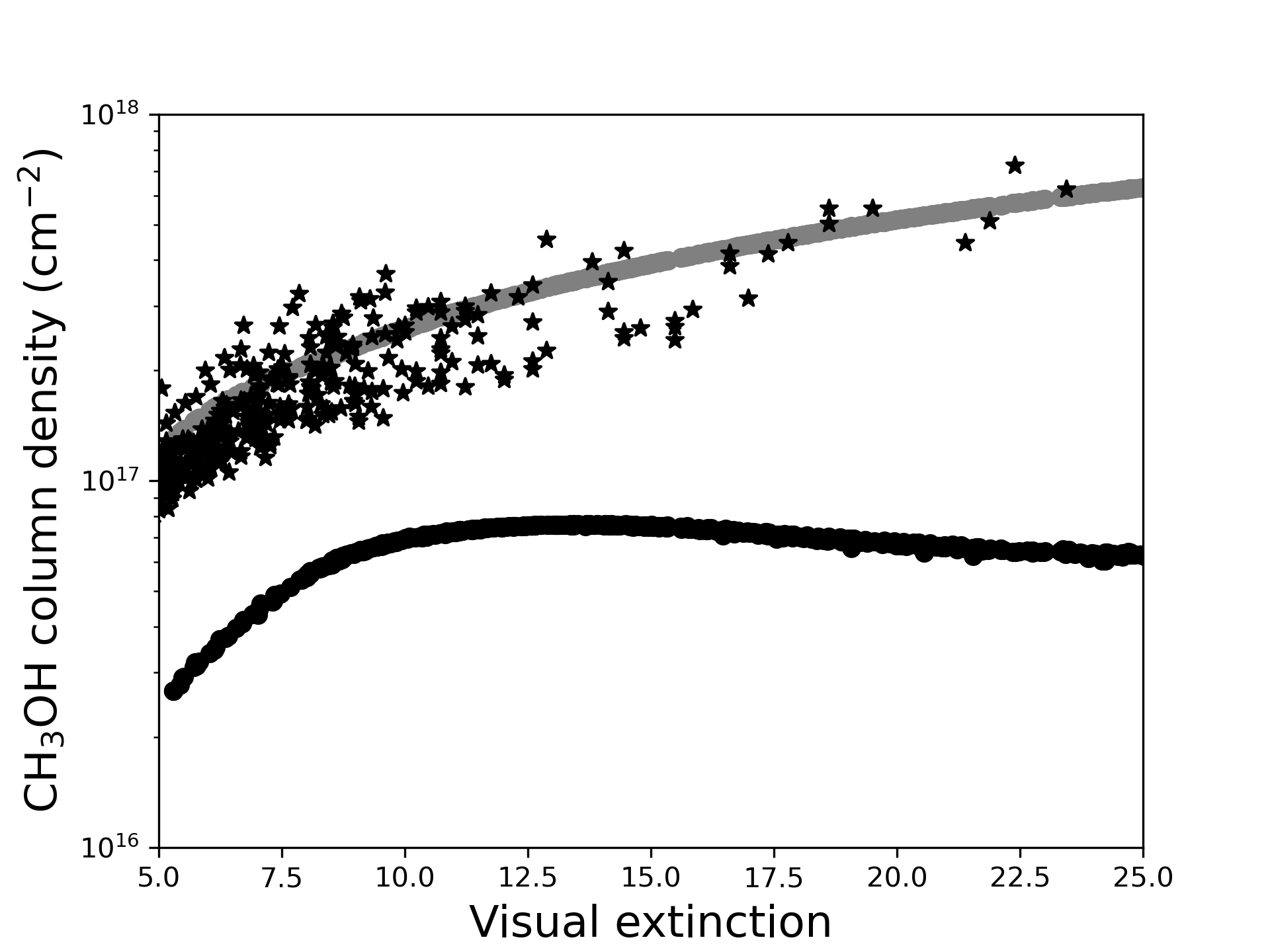}
\caption{Icy molecule column densities as a function of visual extinction for the static model (same as Fig.~\ref{static_model2}, black dots: $10^5$~yr, gray dots: $10^6$~yr) and for the dynamical model of core 0 (same as Fig.~\ref{ice_cold_dens_clump0_all}, black stars). For both simulations, the grain temperature is computed with Hocuk's formula as a function of visual extinction.
\label{coldens-2simus}}
\end{center}
\end{figure*}

In Fig.~\ref{physics-2simus}, we compare the density and gas temperatures used in both sets of simulations (static and dynamical) over the common range of visual extinctions.  
From these simulations, the computed column densities of the main ice components (H$_2$O, CO$_2$, CO, and CH$_3$OH) as a function of A$_{\rm V}$ is shown in Fig.~\ref{coldens-2simus}. Two different times are shown for the static models $10^5$ and $10^6$~yr.

\clearpage
\section{Standard deviation of predicted ice column densities}\label{std-section}

From the dynamical models presented in section~\ref{variability-section} (using the "Hocuk+1" approximation for the dust temperature), we computed, for each core and each of the species H$_2$O, CO$_2$, CO, and CH$_3$OH, the mean column densities and the std for bins of visual extinctions ($<2$, 2-4, 4-6, 6-8, 8-10, 10-15, and $> 15$). In the case of a very large dispersion at a specific A$_{\rm V}$ range, the std is very high and the mean values have no meaning. Some of the large std at high A$_{\rm V}$ are due to a small number of statistical points (for instance core 6 at A$_{\rm V}$ 12.5).

\begin{figure*}[h]
\includegraphics[width=0.33\linewidth]{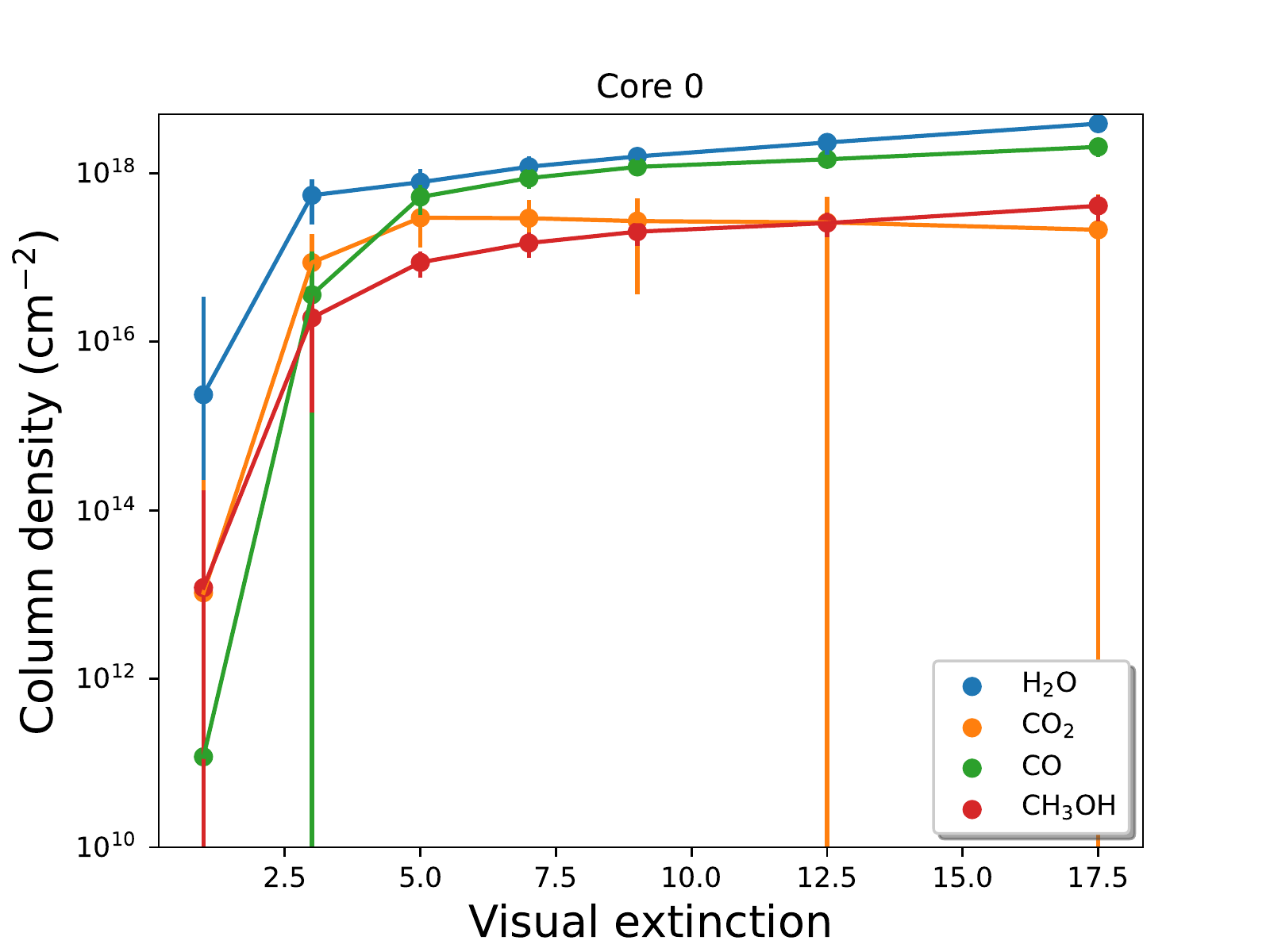}
\includegraphics[width=0.33\linewidth]{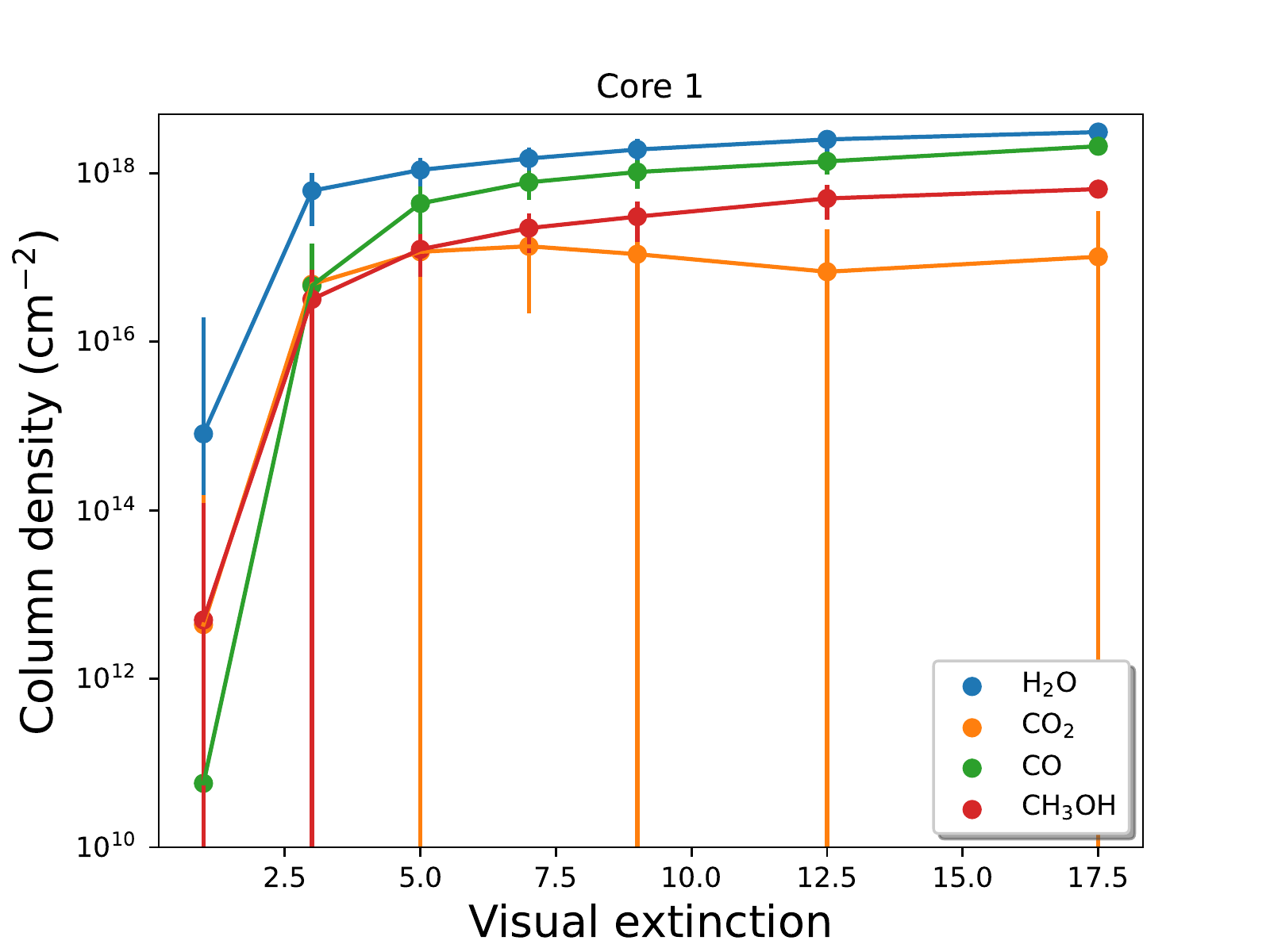}
\includegraphics[width=0.33\linewidth]{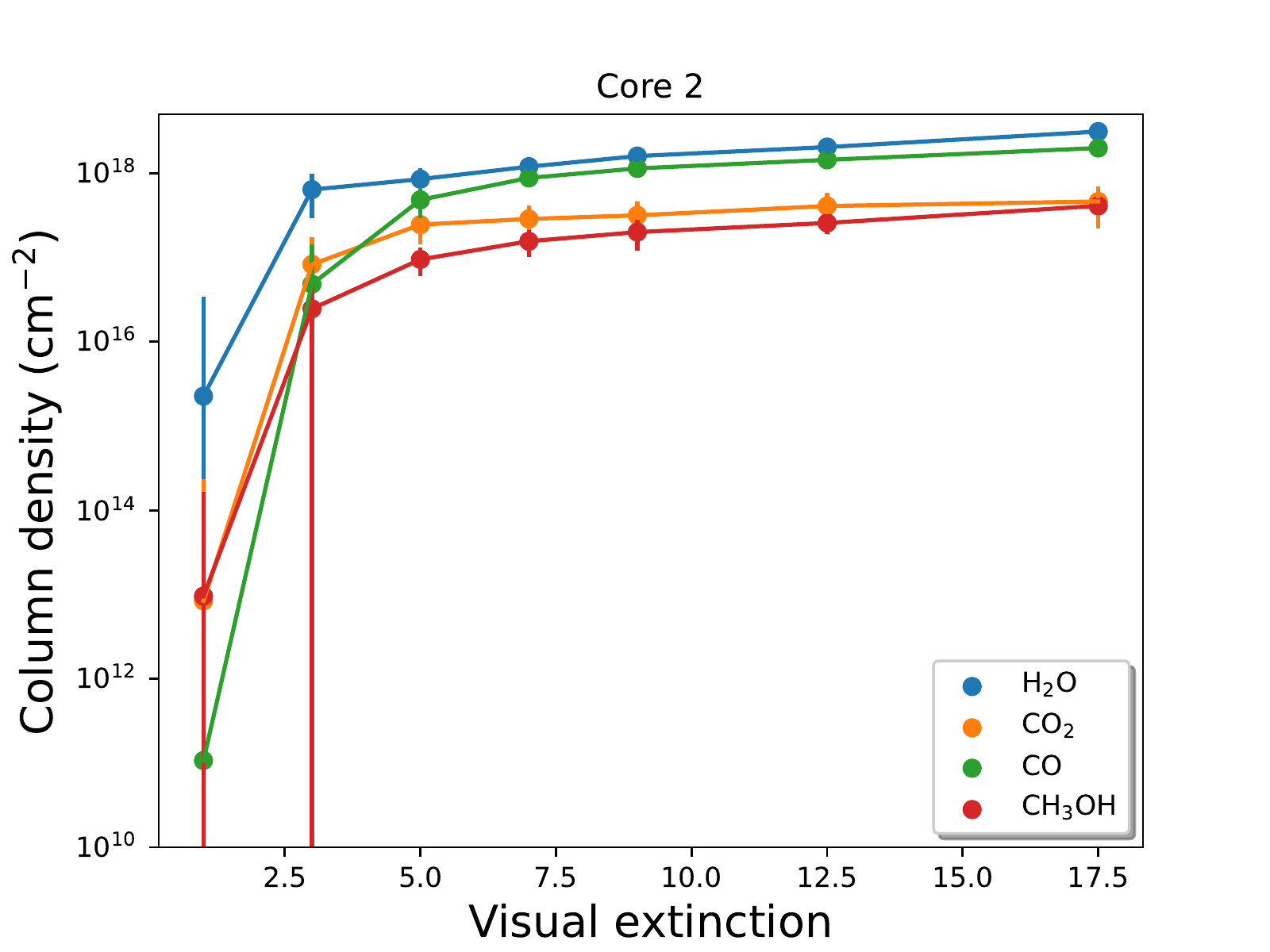}
\includegraphics[width=0.33\linewidth]{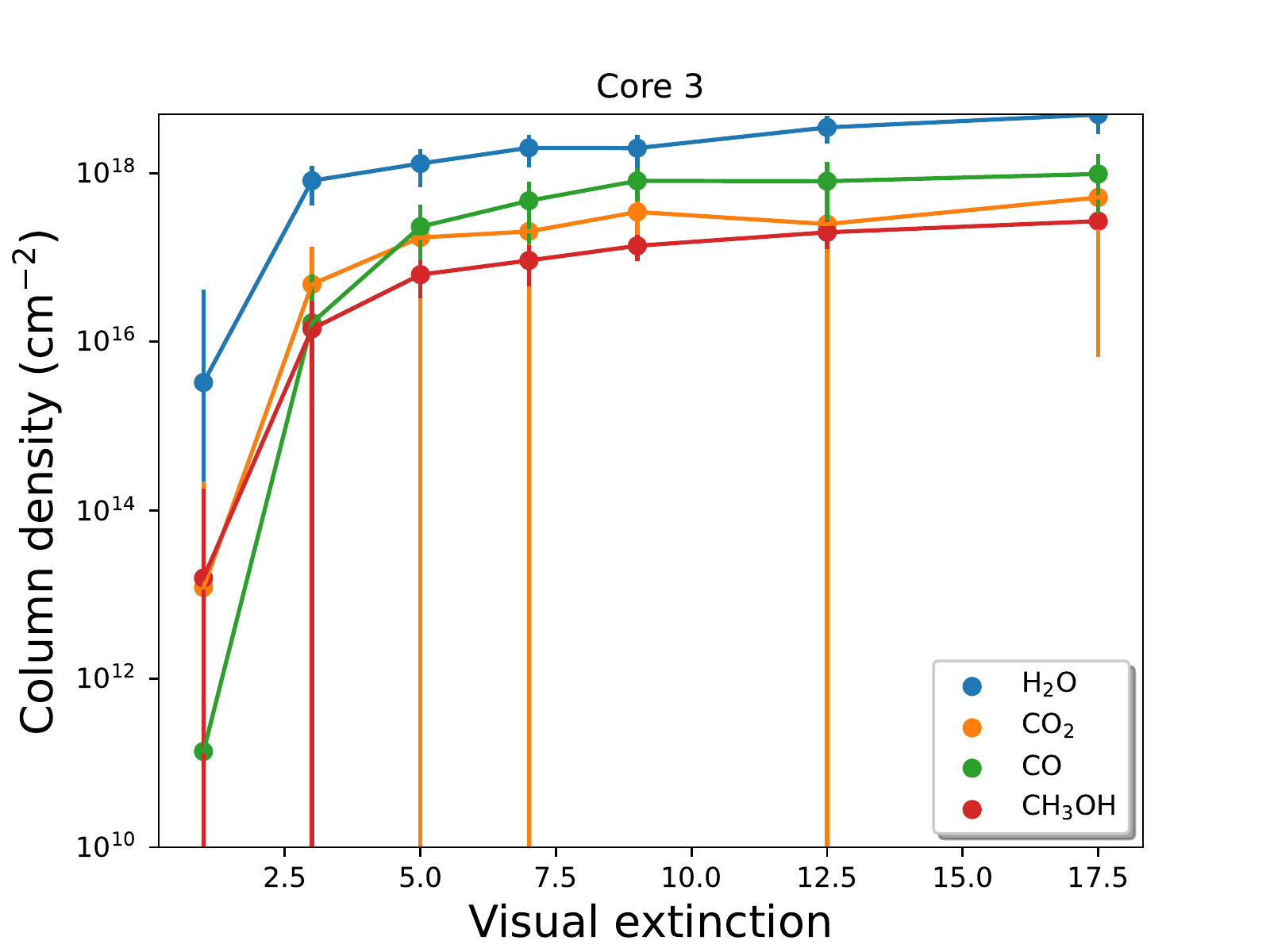}
\includegraphics[width=0.33\linewidth]{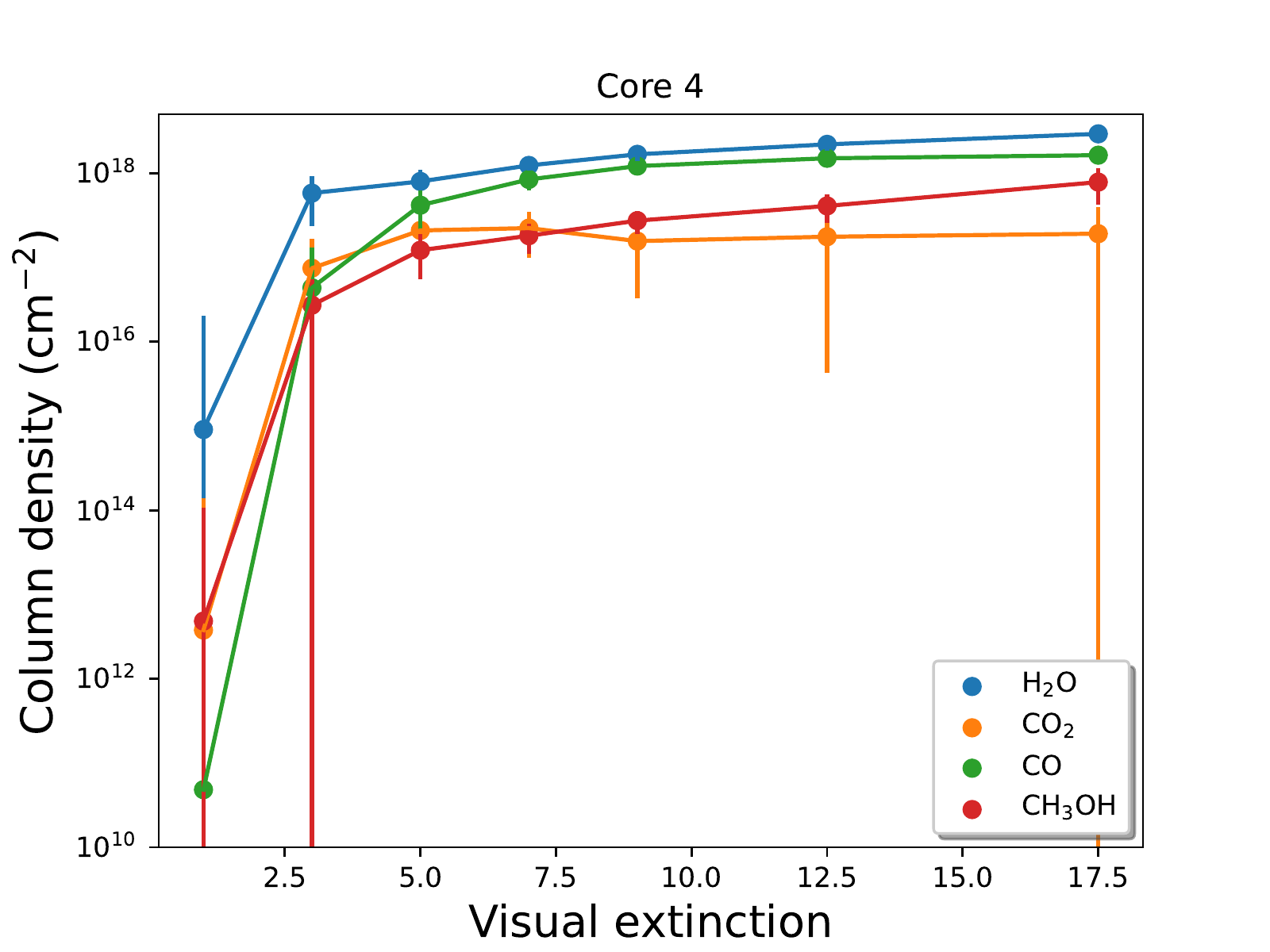}
\includegraphics[width=0.33\linewidth]{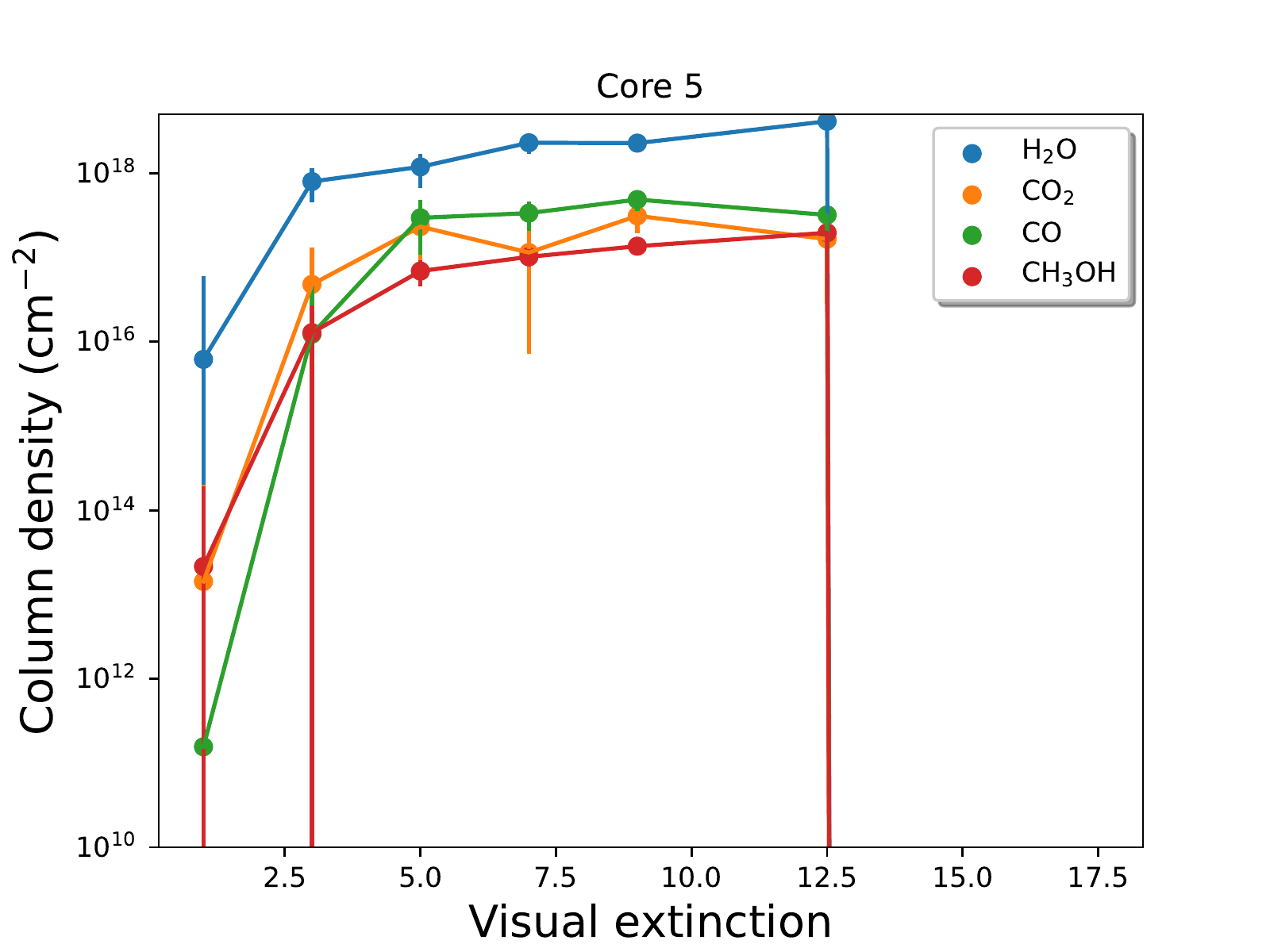}
\includegraphics[width=0.33\linewidth]{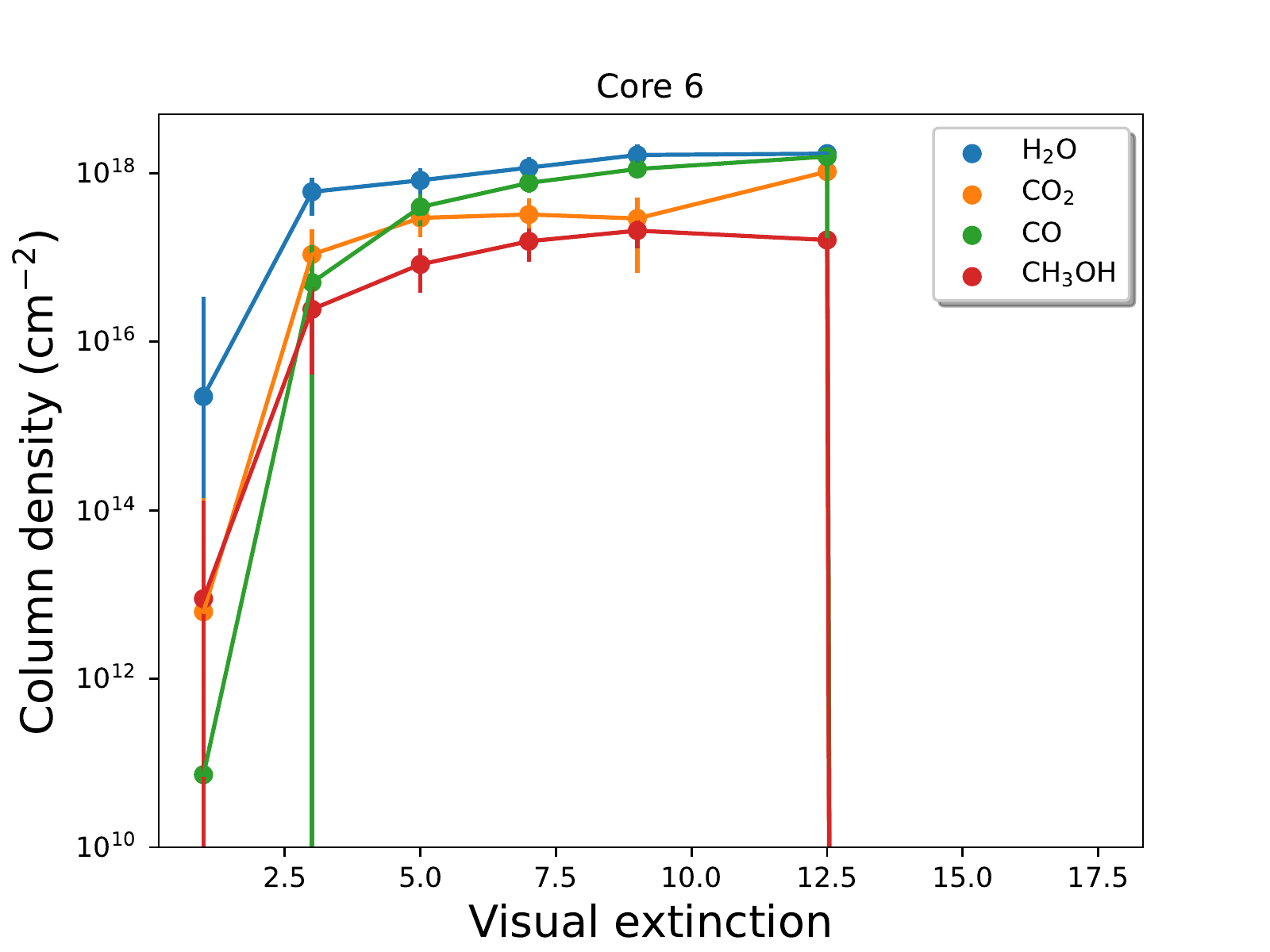}
\includegraphics[width=0.33\linewidth]{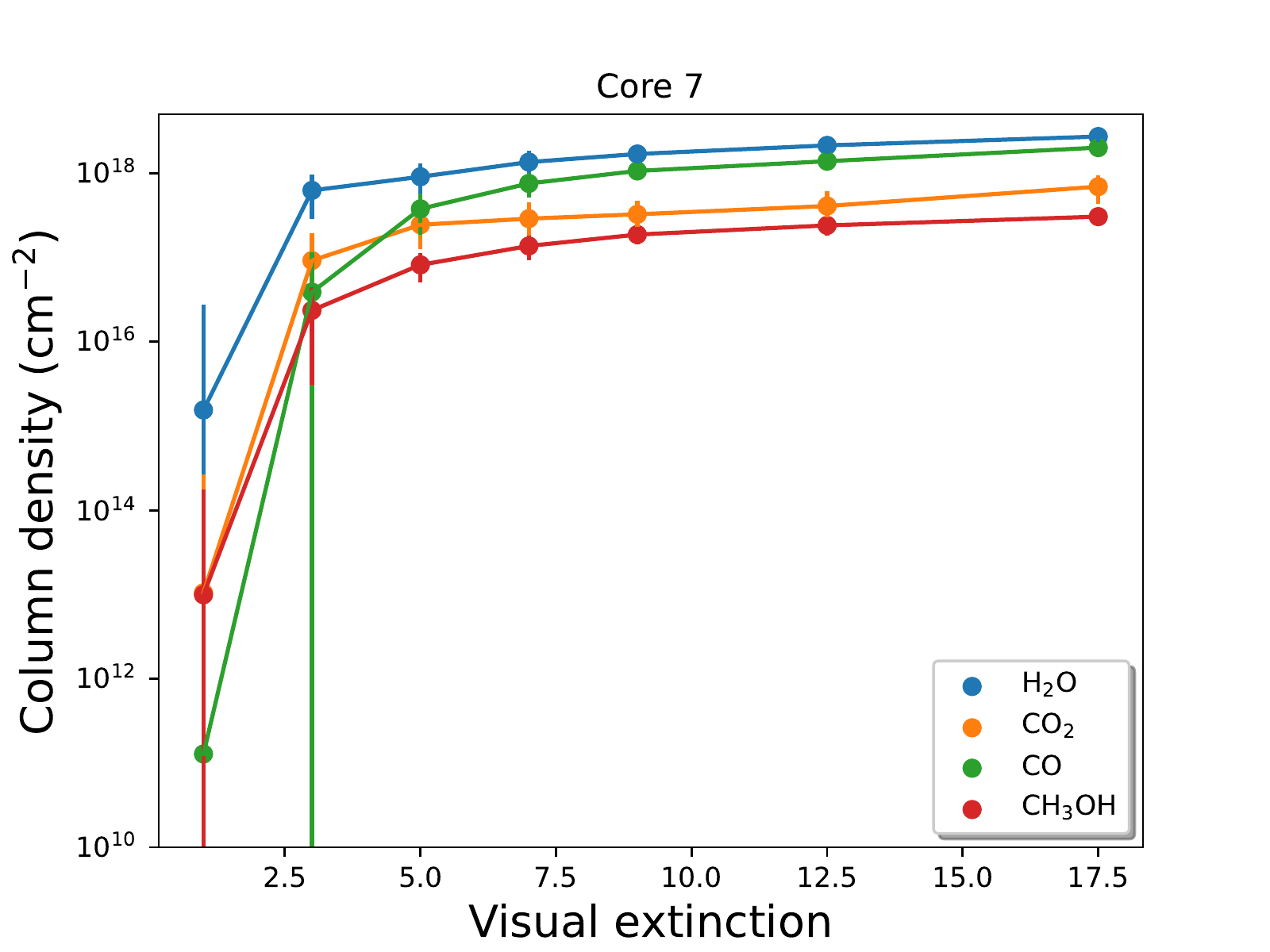}
\includegraphics[width=0.33\linewidth]{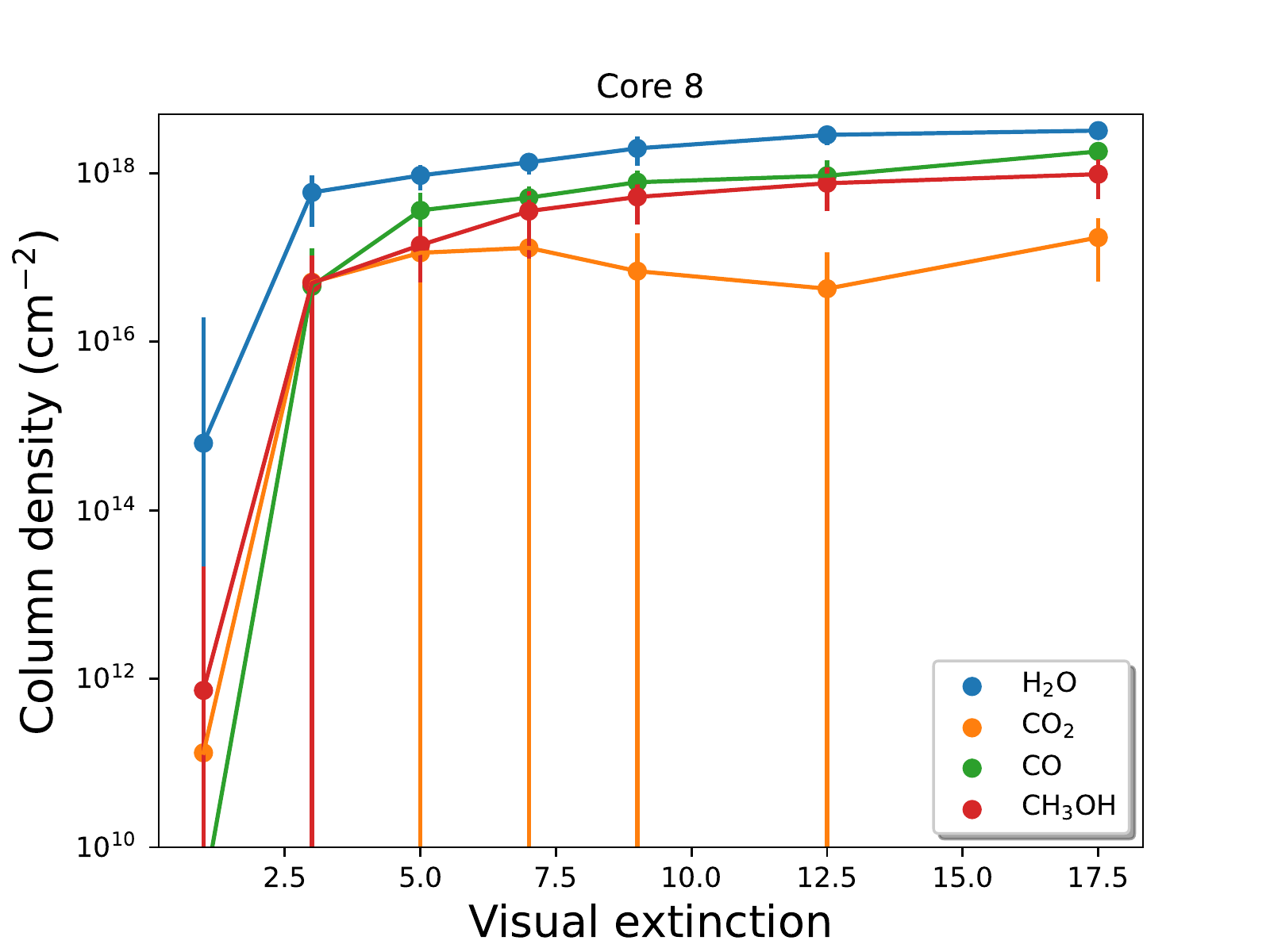}
\includegraphics[width=0.33\linewidth]{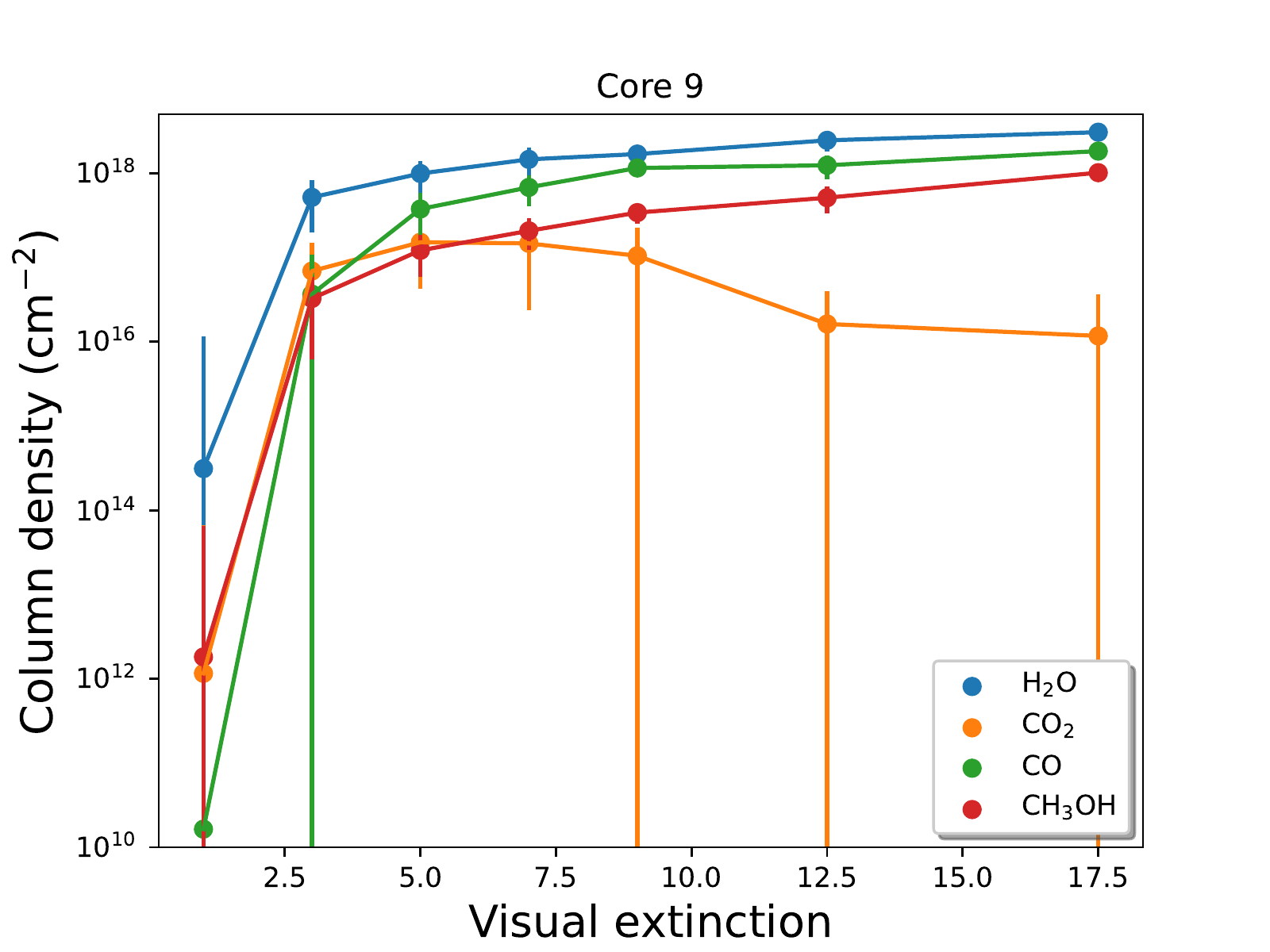}
\includegraphics[width=0.33\linewidth]{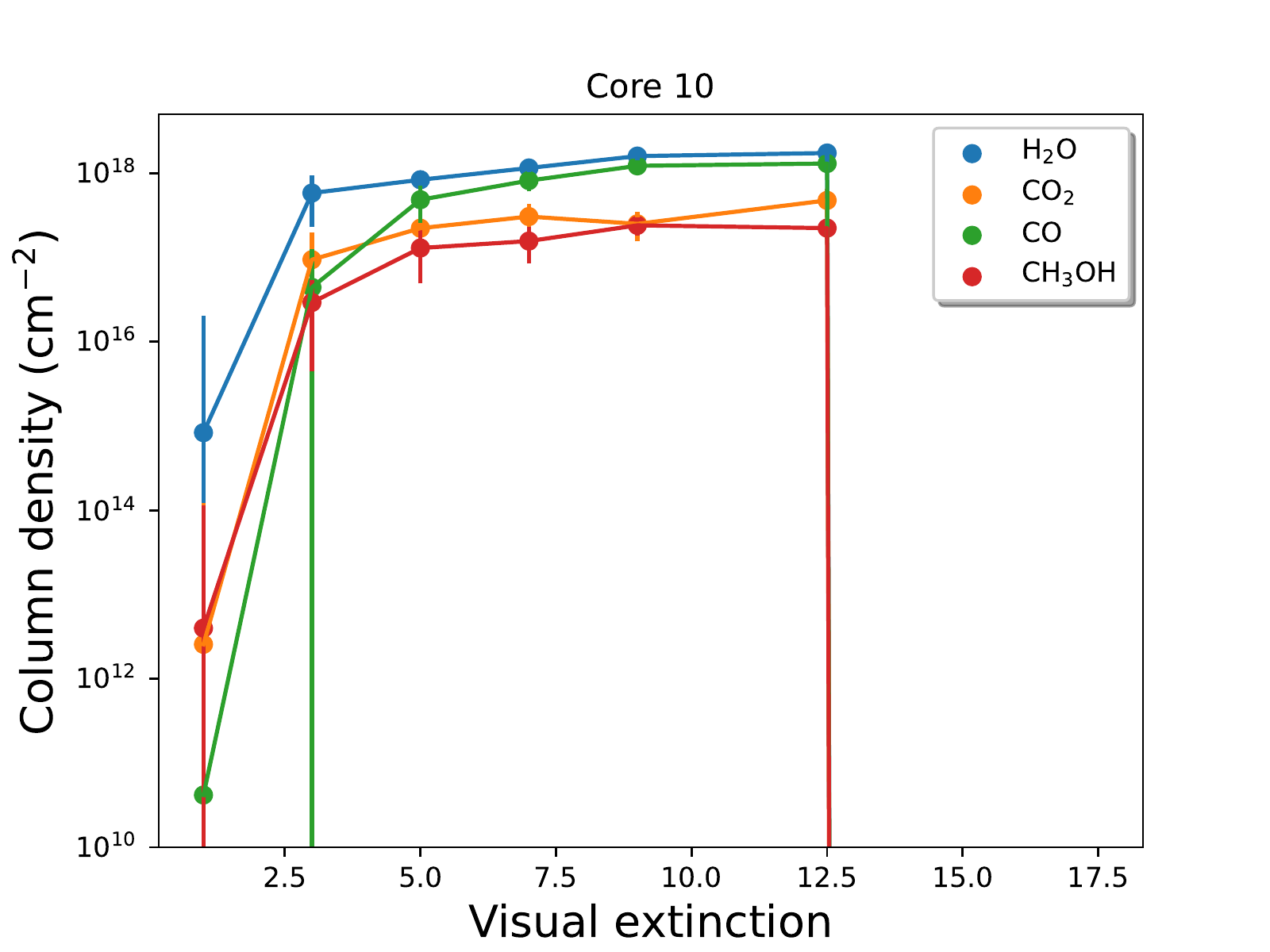}
\includegraphics[width=0.33\linewidth]{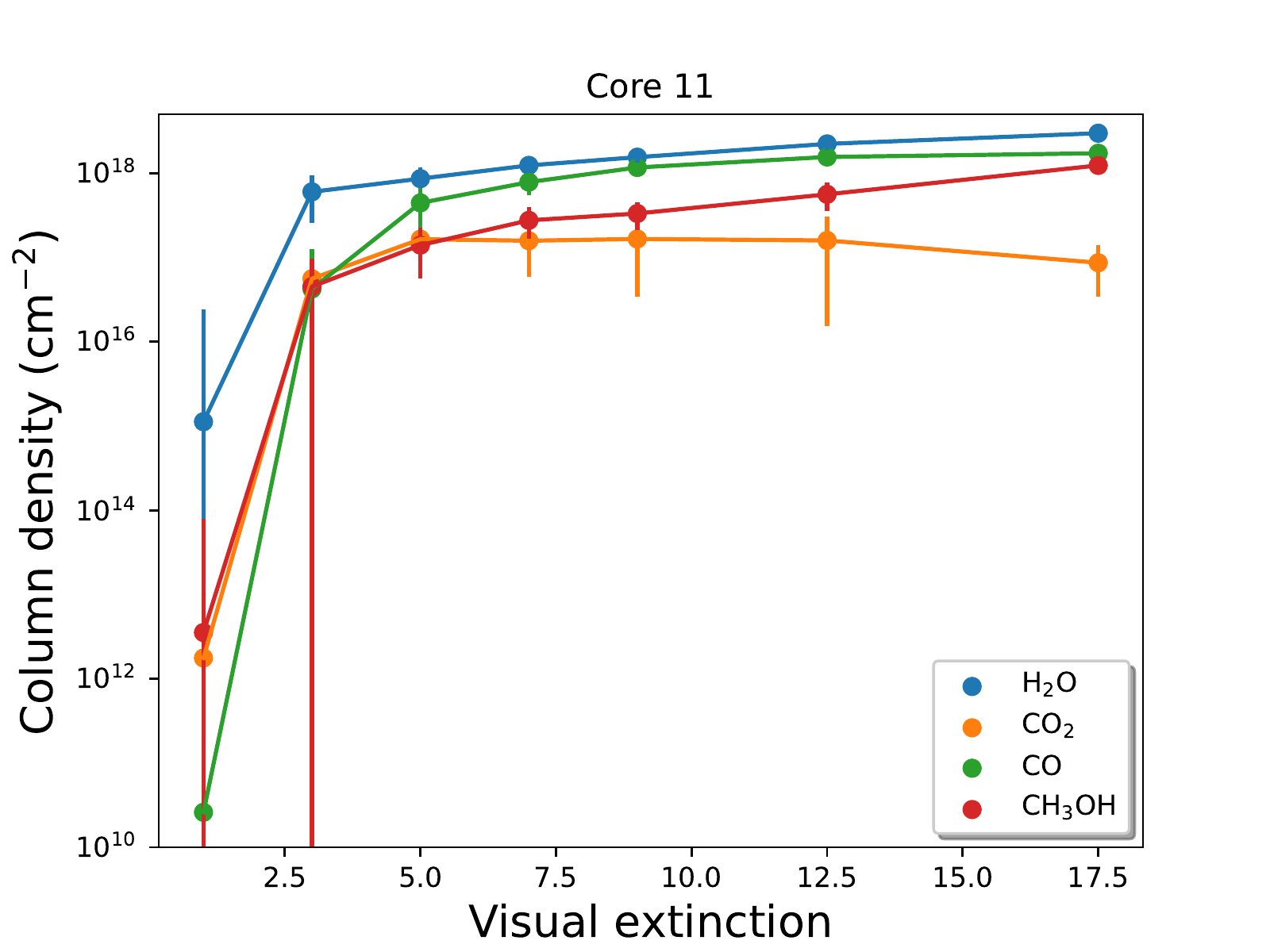}
\caption{Mean column densities and standard deviations as a function of A$_{\rm V}$ for each core and the main ice constituents. 
\label{std}}
\end{figure*}

\end{document}